\documentclass[twocolumn]{aastex7}
\usepackage{amsmath}
\usepackage{upgreek}
\usepackage{multirow}   
\usepackage{booktabs} 

\received{October 29, 2025}
\revised{June 1, 2026}
\submitjournal{ApJ}
\graphicspath{{./}}

\begin{document}

\title{Weak-Lensing Detection of Intercluster Filaments in Three Nearby Cluster Systems}

\correspondingauthor{Rahul Shinde}
\email{rahul\_shinde@brown.edu}

\author[0000-0002-7342-3229]{Rahul Shinde}
\affiliation{Department of Physics, Brown University, 182 Hope Street, Box 1843, Providence, RI 02912, USA}
\email{rahul\_shinde@brown.edu}

\author[0000-0003-0751-7312]{Ian Dell'Antonio}
\affiliation{Department of Physics, Brown University, 182 Hope Street, Box 1843, Providence, RI 02912, USA}
\email{ian\_dellantonio@brown.edu}  

\begin{abstract}
Direct detection of intercluster filaments is challenging due to their low surface density, resulting in a weak deflection field. We present weak-lensing detections of intercluster filaments using wide-field Dark Energy Camera (DECam) observations from the Local Volume Complete Cluster Survey (LoVoCCS). A matched-filter method was applied to identify filamentary structures in three nearby ($z < 0.1$) systems centered on Abell 401, Abell 2029, and Abell 3558. We discover two filaments ($> 3\sigma$) in each system, with the strongest detections ($5.2\sigma - 5.8\sigma$) around Abell 401 and Abell 2029. In particular, we report the first robust weak-lensing detections $(\gtrsim 5 \sigma)$ of the intercluster bridges connecting the cluster pairs Abell 401/399, Abell 2029/2033, Abell 2029/SIG, and Abell 3558/3556. Adopting a filament convergence model motivated by numerical simulations, we infer the maximum convergence ($\kappa_0$) and characteristic width ($h_{\mathrm{c}}$) for all six filaments, yielding  $\kappa_0 \sim 0.016 - 0.040$ and $h_{\mathrm{c}} \sim 0.23 - 0.43 \ \mathrm{Mpc}$. The performance of the matched-filter technique is validated using mock shear catalogs and further tested on a null field around Abell 2351. We explore the potential of using the B-mode lensing signal of filaments to suppress cluster-induced shear contamination. We also quantify the biasing effect of closely separated terminal clusters to the filament signal. These results demonstrate the feasibility of directly mapping dark matter filaments with current and future wide-field weak-lensing datasets.
\end{abstract}




\section{Introduction} 
Predictions from N-body simulations of the standard $\Lambda$CDM cosmological model \citep{springel2005}, supported by observational evidence from spectroscopic surveys \citep{colless2001, jones2009, zehavi2011}, confirm that matter in the universe is organized in a \emph{cosmic web} structure. This hierarchical structure consists of near-empty regions (voids), enclosed by high-density walls (sheets), which intersect to form linear structures (filaments) that transport matter into dense nodes (clusters) \citep{bond1996}. Filaments are estimated to contain roughly half of the matter in the universe \citep{cautun2014}. By regulating cluster growth and sustaining galaxy formation, filaments offer a unique avenue to probe structure formation history, constrain the nature of dark matter, and infer properties of the underlying cosmology.

Cosmic filaments have been studied using various complementary probes. X-ray and Sunyaev-Zel'dovich (SZ) observations trace the diffuse baryonic component -- the warm hot intergalactic medium (WHIM) -- distributed along filaments \citep{werner2008, radiconi2022, hincks2022, mirakhor2022, migkas2025}. Spectroscopic redshift surveys \citep{colless2001, jones2009, zehavi2011}, on the other hand, map the spatial distribution of luminous galaxies that outline the filamentary skeleton. However, both approaches have limitations; redshift surveys are limited to bright galaxies and are affected by galaxy bias, whereas X-ray and SZ analyses rely on assumptions about the thermodynamic state of the filamentary gas. As a result, they only provide partial and biased information about the underlying matter distribution. In contrast to baryonic tracers, gravitational lensing -- particularly through convergence mapping \citep{kaiser2003} -- offers the most unambiguous way to measure the total projected mass distribution, luminous and dark matter alike, in filaments. 

The direct detection of filaments via their lensing signal, however, remains challenging and has only been reported in a small number of cases \citep{dietrich2012, jauzac2012, nature2024}. The intrinsically low-density contrast of filaments results in a weak deflection field. The signal is dominated by shape noise and further obscured by contamination from neighboring clusters and large-scale structure along the line of sight. Consequently, most studies have focused on inferring ensemble properties of filaments through stacking methods \citep{clampitt2016, epps2017, kondo2020, xia2020}. Robust detection of individual filaments, therefore, requires targeted exploitation of their linear morphology and a rigorous treatment of noise. In this paper, we implement the matched-filter technique introduced in \citet{maturi2005} to detect the presence of filamentary structures in three systems centered on (i) Abell 401, (ii) Abell 2029, and (iii) Abell 3558 (Shapley Supercluster Core). 

The remainder of the paper is structured as follows. Section \ref{sec:formalism} reviews the fundamentals of weak-lensing theory and introduces the filament shear model. Section \ref{sec:problem} discusses the challenges associated with detecting filaments via their shear signature and the impact of noise. In Section \ref{sec:data_model}, we present a shear decomposition scheme that leverages filament geometry and motivates a filter-based approach. Section \ref{sec:method} details the construction of the optimal matched filter and examines the impact of noise on our analysis. The efficacy of the matched-filter method is demonstrated in Section \ref{sec:implementation} using mock data. Here, we also explore the potential of using the B-mode signal for filament detection in conjunction with the E-mode signal. Section \ref{sec:data} describes the LoVoCCS dataset and our results are presented in Section \ref{sec:results}. Finally, we summarize our conclusions and outline directions for future work in Section \ref{sec:conclusions}.

Throughout this paper, we assume a flat $\Lambda$CDM cosmology with $H_0 = 71 \, \mathrm{km \ s}^{-1} \, \mathrm{Mpc}^{-1}$ and $\Omega_{\mathrm{m}} = 0.2648$, consistent with the cosmology adopted in the LoVoCCS study \citep{fu2022, fu2024} providing the dataset used in this work.

\section{Weak-Lensing Formalism} \label{sec:formalism}
In this section, we provide a brief overview of the weak lensing theory to establish a connection between lensing quantities and observables (for more details, see \citealt{bartelmann2000}). We consider the 2-D lensing potential $\psi(\boldsymbol{x})$, corresponding to the surface mass density $\Sigma(\boldsymbol{x})$, under the thin-lens approximation:
\begin{equation} \label{eq:lensing_potential}
\psi(\boldsymbol{x}) = \frac{4G}{c^2} \frac{D_{\mathrm{l}}D_{\mathrm{ls}}}{D_{\mathrm{s}}} \int \mathrm{d}^2 {x}' \, \Sigma(\boldsymbol{x}') \ln |\boldsymbol{x} - \boldsymbol{x}'|.
\end{equation}
Here, $D_{l}$, $D_{s}$, and $D_{ls}$ denote the angular diameter distances between the observer and lens, the observer and source, and the lens and source, respectively. $G$ is the gravitational constant and $c$ is the speed of light. ${\boldsymbol{x} =  \{x_1, x_2\}}$ represents the Cartesian angular position vector in the plane of the sky. Let $(r,\phi)$ be the corresponding polar representation.

First-order lensing quantities like the scalar convergence ($\kappa$) and complex shear ($\gamma = \gamma_1 + \mathrm{i} \gamma_2$) can be derived using the complex lensing operator $\partial = \frac{\partial \ \ }{\partial x_1} +  \mathrm{i} \frac{\partial \ \ }{\partial x_2}$.
\begin{align}
\kappa &= \frac{{\partial\partial}^*\psi}{2} \label{eq:convergence}, \\ 
\gamma &= \frac{{\partial\partial}\psi}{2} \label{eq:shear}.
\end{align}
In practice, we can only measure the reduced shear $g = \gamma/(1 - \kappa)$, which, in the weak-lensing limit $\kappa \ll 1$, can be approximated as $ g \approx \gamma$. The shear catalog used in this work is derived from the LoVoCCS Survey \citep{fu2022}, which employs the HSM algorithm \citep{hirata2003, mandelbaum2005, mandelbaum2018a} to measure galaxy shapes and obtain reduced shear estimates. The HSM algorithm adopts the \textit{distortion} definition of the complex ellipticity, 
\begin{align*}
\chi = \frac{1 - q^2}{1 + q^2} e^{2 \mathrm{i} \phi},
\end{align*}
where $q = b/a$ is the ratio of the semi-minor axis to the semi-major axis, and $\phi$ is the position angle of the galaxy. Under a shear, $\chi$ transforms from the unlensed intrinsic source ellipticity $\chi_{\mathrm{s}}$ as 
\begin{equation*}
\chi = \frac{\chi_{\mathrm{s}} + 2g + g^2 \chi_{\mathrm{s}}^*}{1 + |g|^2 + 2\operatorname{Re}(g\chi_{\mathrm{s}}^*)}.
\end{equation*}

In the weak-lensing regime ($\gamma \ll 1$, $\kappa \ll 1$), under the assumption that the source galaxies are randomly oriented, i.e., $ \langle \chi_{\mathrm{s}} \rangle = \langle \chi_{\mathrm{s}}^2 \rangle = 0 $, it follows to first order that $\frac{\langle \chi \rangle}{2} \approx g \approx \gamma$. Thus, the observed image ellipticity provides an unbiased, albeit noisy, estimate of the local shear. Henceforth, we drop the distinction between the true shear and the observed shear estimate, using $\gamma$ to denote both and $\sigma_{\gamma}$ for the corresponding per component uncertainty.

\subsection{Shear Model} \label{sec:model}
We can now proceed to evaluate the lensing quantities described above for typical cluster--filament configurations. In this work, we study a mass model consisting of an isotropic central cluster (primary) with radially emanating filament(s). We also include possible contamination from neighboring isotropic clusters (secondary). For simplicity, we make the following assumptions: 

\begin{figure*}[htb!]
\gridline{\fig{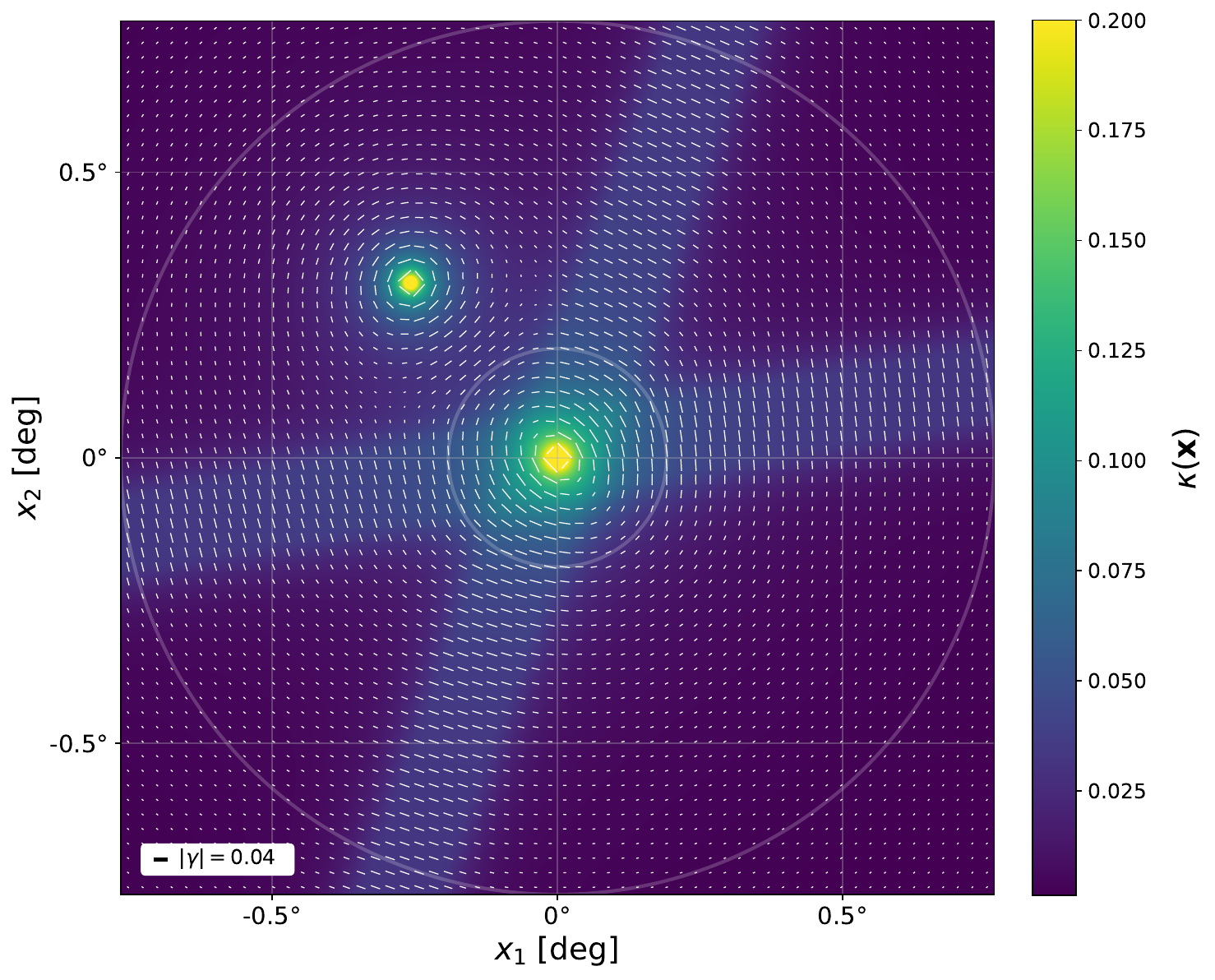}{0.49\textwidth}{(a) Without Shape Noise}
          \fig{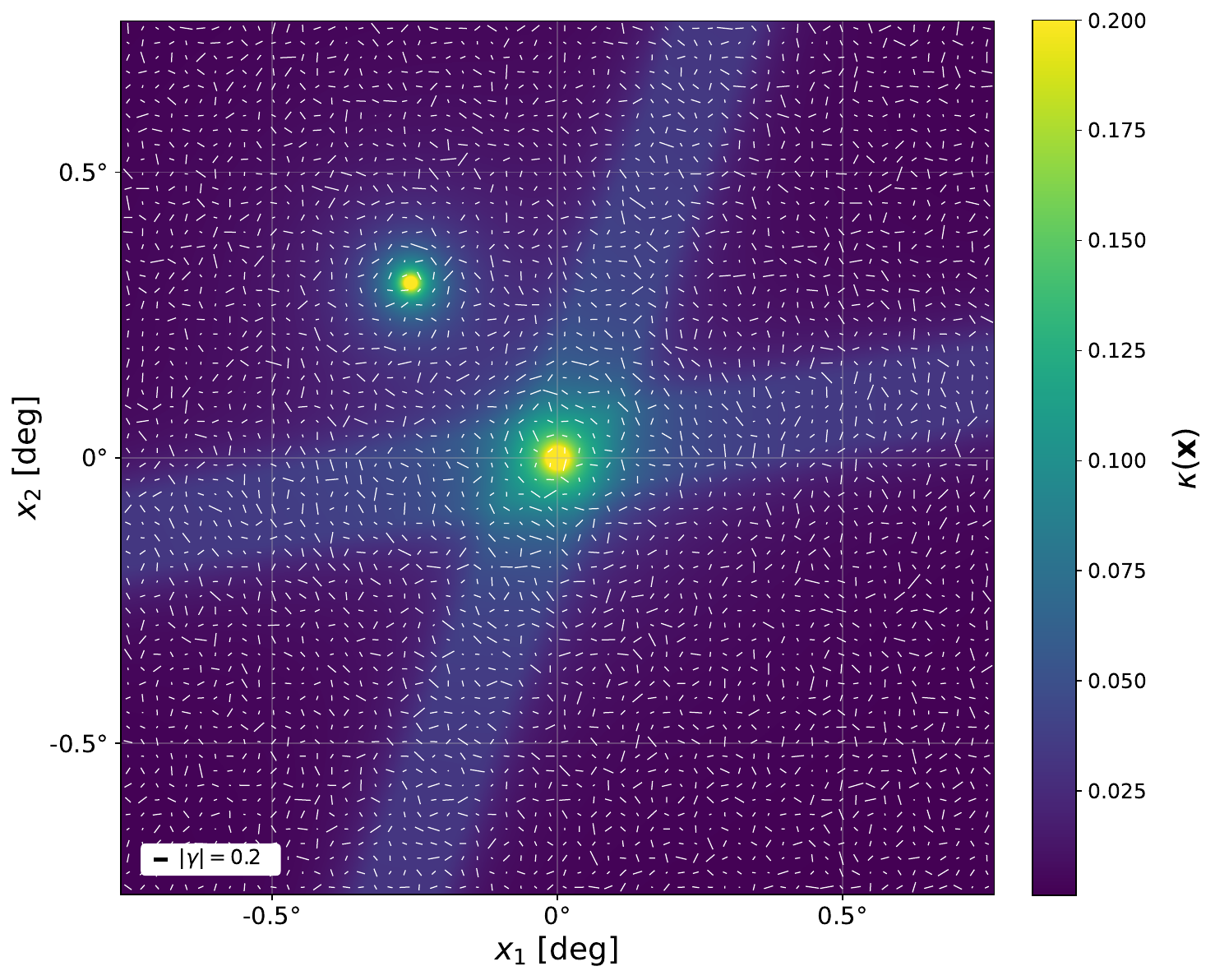}{0.49\textwidth}{(b) With Shape Noise}}
\caption{Convergence field, $\kappa(\boldsymbol{x})$, and binned shear pattern, $\langle\gamma(\boldsymbol{x})\rangle_{\Delta x}$, for the mock catalog. \textit{Left:} Convergence map of the mock model overlaid with the binned shear pattern, computed using a pixel size of $\Delta x = 1.53 \ \mathrm{arcmin}$. White circles indicate the radial cutoffs $(r_1 = 0.91 \ \mathrm{Mpc} $, $r_2 = 3.64 \ \mathrm{Mpc})$ used to restrict the filter. The legend in the bottom-left corner provides a mapping between the shear length in the plot to the physical shear value. \textit{Right:} Same as left panel, but with shape noise ($\sigma_\gamma = 0.32$) added to the mock shear field. In both panels, convergence values are truncated at $\kappa < 0.2$ to emphasize filamentary structures.}
\label{fig:mock_shear_pattern}
\end{figure*}

\begin{enumerate}
    \item Each filament lies in the plane of the sky and extends infinitely away from the central cluster
    \item Far removed from adjacent clusters, the filament surface density remains constant along its axis and varies only in the orthogonal direction. 
    \item The entire system of clusters and filaments is assumed to lie at the same redshift as the primary cluster
\end{enumerate}

Given the limited sky coverage of each system studied in this work $(<2^{\circ})$, we adopt the flat-sky approximation and place the origin of the Cartesian coordinate system at the primary cluster, typically the most massive cluster in the field. This choice is motivated by the empirical findings that (1) the surface mass density of a filament increases towards the host cluster, and (2) massive clusters are connected to a greater number of filaments than their less massive counterparts \citep{bond1996, colberg2005, cautun2014}.

From assumption (2) and equations~\eqref{eq:lensing_potential}, \eqref{eq:convergence}, and \eqref{eq:shear}, it follows that, relative to the filament axis, the shear is purely orthogonal; that is, $\gamma_{\mathrm{f},1'}$ =  - $\kappa$ and $\gamma_{\mathrm{f},2'}$ =  0, where the primed axes $(1',2')$ denote the rotated frame aligned with the filament. If the filament subtends a counterclockwise angle $\theta_{\mathrm{f}}$ with the positive $x_1$ axis, we can decompose the shears in the original frame to get 
\begin{align*}
    \gamma_{\mathrm{f},1} =  - \kappa(h) \cos(2\theta_{\mathrm{f}}), \\
    \gamma_{\mathrm{f},2} = - \kappa(h) \sin(2\theta_{\mathrm{f}}),   
\end{align*}
where $h$ is the perpendicular distance from the filament axis. 

\citet{colberg2005} examined the density profiles of straight filaments in N-body simulations and found that they exhibit well-defined edges, described by the characteristic width $h_{\mathrm{c}}$, within which the density remains approximately constant. Beyond this edge, the density falls off as $h^{-2}$. Motivated by this behavior, we adopt the following model for the filament convergence:
\begin{equation} \label{eq:convergence_model}
\kappa(h)=
\left\{
\begin{array}{@{}c@{\quad}l@{}}
  \kappa_{0}, &
    \text{if } h \le h_{\mathrm{c}}, \\[8pt]
  \dfrac{\kappa_{0}}{1+\left(\dfrac{h-h_{\mathrm{c}}}{h_{\mathrm{c}}}\right)^{2}}, &
    \text{if } h > h_{\mathrm{c}}.
\end{array}
\right.
\end{equation}
where $\kappa_0$ is the maximum convergence at the filament axis. While our model is similar to that implemented in \citet{maturi2013}, \citet{mead2010}, and \citet{nature2024}, we introduce a piecewise function, defined in Eq.~\eqref{eq:convergence_model}, to better capture the flat-top morphology of filaments. However, because of the limited angular resolution of the optimal matched filter (see Figure \ref{fig:mock_filter}), switching between these model choices has a negligible impact on our results $(\Delta \mathrm{S/N} \leq 0.01)$. 

We model clusters in the field as spherically symmetric NFW halos \citep{navarro1997}, which produce a purely tangential shear relative to the radial direction at each point $(r, \phi)$. Denoting the radial profile of the cluster shear amplitude by $\gamma_{\mathrm{NFW}}(r)$, the cluster shear components in the original frame can be expressed as  
\begin{align*}
    \gamma_{\mathrm{c},1}(r,\phi) = - \gamma_{\mathrm{NFW}}(r) \cos(2\phi), \\ \gamma_{\mathrm{c},2}(r,\phi) = - \gamma_{\mathrm{NFW}}(r) \sin(2\phi). 
\end{align*}
Figure~\ref{fig:mock_shear_pattern}(a) illustrates the shear pattern of an example field comprising a primary cluster, two filaments, and a secondary cluster. A detailed description of the configuration employed and the construction of the shear catalog is presented in Section~\ref{subsec:mock_catalog}.

\section{Challenges} \label{sec:problem}
The low-density contrast of filaments results in a weak shear field that is difficult to distinguish from the shear field induced by adjacent host clusters. Simulations show that the shear value decreases from $\sim 0.05 - 0.2$ in the outskirts of adjacent clusters to $\sim 0.01 - 0.02$ along the bulk of the filament \citep{dolag2006}. For low-redshift clusters $(z \lesssim 0.1)$ in the mass range $10^{14} \, \mathrm{M}_{\odot} - 10^{15} \, \mathrm{M}_{\odot}$, the cluster-induced shear can reach $\sim 0.005 - 0.01$ even at projected separations comparable to $R_\mathrm{200c}$. The filament signal is further obscured by various sources of shear noise. N-body simulations by \citet{higuchi2014} show that only $4\%$ of intercluster filaments are detectable with a lensing signal-to-noise ratio $\mathrm{S/N} \geq 2 $ under ideal conditions (source density $n_{\mathrm{g}} \sim 30 \  \mathrm{arcmin}^{-2}$). \citet{higuchi2014} used a convergence profile fitting approach to measure the filament signal-to-noise ratio. Thus, an effective detection strategy necessitates a filter capable of excluding shear contamination from the central and secondary clusters while adequately suppressing noise.

Apart from the biasing effect of nearby clusters (mitigated passively via radial cutoffs; See Appendix~\ref{sec:appendix_c}), we account for two primary sources of random shear noise in our analysis. The first is the shape noise arising from the intrinsic ellipticity of source galaxies, as well as their finite number density. The intrinsic ellipticity contribution can be modeled as Poisson noise with the flat power spectrum, $P_{\mathrm{g}}(k) = \sigma_{\mathrm{\gamma}}^2/2n_{\mathrm{g}}$. The finite sampling of galaxies, on the other hand, imposes a constraint on the smallest spatial scales that can be reliably probed (see Section \ref{subsubsec:shape_noise}). Second, we consider lensing contamination from uncorrelated large-scale structure (LSS) along the line of sight (LOS). \citet{hoekstra2001} showed that distant LSS does not bias mass measurements and can be treated as an additional source of noise. We quantify this effect in the context of our survey in Section \ref{subsubsec:lssnoise}. Thus, the total shear noise, $\gamma_{\mathrm{n}}(\boldsymbol{x}) = \gamma_{\mathrm{g}}(\boldsymbol{x}) + \gamma_{\mathrm{LSS}}(\boldsymbol{x})$, can be modeled as a zero-mean isotropic Gaussian field.

\section{Data Model} \label{sec:data_model}
Following the prescription of \citet{maturi2013}, under the weak-lensing approximation $(\kappa \ll 1)$, the linearity of the lensing potential $\psi(\boldsymbol{x})$, together with Eq.~\eqref{eq:shear}, implies that the observed shear at any point $\boldsymbol{x}$ in the field can be approximated as the sum of the individual shear components. Therefore, the observed shear $\gamma(\boldsymbol{x})$ can be written as
\begin{equation*} \label{eq:shear_addition}
    \gamma(\boldsymbol{x}) = \gamma_{\mathrm{c_k}}(\boldsymbol{x})  + \gamma_{\mathrm{f_l}}(\boldsymbol{x})  + \gamma_{\mathrm{n}}(\boldsymbol{x}), 
\end{equation*}
where identifiers $k = 1,2, ...$ and $l = 1, 2, ...$ represent all clusters and filaments in the field. Or, in terms of the tangential and cross components,
\begin{equation*}
      \gamma_{(+,\times)}(\boldsymbol{x}) =
      \gamma_{\mathrm{c_k},(+,\times)}(\boldsymbol{x}) + 
      \gamma_{\mathrm{f_l},(+,\times)}(\boldsymbol{x}) + 
      \gamma_{\mathrm{n},(+,\times)}(\boldsymbol{x}). 
\end{equation*}
Here ``tangential'' and ``cross'' labels denote components defined with respect to the chosen reference frame or direction (e.g., the filament axis), not relative to the radial direction at each point. This convention is adopted throughout the paper.
 
\subsection{Shear Decomposition}
Given that filaments exhibit a purely orthogonal shear profile relative to their axes, we are motivated to examine how the tangential and cross shear components of both filaments and clusters vary when decomposed along different directions.

\begin{figure}[htb!]
\includegraphics[width=\linewidth]{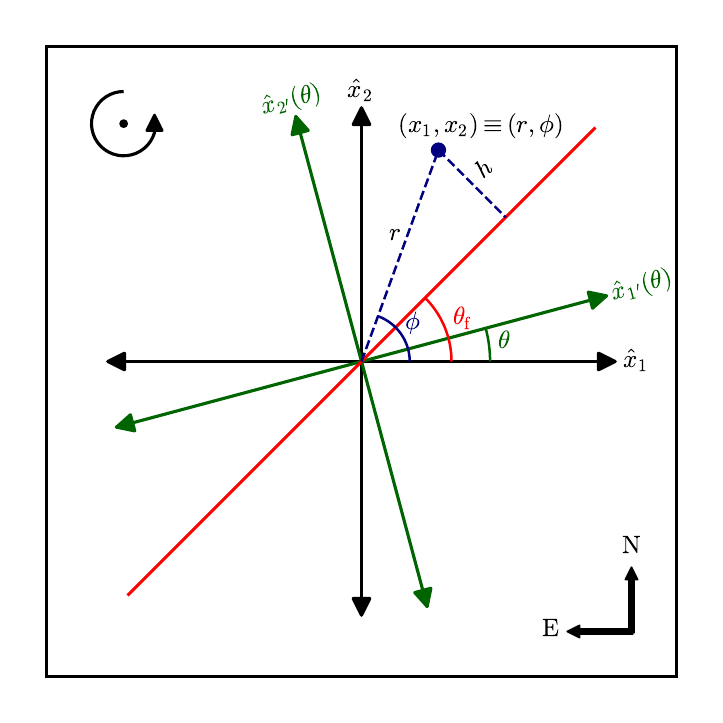}
\caption{Schematic of the coordinate system and reference frames used in this work. The original frame (black) $\hat x_1$--$\hat x_2$ is centered on the primary cluster. The decomposition axis or rotated frame (green) $\hat x_{1'}$--$\hat x_{2'}$ subtends a CCW angle $\theta$ with the positive $\hat x_1$ axis. The filament (red) is oriented at angle $\theta_{\mathrm{f}}$. Any point in the field (blue) can be represented by its Cartesian angular position $\boldsymbol{x} = \{x_1, x_2\}$ or corresponding polar coordinates $(r, \phi)$. The perpendicular distance of the point from the filament axis is marked by $h = r |\sin(\phi - \theta_{\mathrm{f}})|$. The entire coordinate system is right-handed. The compass in the lower-right corner provides the cardinal directions in the image frame.}
\label{fig:coordinate_plot}
\end{figure}

Consider a frame obtained by rotating the original frame counterclockwise (CCW) by an angle $\theta$, as shown in Figure~\ref{fig:coordinate_plot}. The filament shear components, decomposed in the rotated frame, can then be written as
\begin{align*}
    \gamma_{\mathrm{f},+}(r, \phi; \theta) = \kappa(h) \cos(2(\theta_{\mathrm{f}} - \theta_{\mathrm{}})),   \\ 
    \gamma_{\mathrm{f},\times}(r, \phi; \theta) = \kappa(h) \sin(2(\theta_{\mathrm{f}} - \theta_{\mathrm{}})),
\end{align*}
where $h = r |\sin(\phi - \theta_{\mathrm{f}})|$.
Similarly, the cluster shear components in the rotated frame are
\begin{align*}
    \gamma_{\mathrm{c},+}(r, \phi; \theta) &= \gamma_{\mathrm{NFW}}(r) \cos(2(\phi - \theta)),    \\
    \gamma_{\mathrm{c},\times}(r, \phi; \theta) &= \gamma_{\mathrm{NFW}}(r) \sin(2(\phi - \theta)).    
\end{align*}
Hereafter, we suppress the explicit spatial arguments $(r, \phi)$ and denote the tangential and cross shear fields relative to the rotated frame as $\gamma_{+}(\theta)$ and $\gamma_{\times}(\theta)$.

The animations\footnote{Available at \href{https://aas245-aas.ipostersessions.com/default.aspx?s=EE-28-D4-78-A1-1E-7E-7F-6E-87-78-2D-3F-44-32-FA&guestview=true}{https://aas245-aas.ipostersessions.com}} in \citet{shinde2025} illustrate this decomposition scheme for a system consisting of two filaments and two clusters. The decomposed tangential shear for filaments $\gamma_{\mathrm{f},+}(\theta)$ reaches a maximum when the decomposition axis $(\theta)$ is aligned with the filament axis $\theta_{\mathrm{f}}$, and vanishes when they are separated by $45^\circ$. This behavior motivates the use of the observed tangential shear $\gamma_+(\theta)$, as a diagnostic for identifying filamentary structures.

On the contrary, the decomposed tangential shear for clusters, $\gamma_{\mathrm{c},+}(\theta)$, exhibits a quadrupolar pattern with four alternating lobes that rotate in phase with the decomposition axis $(\theta)$. Consequently, the central cluster produces no directional signal, contributing only a positive bias. The secondary cluster, however, can locally resemble a filament. In Section \ref{sec:combined_statistic} we present a method to break this degeneracy. 

\section{Method} \label{sec:method}
We use the decomposition scheme described in the previous section, along with a filter-based approach to extract the lensing signature of any radially emanating filamentary structures present in the field. To this end, we introduce an optimal matched filter designed to exploit the characteristic geometry of filaments and their influence on local shear patterns, while simultaneously suppressing contamination from surrounding halos and other sources of noise. We follow the recipe first described in \citet{maturi2013} and recently applied to the Coma cluster field by \citet{nature2024} to construct a matched-filter statistic that probes the presence of filaments in all directions. 

\subsection{Optimal Matched Filter} 
An optimal matched filter refers to a kernel function that guarantees the highest possible signal-to-noise ratio (S/N) when correlated with the signal. Effective filter design requires careful consideration of both the signal shape and the spectral characteristics of the underlying noise. 

We use the discretized form of the matched-filter statistics provided by \citet{maturi2013}, but adopt the normalization of \citet{nature2024}. 
\begin{align}
    \Gamma_{+}(\theta) &= \frac{1}{\sum_{i} \Psi_i} \sum_{i} \gamma_{+,i}\,\Psi_i 
    \label{eq:tangential_statistic}, \\
    \Gamma_{\times}(\theta) &= \frac{1}{\sum_{i} \Psi_i} \sum_{i} \gamma_{\times,i}\,\Psi_i,
    \label{eq:cross_statistic}
\end{align}
where, $\gamma_{+,i}$ and $\gamma_{\times,i}$ are the tangential and cross components of the complex shear $\gamma_i$ measured for the $i$-th galaxy in the field, decomposed along the $\theta$ direction. $\Psi_i$ is the value of the filter corresponding to the search angle $\theta$, defined at the position of the $i$-th galaxy:  $\Psi_i(\boldsymbol{x}_i; \theta)$. 

We use the expression for the variance of an optimal filter provided in \citet{schirmerthesis2004} and implement the following form for the tangential and cross statistics: 
\begin{align}
    \sigma_{+, \times}^2(\theta) = \frac{1}{2(\sum_{i} \Psi_i)^2} \sum_{i} &|\gamma_{i}|^2 \ \Psi_i^2, \label{eq:standard_deviation} \\ 
    \text{where} \ |\gamma_{i}|^2 = \gamma_{+,i}^2 \ +  \  &\gamma_{\times, i}^2. \nonumber
\end{align}
This expression does not include the noise contribution from large-scale structure, which is evaluated separately in Section~\ref{subsubsec:lssnoise}.

\citet{maturi2005} demonstrated that the optimal filter can be conveniently derived in Fourier space, where it is proportional to the signal shape and inversely weighted by the noise power spectrum. Following this approach, we define the filter in real space to match the tangential shear profile of a filament aligned with it:
\begin{equation*}
    \tau(\boldsymbol{x}) = \gamma_{f,+}(\boldsymbol{x}; \theta_{\mathrm{f}} = \theta_{\mathrm{}}) = \kappa(h),
\end{equation*}
and suppress the noise to obtain the optimized filter in Fourier space:
\begin{align}\label{eq:optimized_filter_fourier}
    \hat \Psi (\boldsymbol{k}) = \frac{ \hat \tau (\boldsymbol{k})}{P_{\mathrm{n}}(\boldsymbol{k})},
\end{align}
\begin{equation*}
    \text{with} \ \hat \tau (\boldsymbol{k}) = \mathcal{F} \{\tau(\boldsymbol{x})\}, 
\end{equation*}
where $\hat \tau (\boldsymbol{k})$ is the Fourier transform of the filament tangential shear profile. Following the convention of \citet{nature2024}, we omit the normalization constant in Eq.\eqref{eq:optimized_filter_fourier}, since the matched-filter statistic in Eq.\eqref{eq:tangential_statistic} is defined with appropriate normalization. Finally, the optimized filter can be expressed in real space through the inverse Fourier transform as 
\begin{align}\label{eq:optimized_filter_inversefourier}
   \Psi (\boldsymbol{x}) = \mathcal{F}^{-1} \{\hat \Psi(\boldsymbol{k})\}.
\end{align}

\subsection{Noise Power Spectrum}
As mentioned in Section \ref{sec:problem}, we consider two sources of shear noise\,---\,(1) shape noise and (2) LSS noise. The total noise can be modeled as an isotropic Gaussian random field with zero mean, described by the power spectrum
\begin{equation*}
    P_{\mathrm{n}}(k) = P_{\mathrm{g}}(k) + P_{\mathrm{LSS}}(k). 
\end{equation*}

\subsubsection{Shape Noise} \label{subsubsec:shape_noise}
For the shape noise, \citet{maturi2013} use a k-dependent exponential correction to the shot noise power spectrum $P_{\mathrm{shot}}(k) = \frac{\sigma_{\mathrm{\gamma}}^2}{2n_{\mathrm{g}}}$, given by
\begin{equation}\label{eq:maturi_power_spectrum}
    P_{\mathrm{g}}(k) = \frac{\sigma_{\mathrm{\gamma}}^2}{2n_{\mathrm{g}}} \exp(\frac{k^2}{n_{\mathrm{g}} \ln2}). 
\end{equation}
The noise power spectrum $P_g(k)$ accounts for the intrinsic shear dispersion ($\sigma_\gamma$), finite source density ($n_g$), and the limited angular resolution set by the average separation of source galaxies ($\exp(\frac{k^2}{n_{\mathrm{g}} \ln2})$). The exponential term arises from modeling the low-pass filtering effect of finite sampling as a Gaussian frequency response, $\hat{W}(k) = \exp(-\frac{k^2}{k_{\mathrm{cut}}^2})$, with $k_{\mathrm{cut}} = \sqrt{n_g \ln 2} $ \citep{maturi2010}. Multiplying this response with the filter in Fourier space yields the effective filter:
\begin{equation}
\hat \Psi (\boldsymbol{k}) = \hat \tau (\boldsymbol{k}) \cdot \hat{W}(k).
\label{eq:W_k_optimize}
\end{equation}
The effect of the Gaussian frequency response, $\hat{W}(k)$, can be equivalently accounted for by modifying the shot noise power spectrum $P_{\mathrm{shot}}(k)$, resulting in Eq.~\eqref{eq:maturi_power_spectrum}. In applying Eq.~\eqref{eq:W_k_optimize} to our analysis, however, we find that $k_{\mathrm{cut}} = \sqrt{n_g \ln 2} $ does not adequately capture the excess noise power at low k-modes (or large angular scales), which leads to a suboptimal filter. To address this, we rely on mock data to calibrate the filter (see Appendix \ref{sec:appendix_a}). We generate multiple realizations of signal--plus--noise shear fields that closely mimic the statistical properties of our observations $(\sigma_{\gamma} = 0.32)$ and use them to determine the optimal cutoff frequency that maximizes S/N. As illustrated in Figure~\ref{fig:optimal_cutoff_frequency}(a), a cutoff frequency of $k_{\mathrm{cut}} \sim \frac{\sqrt{n_{\mathrm{g}} \ln 2}}{10} = 0.21 \ \mathrm{arcmin}^{-1}$ provides the optimal filter for our dataset, yielding an improvement of $\Delta \mathrm{S/N} \sim 1$ over $k_{\mathrm{cut}} = \sqrt{n_g \ln 2}$.

\subsubsection{LSS Noise}\label{subsubsec:lssnoise}
To estimate the uncertainty contributed by LSS, we follow the procedure outlined in \citet{nature2024}. We use mock weak-lensing maps from the kappaTNG dataset\footnote{\url{https://columbialensing.github.io/}} \citep{kappaTNG}, constructed from IllustrisTNG  (TNG300-1) simulations \citep{TNG1, TNG2, TNG3, TNG4, TNG5}. We select shear datasets with a source redshift of $z_{\mathrm{s}} = 0.5064$, corresponding to the average effective source redshift in our data derived from photometric estimates, $\langle z_{\mathrm{eff}} \rangle = 0.51$. Here, $z_{\mathrm{eff}}$ denotes the source redshift that yields the same lensing efficiency\footnote{The effective lensing efficiency $\langle \beta \rangle = \int_{z_\mathrm{l}}^{\infty} p(z_\mathrm{s})\, \frac{D_{\mathrm{ls}}(z_\mathrm{l}, z_\mathrm{s})}{D_\mathrm{s}(z_\mathrm{s})}\, dz_\mathrm{s}$, where $p(z_\mathrm{s})$ is the normalized distribution of source redshifts.} as the entire source population. The $\kappa$TNG maps span a $5^\circ \times 5^\circ$ field with a resolution of  $1024 \times 1024$ pixels, yielding a pixel scale of 0.29 arcmin/pixel. Thus, we divide the $\kappa$TNG field into 4 to 9 patches to match the field of view of the systems used in our analysis ($1.5^\circ \times1.5^\circ$ for A401, A2029, A2351, and $2^\circ \times2^\circ$) and interpolate each $\kappa$TNG patch to obtain mock shear catalogs corresponding to the source galaxy positions for each system. Finally, we apply the optimal matched-filter described in Section~\ref{subsec:filter} to a total of $900$ such patches to estimate the mean standard deviation of the tangential statistic at each filter angle. The standard deviations from LSS noise ($\sigma_{\mathrm{LSS}}$) and shape noise ($\sigma_{\mathrm{shape}}$) can then be added in quadrature to obtain the total standard deviation ($\sigma_{\mathrm{}} = \sqrt{\sigma_{\mathrm{LSS}}^2 + \sigma_{\mathrm{shape}}^2}$). On average, across all four systems, we find that the LSS noise contribution ($\langle\sigma_{\mathrm{LSS}}\rangle = 1.21 \times 10^{-4}$) is about 1.56 times smaller than the shape noise contribution ($\langle\sigma_{\mathrm{shape}}\rangle = 1.88 \times 10^{-4}$), implying that while shape noise is dominant, the addition of LSS noise leads to a 19\% increase in the total standard deviation. Our conclusion differs slightly from that of \citet{nature2024}, who found a higher LSS noise contribution relative to the shape noise contribution. We suspect this is due to the relatively low source density in our survey ($n_g \approx 7 - 12 \ \mathrm{arcmin}^{-2}$), which results in higher effective shot noise. To validate our results, we estimate the LSS uncertainty for each system analytically by integrating the shear power spectrum of the large-scale structure. We find both estimates, numerical and analytical, to be in close agreement. Further details are presented in Appendix~\ref{sec:appendix_b}.

\section{Implementation} \label{sec:implementation}
\begin{figure*}[htb!]
\gridline{\fig{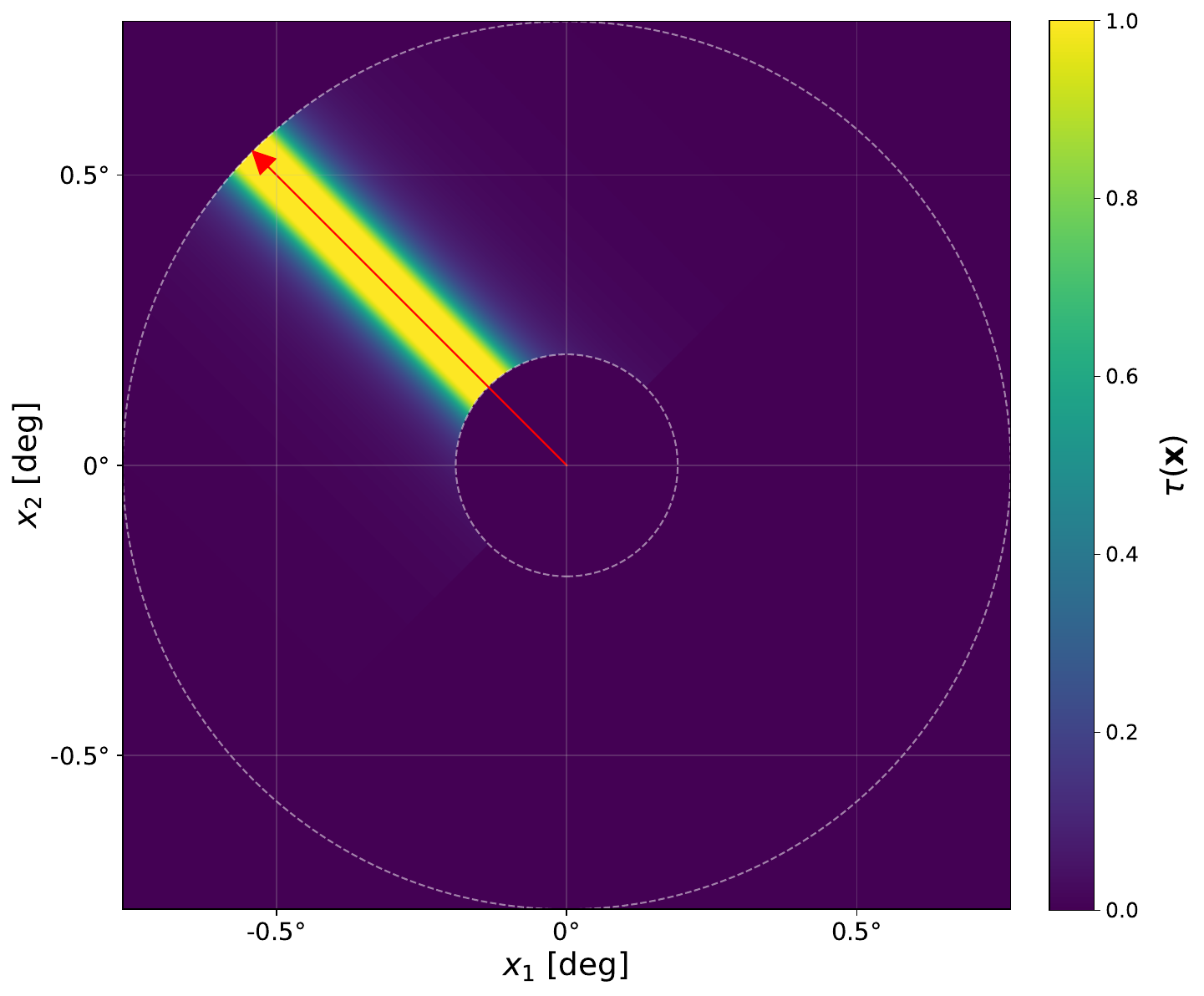}{0.33\textwidth}{(a) Filter}
          \fig{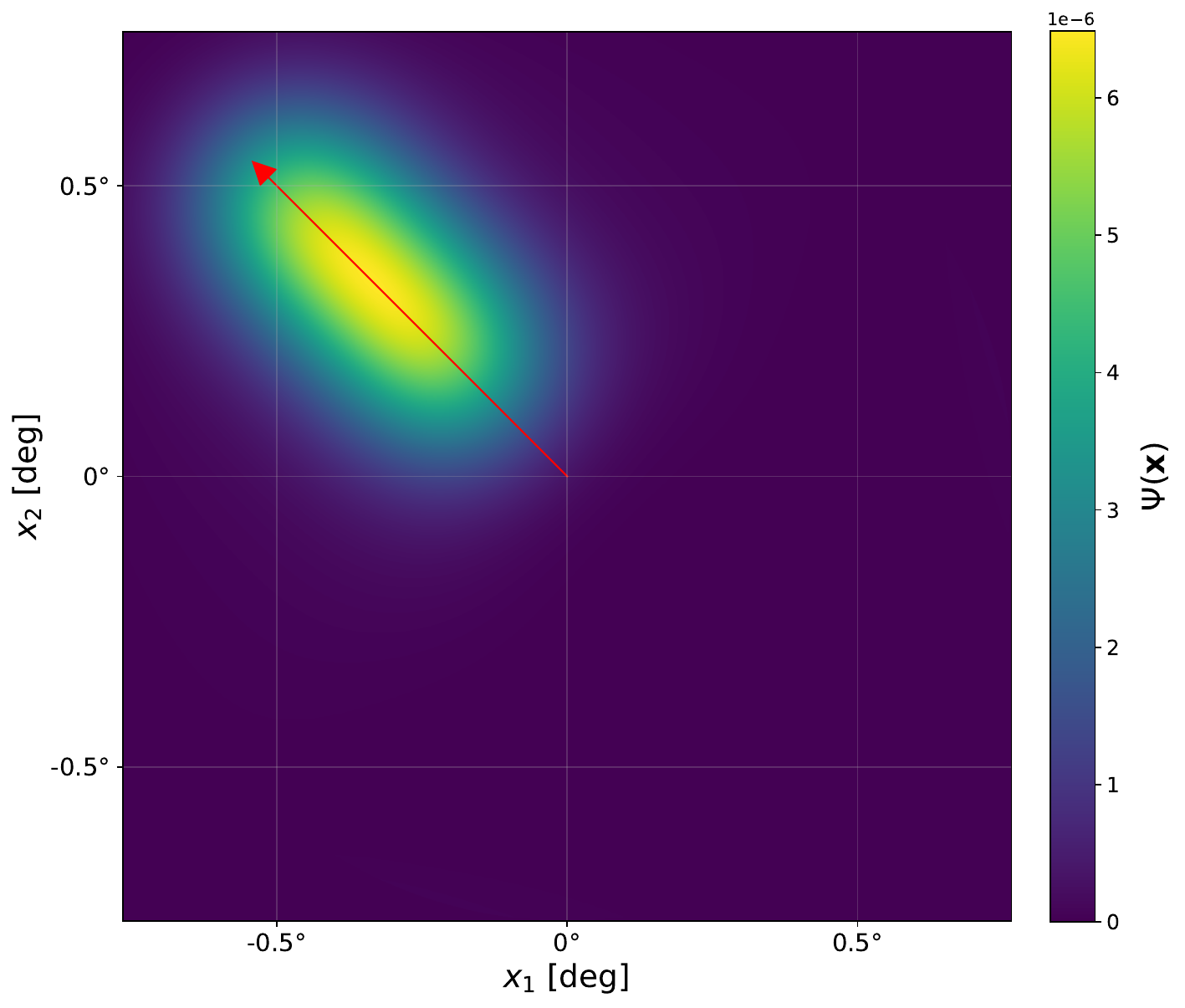}{0.33\textwidth}{(b) Optimal Filter}
          \fig{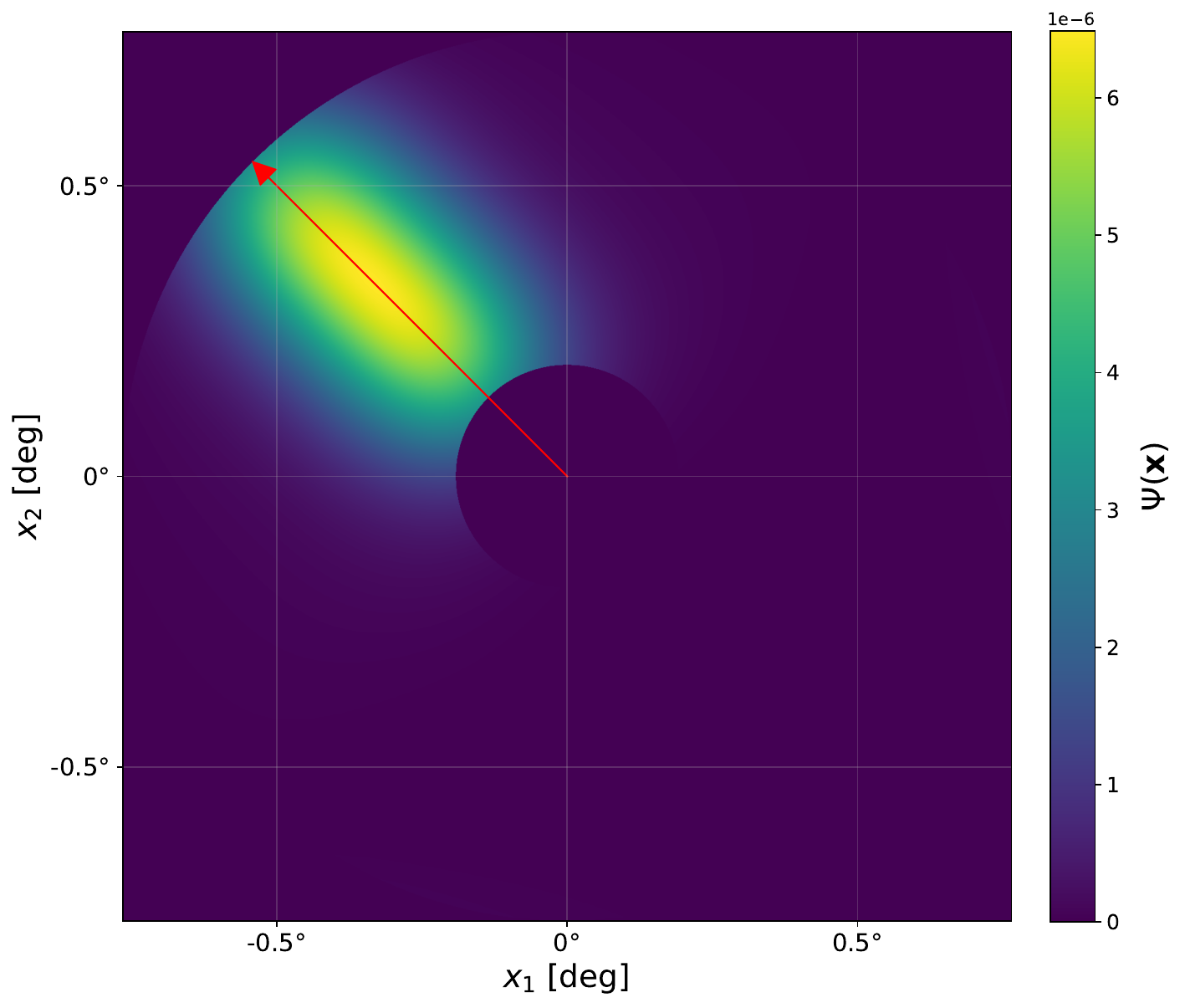}{0.33\textwidth}{(c) Optimal Filter with Radial Cuts}}
\caption{Constructed matched filter and optimal matched filter in real space. \textit{Left:} Matched filter modeled after a template filament with characteristic width $h_{\mathrm{c, filter}} = 0.15 \ \mathrm{Mpc}$ and normalization $\kappa_0 = 1$. We note that the choice of $\kappa_0$ does not affect the analysis. White dashed circles indicate the radial cutoffs $(r_1 = 0.91 \ \mathrm{Mpc} $, $r_2 = 3.64 \ \mathrm{Mpc})$, and the red arrow marks the filter orientation $(\theta = 135^\circ)$. \textit{Middle:} Corresponding filter optimized using $\hat{W}(k)$ with a cutoff frequency of $k_{\mathrm{cut}} = 0.21 \ \mathrm{arcmin}^{-1}$. \textit{Right:} Final optimal matched filter with reapplied radial cuts.}
\label{fig:mock_filter}
\end{figure*}

\subsection{Mock Catalog} \label{subsec:mock_catalog}
In this section, we demonstrate the efficacy of the matched-filter method described above using mock data. We generate a mock shear catalog for a cluster--filament configuration closely resembling the system described in \citet{nature2024}, placed at redshift $z = 0.07$ (typical of cluster pairs studied in this work). We consider a primary central cluster ($M_{\mathrm{200c}} = 8 \times 10^{14}\,\mathrm{M}_\odot$) and two identical filaments intersecting the primary cluster at angles $\theta_{\mathrm{f_1}} = 10 ^ \circ$ and  $\theta_{\mathrm{f}_2} = 70 ^ \circ $. We model the filaments using Eq.~(\ref{eq:convergence_model}) and set $\kappa_0 = 0.03$ and $h_{\mathrm{c}} = 0.25 \ \mathrm{Mpc}$. A secondary cluster, identical to the primary, is placed  at $\phi = 130 ^ \circ$ and $r = 1.90 \ \mathrm{Mpc}$ (or $24 \ \mathrm{arcmin}$) away to serve as a contaminant. We use the lensing engine \texttt{galsim.NFWHalo} \citep{galsim2015} to compute shear contributions from the primary and secondary clusters. To mimic our observations, we adopt a background source density of $n_g = 8 \ \mathrm{arcmin}^{-2}$. We place all background galaxies at the fixed redshift $z_{\mathrm{source}} = 0.51$, corresponding to the average effective source redshift for our dataset. We neglect contamination from foreground galaxies, as all clusters considered in this work are located at very low redshifts $(z <  0.1)$. Because our data is largely dominated by shape noise, we model each component of the total shear noise as a Gaussian random field with zero mean and $\sigma_{\gamma_1} = \sigma_{\gamma_2} = 0.32$, consistent with our dataset. Figure~\ref{fig:mock_shear_pattern} illustrates the convergence field and the resulting shear pattern, both with and without noise. To improve interpretability, we display the shear field binned using a grid with pixel size $\Delta x = 1.53 \ \mathrm{arcmin}$. This binning scheme is applied solely for visualization; all analyses reported in this paper use the unbinned shear catalog.

\subsection{Filter} \label{subsec:filter}

To construct the unoptimized filter (prior to noise suppression via $\hat{W}(k)$) that follows the filament template, we assume a characteristic width, $h_{\mathrm{c}}$, which represents the typical transverse scale of the filaments of interest. In this work, we use a filter with $h_{\mathrm{c, filter}} = 0.15$ Mpc. Varying the filter width within the range $0.05 \ \mathrm{Mpc}<h_{\mathrm{c, filter}} < 0.25 \ \mathrm{Mpc}$ changes the resulting S/N  only marginally by $\lesssim 3 \%$. To minimize shear contamination from both the primary and secondary clusters, we impose radial cutoffs $r_1$ and $r_2$, restricting the filter to the annulus $r_1<r<r_2$. In practice, these cutoffs are chosen such that the cluster shear contribution within the search space remains below 2\% (see Appendix \ref{sec:appendix_c}). For the mock run, however, we explicitly include the secondary cluster and set $r_1 = 0.91  \ \mathrm{Mpc} \ (11.5 \ \mathrm{arcmin})$ and $r_2 = 3.64  \ \mathrm{Mpc} \ (46 \ \mathrm{arcmin})$ to study its effect on the results. The final parameter that fully specifies the filter is the search angle $\theta$, which defines the orientation of the filter.

The filter can then be optimized in Fourier space according to Eq.~\eqref{eq:W_k_optimize}, and reconstructed in real space via the inverse Fourier transform. Since the window function $\hat{W}(k)$ decays exponentially at high frequencies, the optimization process suppresses Fourier modes at the smallest scales. The result is a smoothed filter with limited angular resolution, as illustrated in Figure~\ref{fig:mock_filter}(b). To prevent the smoothed filter from extending beyond the annular region $r_1<r<r_2$ and to exclude contamination from terminal clusters, we reapply radial cuts to the filter as shown in Figure~\ref{fig:mock_filter}(c). Once constructed, the optimal matched filter can be convolved with the tangential and cross shear fields to scan all directions $\theta$ for the presence of filaments according to Eq.~\eqref{eq:tangential_statistic} and Eq.~\eqref{eq:cross_statistic}.

\subsection{Results}

\begin{figure*}[htb!]
\gridline{\fig{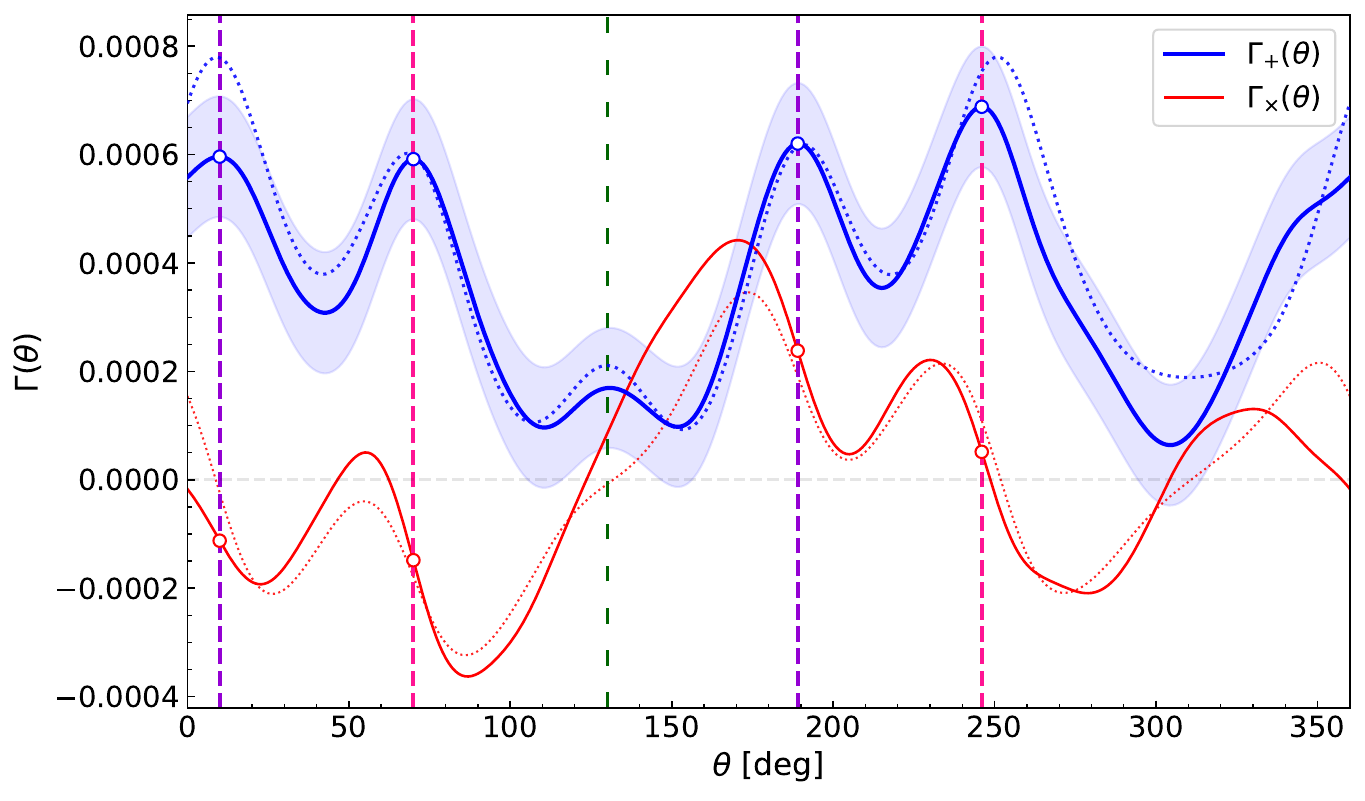}{0.6\textwidth}{}
          \fig{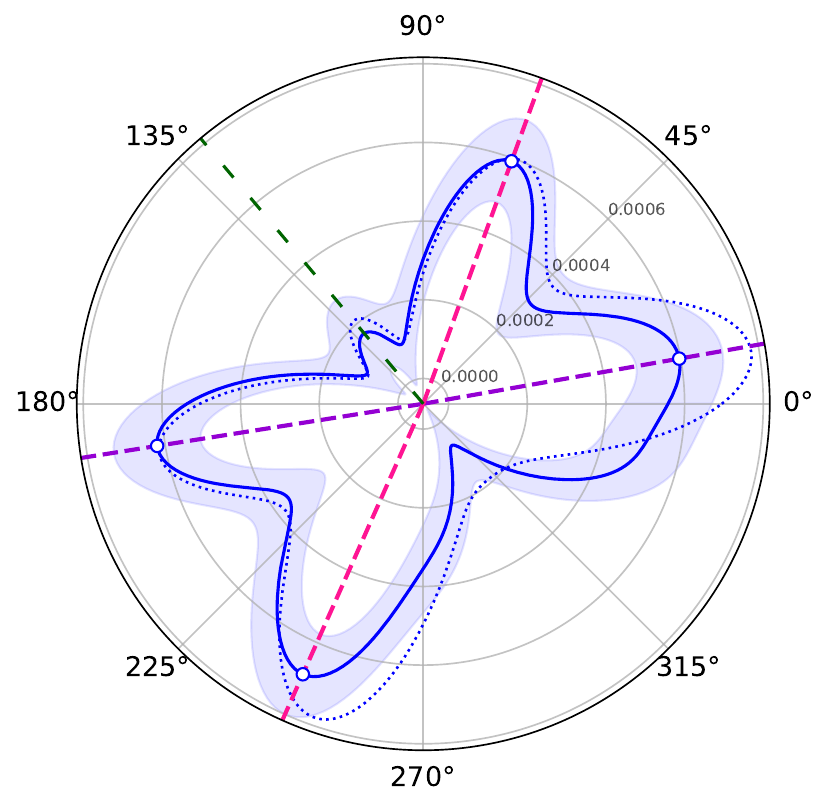}{0.39\textwidth}{}}
\caption{Filament detection results for the mock catalog. \textit{Left:} Matched-filter statistics $\Gamma_{\times}(\theta)$ and $\Gamma_{+}(\theta)$ as a function of the search angle $\theta$. The blue and red solid lines correspond to the tangential and cross components, respectively, with the light blue shade indicating the $1\sigma$ uncertainty in the tangential signal. The uncertainty in the cross signal is omitted for visual clarity. The dotted blue and red lines denote the corresponding statistics in the absence of noise. The thick violet and pink vertical dashed lines represent detections associated with filaments oriented at $\theta_{\mathrm{f_1}} = 10^{\circ}$ and $\theta_{\mathrm{f_2}} = 70^{\circ}$, respectively; the loosely dashed green line marks the orientation of the secondary cluster $(\phi = 130^{\circ})$. \textit{Right:} Polar representation of the left panel}
\label{fig:mock_results}
\end{figure*}

\begin{deluxetable}{ccccc}
\tablecaption{Filament Properties - Mock Catalog \label{tab:mock_results_table}}
\tablehead{
\colhead{Filament Orientation} & \colhead{Detected Orientation} & \colhead{Peak Significance} & \colhead{$\kappa_0$} & \colhead{$h_{\mathrm{c}}$} \\
\colhead{(deg)} & \colhead{(deg)} & \colhead{($\sigma$)} & \colhead{} & \colhead{(Mpc)}
}
\startdata
10  & 10  & 5.4 & $0.025^{+0.006}_{-0.006}$ & $0.33^{+0.07}_{-0.06}$ \\
70  & 70  & 5.3 & $0.031^{+0.007}_{-0.006}$ & $0.26^{+0.07}_{-0.05}$ \\
190 & 189 & 5.6 & $0.033^{+0.007}_{-0.007}$ & $0.26^{+0.06}_{-0.05}$ \\
250 & 246 & 6.2 & $0.034^{+0.007}_{-0.006}$ & $0.28^{+0.07}_{-0.06}$ \\
\enddata
\end{deluxetable}

The results for the mock catalog, both with and without noise, are displayed in Figure~\ref{fig:mock_results}. Since the tangential statistic is expected to vanish in the presence of pure noise $(\langle\Gamma_{+}(\theta)\rangle_{\mathrm{noise}} = 0)$, the significance of each detection is quantified by the signal-to-noise ratio $\mathrm{S/N} = \Gamma_{+}/\sigma_{+}$. As expected, the tangential statistic $\Gamma_+(\theta)$ peaks within $4 ^\circ$ of the true filament orientations with significance $>5\sigma$ (see Table \ref{tab:mock_results_table}). Additionally, a lower-significance ($1.7\sigma$) peak is detected in the direction of the secondary cluster, resulting from the tangential alignment of the cluster shear along the filter axis. This implies that the presence of a secondary cluster in the field can lead to a marginal albeit spurious filament detection. Such contamination from secondary clusters can be preliminarily identified and flagged by applying WL filters optimized for detecting galaxy clusters \citep{maturi2005} or via optical cluster finding algorithms \citep{redmapper2014, amico2018}.
In the next section, we discuss a strategy to eliminate false positives arising from cluster contamination. 

\subsection{Combined Statistic} \label{sec:combined_statistic}

\begin{figure*}[htb!]
\centering
\includegraphics[width=0.75\textwidth]{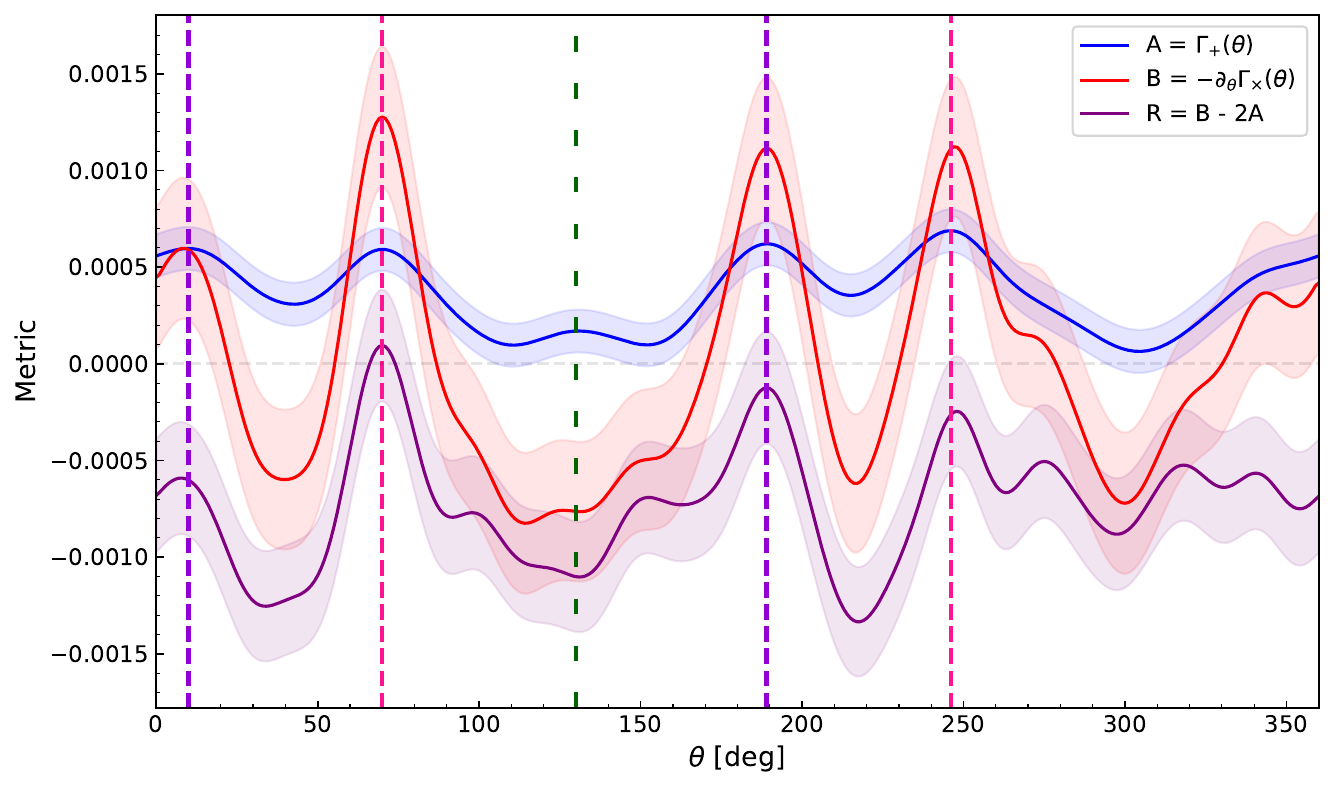}
\caption{Orthogonal decomposition of the negative cross gradient. The blue, red and purple solid lines represent the tangential statistic, $A(\theta)$, negative cross gradient, $B(\theta)$, and the residual, $R(\theta)$. The corresponding shaded regions represent the  $1\sigma$ uncertainty in each statistic.}
\label{fig:mock_signal_ABR}
\end{figure*}

A key observation that resolves the cluster-filament degeneracy is that the cross statistic exhibits a sharp decline in the immediate vicinity of the filament axis, while no such behavior is seen for the secondary cluster. This distinction becomes more evident when the negative gradient of the cross statistic (negative cross gradient hereafter) is plotted alongside the tangential statistic as shown in Figure~\ref{fig:mock_signal_ABR}. This motivates a closer inspection of the negative cross gradient, $- \partial_{\theta} \Gamma_{\times}(\theta)$, to glean additional information that, in conjunction with the tangential statistic, can help distinguish cluster contribution from genuine filament signal. 
 
For the $i$-th galaxy, the tangential and cross shear components are defined as
\begin{align*}
    \gamma_{+,i}(\theta) &= -\,\gamma_{1,i}\cos(2\theta) - \gamma_{2,i}\sin(2\theta),\\[4pt]
    \gamma_{\times,i}(\theta) &= \ \ \ \gamma_{1,i}\sin(2\theta) - \gamma_{2,i}\cos(2\theta).
\end{align*}
Rewriting Eqs.~\eqref{eq:tangential_statistic} and \eqref{eq:cross_statistic} in terms of the normalized filter function $\tilde{\Psi}_{i} = \frac{\Psi_{i}}{{\sum_{i} \Psi_i}}$, we obtain
\begin{align*}
    \Gamma_{+}(\theta) &= \sum_{i} \tilde{\Psi}_{i}(\theta)\,\gamma_{+,i}(\theta),\\
    \Gamma_{\times}(\theta) &= \sum_{i} \tilde{\Psi}_{i}(\theta)\,\gamma_{\times,i}(\theta).
\end{align*}
For ease of notation, we define the statistics 
\begin{align*}
    A(\theta) &= \Gamma_{+}(\theta), \\ 
    B(\theta) &=  - \partial_{\theta}\Gamma_{\times}(\theta).    
\end{align*}
Differentiating $\Gamma_{\times}(\theta)$ with respect to $\theta$ gives 
\begin{align}\label{eq:diff_cross_statistic}
    \frac{\partial}{\partial\theta}\Gamma_{\times}(\theta)
    = \sum_{i} \frac{\partial \tilde{\Psi}_{i}(\theta)}{\partial \theta}\,\gamma_{\times,i}(\theta)
       + \sum_{i} \tilde{\Psi}_{i}(\theta)\,\frac{\partial \gamma_{\times,i}(\theta)}{\partial \theta}.
\end{align}
The derivative of $\gamma_{\times,i}(\theta)$ with respect to $\theta$ is
\begin{align}\label{eq:diff_cross_shear}
    \frac{\partial \gamma_{\times,i}}{\partial \theta}
    = 2\,\gamma_{1,i}\cos(2\theta) + 2\,\gamma_{2,i}\sin(2\theta)
    = -2\,\gamma_{+,i}(\theta).
\end{align}
Substituting Eq.~(\ref{eq:diff_cross_shear}) into  Eq.~(\ref{eq:diff_cross_statistic}) yields 
\begin{align*}
    \partial_{\theta}\Gamma_{\times}(\theta)
    = \sum_{i} \frac{\partial \tilde{\Psi}_{i}}{\partial \theta}\,\gamma_{\times,i}(\theta)
       - \; \; 2 \; \underbrace{ \sum_{i} \tilde{\Psi}_{i}(\theta) \, \gamma_{+,i}(\theta)}_{A(\theta)}.
\end{align*}
Thus, we obtain 
\begin{align*}
    B(\theta) = 2\,A(\theta)\;-\; \sum_{i} \frac{\partial \tilde{\Psi}_{i}}{\partial \theta}\,\gamma_{\times,i}(\theta),
\end{align*}
or equivalently, the compact identity
\begin{align*}
   \boxed{ \; B=2A + R\; }, 
\end{align*}
where the residual term is defined as
\begin{align*}
    R(\theta)= \; -\sum_{i} \frac{\partial \tilde{\Psi}_{i}}{\partial \theta}\,\gamma_{\times,i}(\theta).     
\end{align*}
Since the normalization factor $\sum_{i} \Psi_i$ does not vary with the search angle $\theta$, i.e. $\frac{\partial }{\partial \theta} (\sum_{i} \Psi_i) = 0$, the variance of the residual can be written, analogous to Eq.~\eqref{eq:standard_deviation}, as
\begin{align*}  
    \sigma_{R}^2 = \frac{1}{2}\sum_{i}(\frac{\partial \tilde{\Psi}_{i}}{\partial \theta})^2\,|\gamma_{i}|^2.
\end{align*}
Finally, the covariance between A and R is given by 
\begin{align*}
    \text{Cov}(A, R) = -\sum_{i} \tilde{\Psi}_{i}(\theta) \frac{\partial \tilde{\Psi}_{i}(\theta)}{\partial \theta} \text{Cov}(\gamma_{+,i}, \gamma_{\times,i}).
\end{align*}

Under the assumption of isotropic noise, $\text{Cov}(\gamma_{+,i}, \gamma_{\times,i}) =  0$, and consequently, $\text{Cov}(A, R) = 0$. This implies that the tangential statistic $A$ and the residual $R$ are independent metrics that together form an orthogonal basis for representing the negative cross gradient $B$. Within this basis, the decomposition $B = 2A + R$ highlights that, although $B$ is strongly correlated with $A$, it is precisely the contribution of $R$ in $B$ that enables the distinction between the filament signature and cluster contamination.

\begin{figure*}[htb!]
\centering
\includegraphics[width=0.75\textwidth]{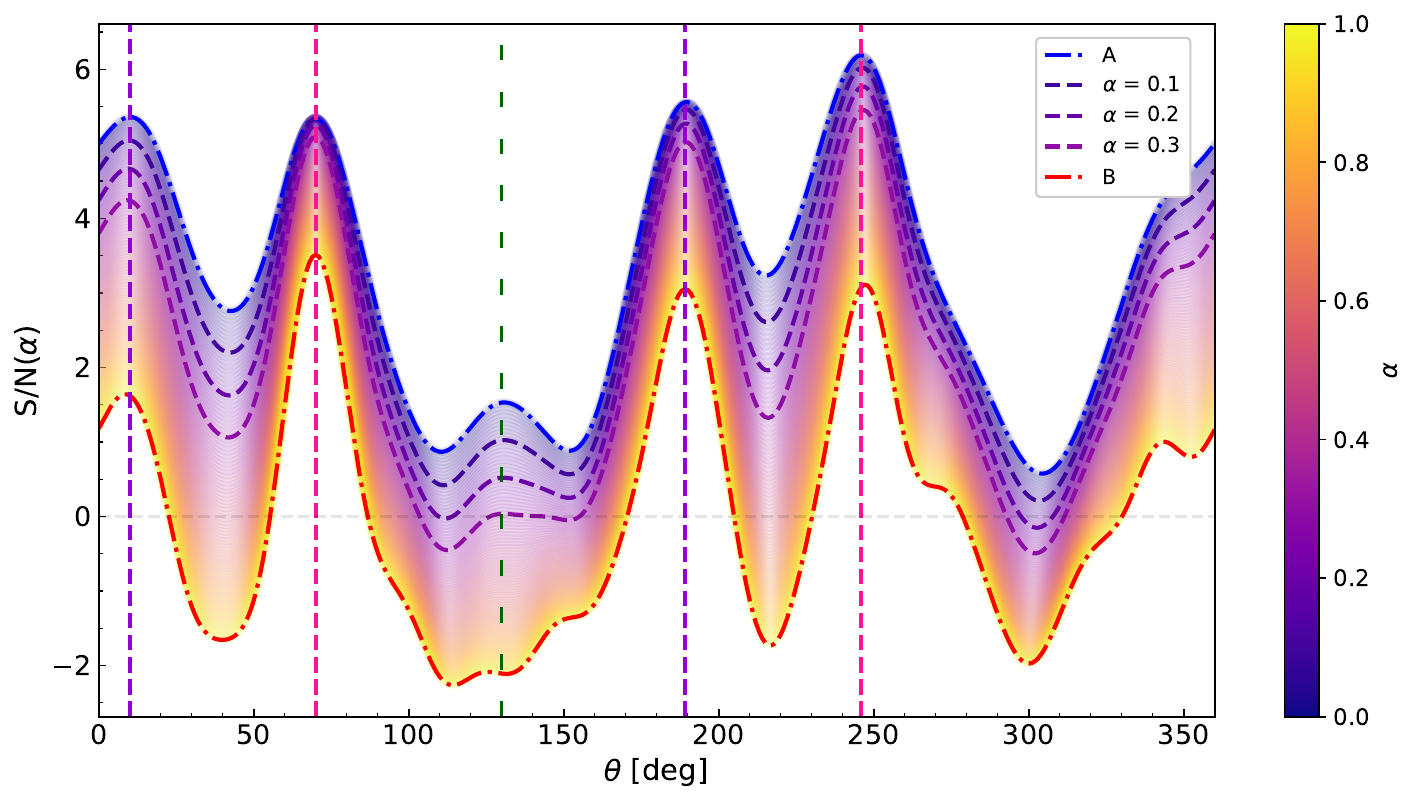}
\caption{Signal-to-noise ratio (S/N) for linear combinations of the tangential statistic, $A(\theta)$, and the residual, $R(\theta)$. The free parameter $\alpha \in [0,1]$ is mapped according to the color bar. The corresponding locus of S/N curves representing the combined statistic $L(\alpha) =2A + \alpha R$ is shown. Three dashed $\mathrm{S/N}(\alpha)$ curves corresponding to $\alpha =\{0.1, \ 0.2, \ 0.3\}$ illustrate the effect of progressively adding the residual term to the tangential statistic. The blue and red dash-dot curves represent the extremes, $\alpha = 0$ or $A(\theta)$, and $\alpha = 1$ or $B(\theta)$, respectively.}
\label{fig:mock_combined_metric}
\end{figure*}

To determine the optimal contribution of $R$ that maximizes the detection S/N, we introduce the combined statistic
\begin{align*}
    L(\alpha) = 2A + \alpha R,
\end{align*}
where $\alpha$ is a free parameter that can be tuned to optimize $L$. Setting $\alpha = 0$ recovers the tangential statistic $A$, while $\alpha = 1$ corresponds to the negative cross gradient $B$. Since both means vanish under noise, i.e., $\langle A \rangle _{\mathrm{noise}} = \langle R \rangle _{\mathrm{noise}}  = 0$, the S/N can be written as
\begin{align*}
    \text{S/N}(\alpha) = \frac{2A+ \alpha R}{\sqrt{4\sigma_{A}^2 + \alpha^2 \sigma_{R}^2}}.
\end{align*}

We find that the residual is inherently more sensitive to noise than the tangential statistic ($\sigma_R \gg \sigma_A$). As a result, introducing a residual-weighted term to the tangential statistic ($\alpha > 0$) does not generally improve the S/N for filament detections. For small values of alpha ($\alpha \sim 0.1 - 0.3$), the combined statistic partially suppresses contamination from the secondary cluster -- reducing its significance from $\sim 2 \sigma$ to $\sim 0 \sigma$ -- while maintaining $\mathrm{S/N}>4$ at all filament peaks. As $\alpha \to 1$, the combined statistic effectively eliminates the cluster contribution altogether, although at the expense of lowering filament significance from $\sim 5-6\sigma$ to $\sim 2-3\sigma$. Figure~\ref{fig:mock_combined_metric} shows $L(\alpha)$ for $\alpha \in [0,1]$, illustrating these trends.

Therefore, despite its ability to distinguish cluster contamination from genuine filamentary signal, the negative cross gradient is unusable in our case due to the relatively low source density of the LoVoCCS shear catalog. However, for deep space-based observations (e.g., JWST) with substantially higher source densities, the corresponding loss in detection significance may be negligible, making the use of the negative cross gradient viable. Thus, in the subsequent analysis, we use the cross statistic solely as a qualitative indicator of potential cluster contamination. 

\subsection{Estimating Filament Properties}

\begin{figure}
\hspace*{-0.75cm} 
\includegraphics[width=\linewidth]{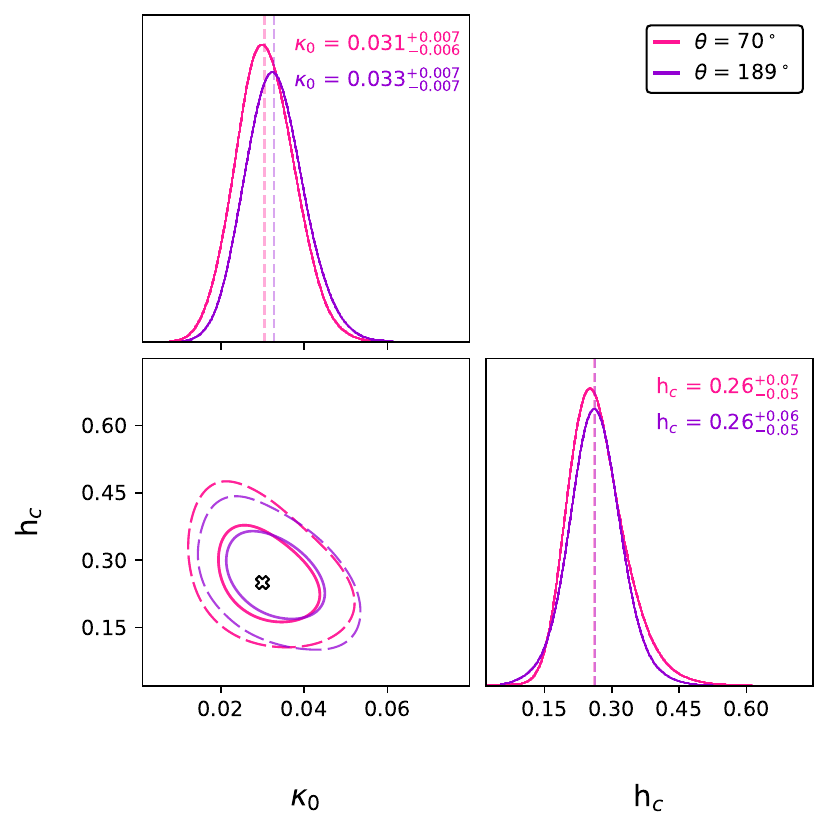}
\caption{Mock Catalog: Posterior distributions of the filament parameters obtained via MCMC sampling. The inner (solid) and outer (dashed) contours in the joint distribution enclose the  68\% $(1\sigma)$ and 95\% $(2\sigma)$ credible regions, respectively. Dashed vertical lines indicate the median values of the marginalized distributions. The black-outlined cross in the contour plot marks the input filament parameters, $\kappa_0 = 0.03$ and $h_{\mathrm{c}} = 0.25 \ \mathrm{Mpc}$. The legend in the upper-right corner shows the color coding for the two detections.}
\label{fig:mock_mcmc}
\end{figure}

We use the Markov Chain Monte Carlo (MCMC) sampler \textit{emcee} \citep{emcee} to infer filament properties -- $\kappa_0$ and $h_{\mathrm{c}}$. For each detected filament, we fit the model described in Eq.~\eqref{eq:convergence_model} to the shear data within a fixed window of width $h_{\mathrm{ref}} = 1.5 \ \mathrm{Mpc}$ about the detection axis. This window size allows the sampler to explore all plausible filament widths $(h_{\mathrm{c}} < 0.75 \ \mathrm{Mpc})$, resolving both the constant $\kappa_0$ region up to $h_{\mathrm{c}}$ and the $\propto h^{-2}$ decline, while also limiting contamination from adjacent filament(s). The resulting posterior distributions for $\kappa_0$ and $h_{\mathrm{c}}$ are displayed in Figure~\ref{fig:mock_mcmc}, with the corresponding median values summarized in Table \ref{tab:mock_results_table}. To avoid clutter and improve interpretability, we present the posterior distribution for one branch per filament $(\theta = 70 ^\circ, \, \theta = 189^\circ)$. For reference, we mark the input parameters $(\kappa_0 = 0.03 ,\, h_{\mathrm{c}} = 0.25 \ \mathrm{Mpc})$ with a `$\times$' in the contour plot. We find all model parameters to be well constrained. The maximum deviations of the estimated median values from the input parameters are $\Delta \tilde{\kappa}_{0}  = 0.005 \  (17 \%)$ and $\Delta \tilde{h}_{\mathrm{c}} = 0.08 \  (32 \%)$. While most of this scatter can be attributed to shape noise, comparison with the noise-free case indicates that the estimated filament parameters are also slightly biased by other structures in the field. 

\section{Data} \label{sec:data}
We used observations from the Dark Energy Camera (DECam) \citep{flaugher2015} on the 4-m Blanco Telescope at the Cerro Tololo Inter-American Observatory (CTIO) as part of the Local Volume Complete Cluster Survey (LoVoCCS; Proposal ID: 2019A-0308; PI: Ian Dell’Antonio), an NSF NOIRLab survey of 107 nearby $(0.03<z<0.12)$ X-ray luminous clusters $([0.1 - 2.4 \ \mathrm{keV}] \ L_\mathrm{X500c}>10^{44} \ \mathrm{erg} \ \mathrm{s}^{-1})$ in \textit{ugriz} bands \citep{fu2022}. The LoVoCCS targets are imaged to depths comparable to the upcoming Vera C. Rubin Observatory’s Legacy
Survey of Space and Time (LSST) Year 1–2 data, reaching a 5$\sigma$ coadded median depth of $\sim 25-26$ magnitude for point sources and about $\sim 0.3$ mag shallower for extended objects in each band \citep[Table 1 \& 2;][]{fu2024}. We select the \textit{r} band for lensing measurements and require seeing (median FWHM) $\lesssim 0.9''$ for high-quality shape information. We use the LSST Science Pipeline in conjunction with the LoVoCCS data analysis pipeline to process raw exposures and obtain a uniformly processed catalog of coordinates, magnitudes, shapes, photometric redshifts, and other associated quantities of all deblended and well-measured coadd objects within each cluster. We use the HSM algorithm \citep{hirata2003, mandelbaum2005} integrated within the LSST Science Pipeline to measure galaxy shapes. We then apply the quality cuts described in \citet[Section~3.2; \texttt{mass\_map}]{fu2022}, with some modifications, to extract a clean sample of galaxy shapes. We made the following modifications to the quality cuts: 

(1) Expand Background Population: \citet{fu2022} applied a cut on $z_{\mathrm{b}}$ (most probable redshift); $0.15 < z_{\mathrm{b}} < 1.4$ to avoid contamination from cluster members and exclude unreliable photo-$z$ at high redshift. We use the updated criteria in \citet{fu2024} and require that $0.1 + z_{\mathrm{l}} < z_{\mathrm{b}} < 1.4$ where $z_{\mathrm{l}}$ is the cluster redshift. The introduction of a $z_{\mathrm{l}}$-based variable limit has been reported to increase lensing S/N by $\leq 10 \%$.

(2) Lower BPZ \texttt{odds}: Within the framework of Bayesian Photometric Redshift estimation \citep[BPZ;][]{benitez2004}, \texttt{odds} represents the likelihood of the estimated redshift, $z_{\mathrm{b}}$, being correct. We relax the minimum \texttt{odds} requirement from $>0.95$ to $>0.8$ to admit sources with moderately uncertain photo-$z$ estimates. 
The resulting increase in true background sources outweighs the impact of misclassified foreground galaxies, thereby boosting the lensing S/N.

(3) Position Cut: Finally, since we are only concerned with filament detection in the intercluster region between clusters, we limit the catalog to sources within a radius of 1.25 times the intercluster distance between the primary cluster and the outermost secondary cluster under consideration. This cut does not affect the outcome, since the filter itself is spatially constrained, but it helps reduce computational overhead.

We use the shear calibration script\footnote{\url{https://github.com/PrincetonUniversity/hsc-y1-shear-calib}} for the HSM algorithm to obtain the shape weights and biases required to convert galaxy shapes into reduced shear estimates $\hat{\mathrm{g}}_{1,2}$ \citep{mandelbaum2018a}. The shear calibration parameters are derived from HSC image simulations \citep{mandelbaum2018b}. While the original source catalog contains galaxies with a number density of $n_{\mathrm{g}}\sim50 \ \mathrm{arcmin}^{-2}$, our selection criteria yields a shear catalog with an effective number density of $n_{\mathrm{g}} \sim7-12 \ \mathrm{arcmin}^{-2}$.

N-body simulations show that clusters separated by $\leq 5\,h^{-1}\mathrm{Mpc}$ are invariably connected by at least one filament \citep{colberg2005}. Furthermore, massive clusters are typically embedded within rich filamentary networks with high-density contrast, thereby producing a strong lensing signal. Motivated by these considerations, we select the most massive and X-ray luminous cluster pairs in the LoVoCCS survey as prime targets for filament detection. In this work, we focus on three such systems: (i) Abell 401–399, (ii) Abell 2029–2033-SIG, and (iii) Abell 3558–3556–3562 (Shapley Supercluster Core). Additionally, to build confidence in our analysis, we include a lower-mass system, Abell 2351, as a null test to demonstrate that our method does not produce spurious filament detections. Hereafter, we denote each system by the primary cluster around which the analysis is centered. 

\section{Results} \label{sec:results}
Following the procedure outlined in Section~\ref{sec:implementation}, we apply the matched-filter method to the dataset described above. We use the coordinates of the brightest cluster galaxy (BCG) to define the centers of all clusters in the field. Cluster redshifts are sourced from the MCXC catalog \citep{piffaretti2011}. For each system, we determine the radial filter cutoffs, $r_1$ and $r_2$, by requiring that the shear contribution from clusters in the field stays below 2\% within the annular search space. A full description is provided in Appendix~\ref{sec:appendix_c}; the resulting cutoff values for each system are listed in Table \ref{tab:radial_cutoffs_table}.

\subsection{Abell 401}

\begin{figure*}[htb!]
\centering
\hspace*{-0.75cm} 
\gridline{\fig{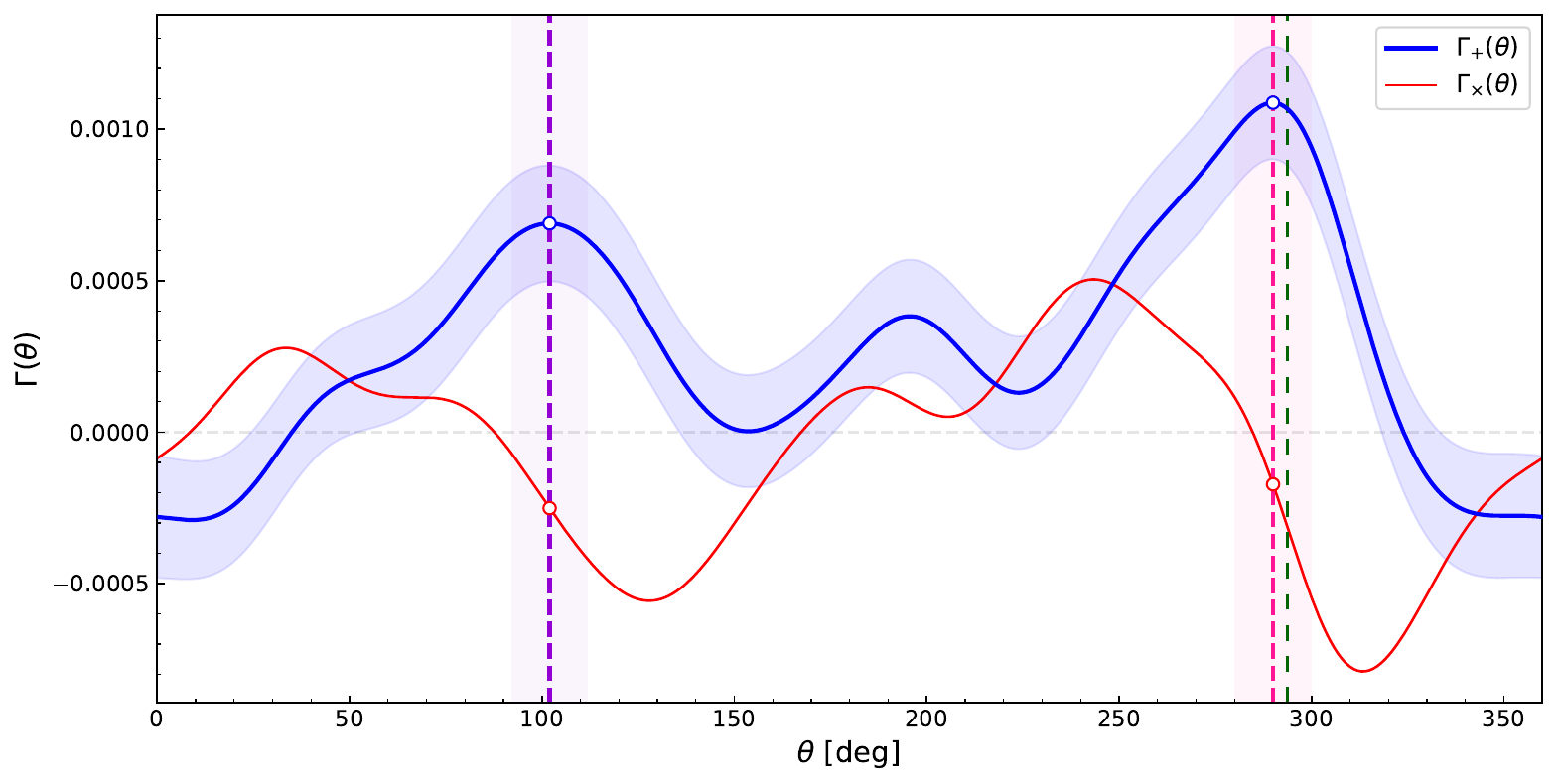}{0.92\textwidth}{(a) Linear Plot}}
\gridline{\fig{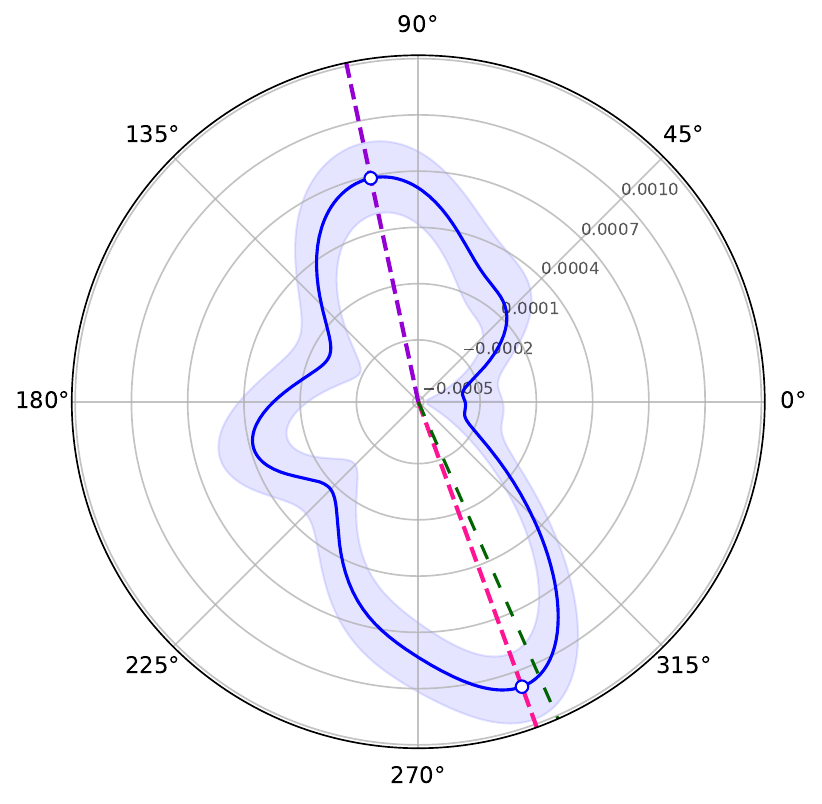}{0.49\textwidth}{(b) Polar Plot}
         \fig{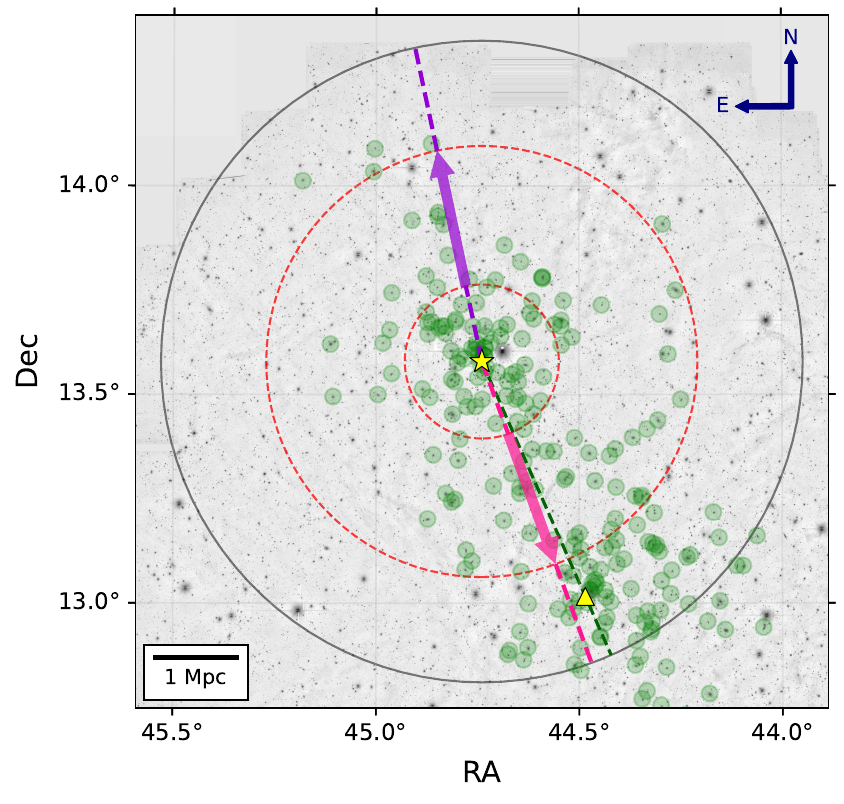}{0.49\textwidth}{(c) Overlaid Plot}}
\caption{Filament detection results for the A401 system. \textit{Top:} Matched-filter statistics $\Gamma_{+}(\theta)$ and $\Gamma_{\times}(\theta)$ as a function of the search angle $\theta$. The blue and red solid lines correspond to the tangential and cross components, respectively, with the light blue shade indicating the $1\sigma$ uncertainty in the tangential signal. The violet and pink vertical dashed lines mark the tangential signal peaks detected at $\theta = 102^{\circ}$ and $\theta = 290^{\circ}$, respectively. The corresponding light shade indicates a $\pm10^{\circ}$ window centered on the detected peak, provided for comparison with orientations of other clusters in the field. The loosely dashed green line marks the orientation of A399 relative to A401. \textit{Bottom-left:} Polar representation of the top panel. \textit{Bottom-right:} Detected filament orientations overlaid on the inverted $r$-band coadded image used for shape measurement. The dashed red circles represent the radial cutoffs $r_1 = 0.92 \ \mathrm{Mpc}$ and $r_2 = 2.58 \ \mathrm{Mpc}$. The violet and pink arrows extending from $r_1$ to $r_2$ indicate the spatial orientation of the detected filaments. A401 (primary) and A399 (secondary) are marked with a yellow star and triangle, respectively. Spectroscopic member galaxies $(0.06 < z < 0.085)$ in the field are marked with green circles. The gray circle represents the position cut applied to the source catalog used for filament detection.}
\label{fig:A401_results}
\end{figure*}

Abell 401 (A401; $z =  0.0739$) and Abell 399 (A399; $z = 0.0722$) are separated approximately by $3.07 \  \mathrm{Mpc}$ (or $\sim37 \ \mathrm{arcmin}$) and are widely regarded to be in a pre-merger phase. X-ray and SZ observations provide strong evidence for a hot gas filament connecting the two clusters \citep{akamatsu2017, bonjean2018, hincks2022, radiconi2022}. Analysis of the galaxy populations shows no significant difference in star-formation activity between the clusters and the intercluster filament \citep{bonjean2018}. \citet{fu2024} confirms this by identifying a bridge of red-sequence galaxies connecting the two clusters.
The weak-lensing mass map presented in \citet{fu2024} reveals a stronger lensing signal for A401 than for A399; however, no clear lensing signal is seen from the intercluster filament. We report the \emph{first} weak-lensing (WL) detection of the intercluster bridge connecting  A401 and A399.

\begin{figure}[htb!]
\hspace*{-0.75cm} 
\includegraphics[width=\linewidth]{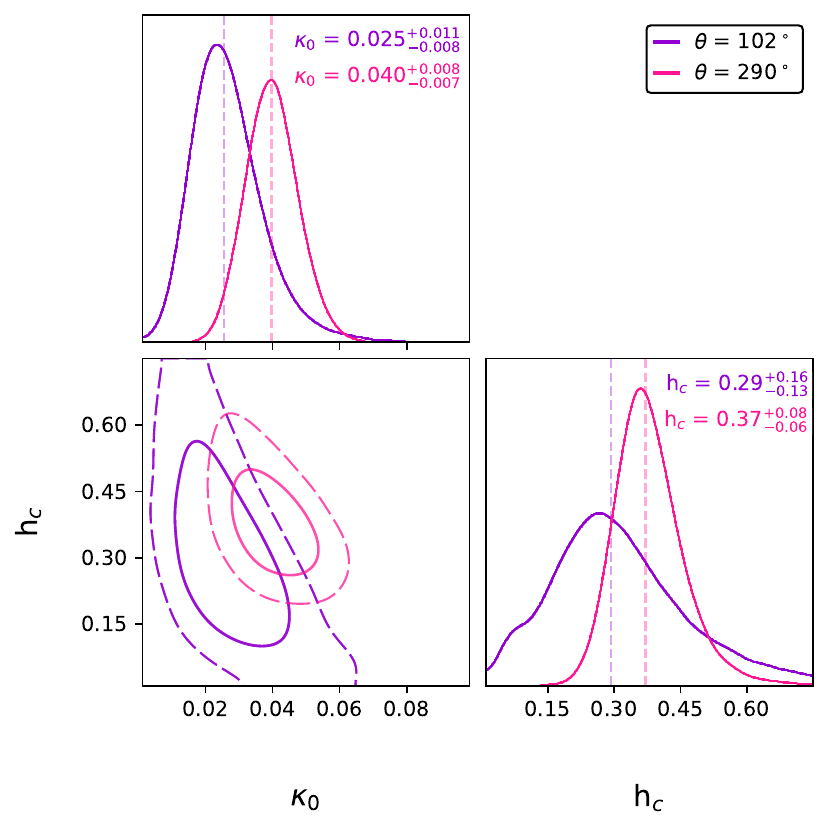}
\caption{A401: Posterior distributions of the filament parameters obtained via MCMC sampling. The inner (solid) and outer (dashed) contours in the joint distribution enclose the  68\% $(1\sigma)$ and 95\% $(2\sigma)$ credible regions, respectively. Dashed vertical lines indicate the median values of the marginalized distributions. The legend in the upper-right corner shows the color coding for the two detections.}
\label{fig:A401_mcmc}
\end{figure}

We present our results in the composite Figure~\ref{fig:A401_results}. Panel~(a) displays the statistics $\Gamma_{+}(\theta)$ and $\Gamma_{\times}(\theta)$, computed using the optimal matched filter with $r_1 = 0.92 \ \mathrm{Mpc}$ and $r_2 = 2.58  \ \mathrm{Mpc}$. The tangential signal $\Gamma_{+}$ exhibits two prominent peaks at $\theta=290^{\circ}$ (south; S) and $\theta=102^{\circ}$ (north; N), with significances of $5.8\sigma$ and $3.6\sigma$, respectively. A lower-significance ($2\sigma$) peak is detected towards the east; however, the absence of a corresponding sharp decline in the cross component $\Gamma_{\times}$ suggests that this feature may instead be associated with a halo. Panel~(c) displays the spatial distribution of spectroscopic members drawn from the NASA/IPAC Extragalactic Database \citep{nasa_ned, nasa_ned2} and DESI DR1 \citep{desi_dr1} in the A401-399 system (redshift criteria: $ 0.06 <z < 0.085$). Both filaments are aligned with overdensities in the spec-$z$ member distribution as well as the red-sequence galaxy distribution reported by \citet{fu2024}. Figure~\ref{fig:A401_mcmc} presents the posterior distributions for the maximum convergence $\kappa_0$ and characteristic width $h_{\mathrm{c}}$ for both filaments (results summarized in Table \ref{tab:A401_results_table}). The S filament is well constrained, with $\kappa_0 = 0.040^{+0.008}_{-0.007}$ and  $h_{\mathrm{c}} = 0.37^{+0.08}_{-0.06}$, whereas the N filament is less tightly constrained, with $\kappa_0 = 0.025^{+0.011}_{-0.008}$ and  $h_{\mathrm{c}} = 0.29^{+0.16}_{-0.13}$.

\begin{deluxetable}{cccc@{\hspace{15pt}}c}[h!]
\tablecaption{Filament Properties - A401 \label{tab:A401_results_table}}
\tablehead{
\colhead{Direction} & \colhead{Orientation} & \colhead{Significance} & \colhead{$\kappa_0$} & \colhead{$h_{\mathrm{c}}$} \\
\colhead{} & \colhead{(deg)} & \colhead{($\sigma$)} & \colhead{} & \colhead{(Mpc)}
}
\startdata
N & 102  & 3.6 & $0.025^{+0.011}_{-0.008}$ & $0.29^{+0.16}_{-0.13}$ \\
S & 290  & 5.8 & $0.040^{+0.008}_{-0.007}$ & $0.37^{+0.08}_{-0.06}$ \\
\enddata
\end{deluxetable}

\subsection{Abell 2029}

\begin{figure*}[htb!]
\centering
\hspace*{-0.75cm} 
\gridline{\fig{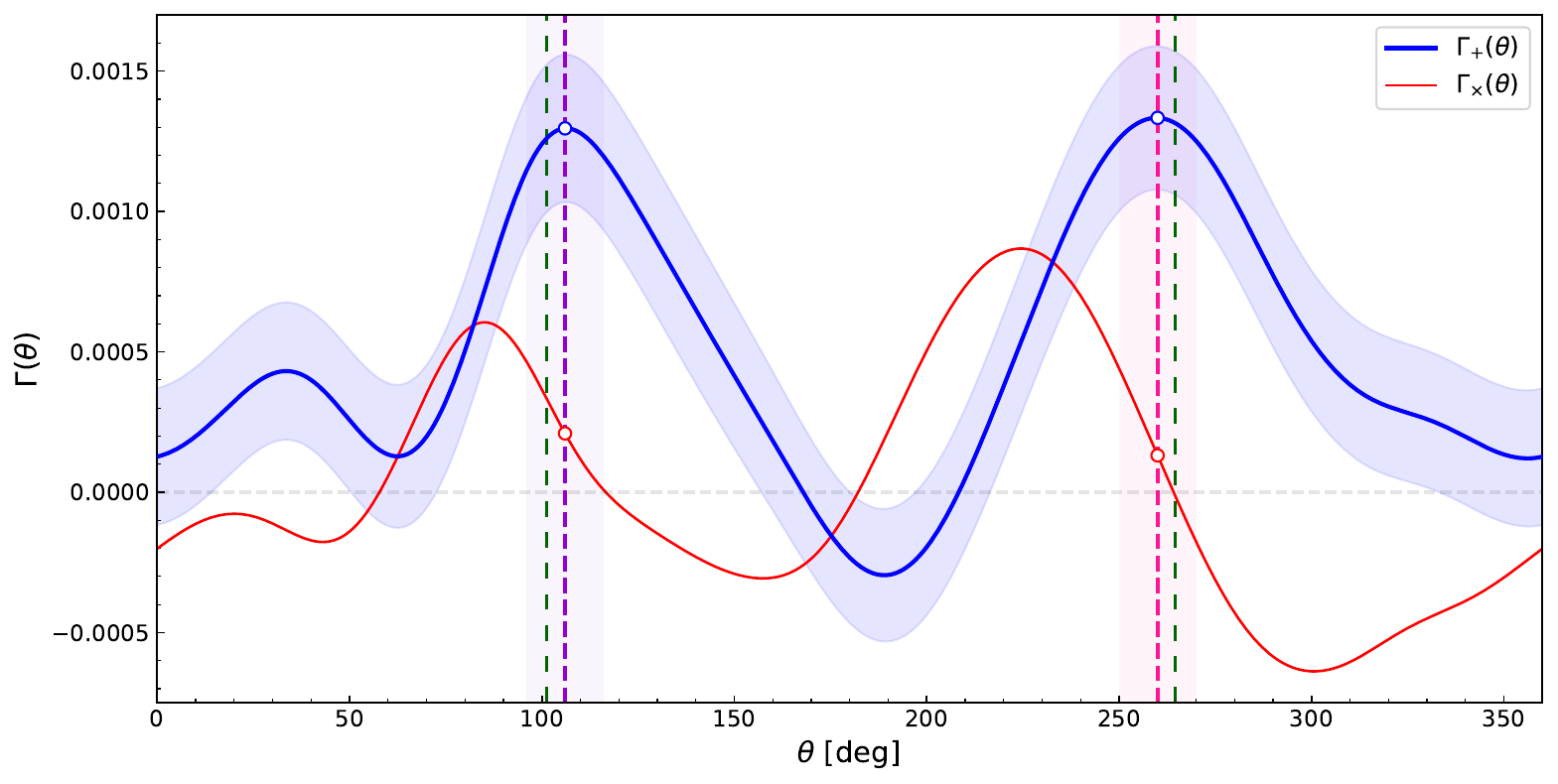}{0.92\textwidth}{(a) Linear plot}}
\gridline{\fig{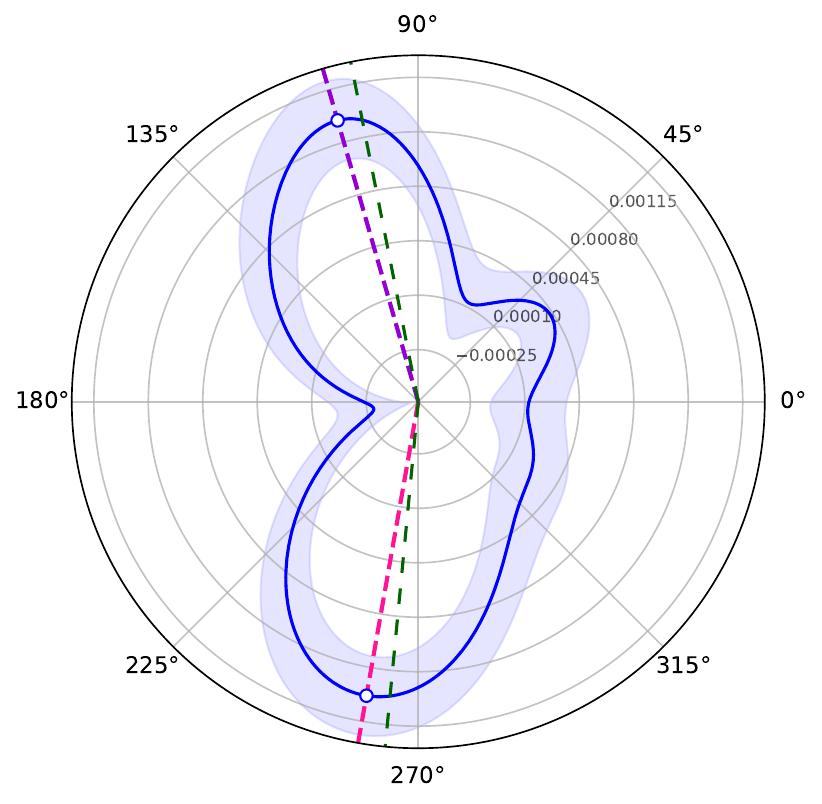}{0.49\textwidth}{(b) Polar plot}
         \fig{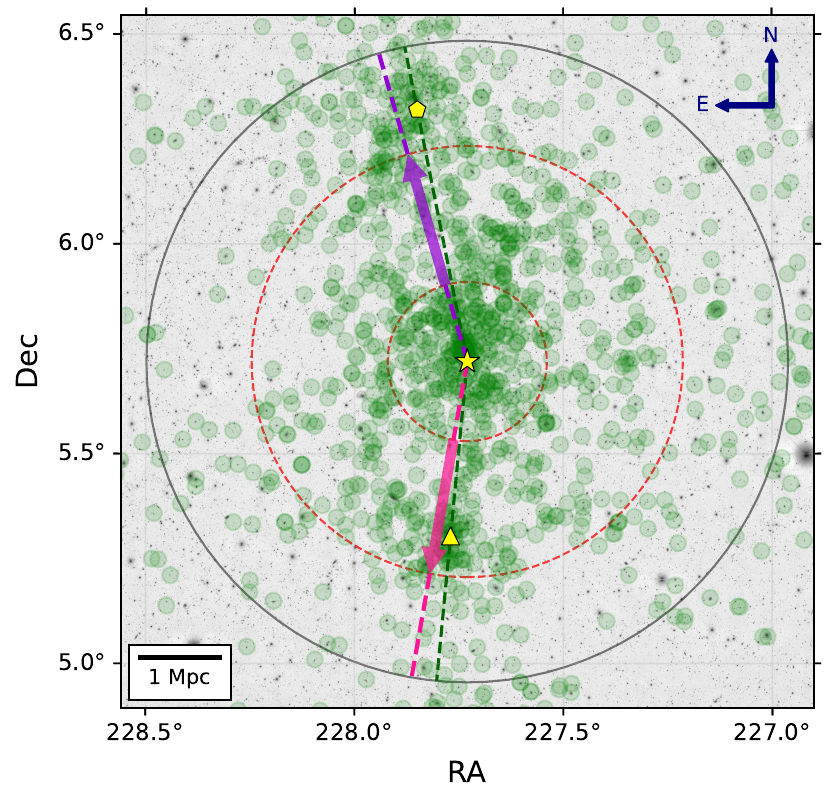}{0.49\textwidth}{(c) Overlaid plot}}
\caption{Filament detection results for the A2029 system. \textit{Top:} Matched-filter statistics $\Gamma_{+}(\theta)$ and $\Gamma_{\times}(\theta)$ as a function of the search angle $\theta$. The violet and pink vertical dashed lines mark the tangential signal peaks detected at $\theta = 106^{\circ}$ and $\theta = 260^{\circ}$, respectively. The corresponding light shade indicates a $\pm10^{\circ}$ window centered on the detected peak, provided for comparison with orientations of other clusters in the field. The loosely dashed green lines mark the orientations of A2033 and SIG relative to A2029, ordered counterclockwise from $0^\circ$ to $360^\circ$. \textit{Bottom-left:} Polar representation of the top panel. \textit{Bottom-right:} Detected filament orientations overlaid on the inverted $r$-band coadded image used for shape measurement. The dashed red circles represent the radial cutoffs $r_1 = 0.98\ \mathrm{Mpc}$ and $r_2 = 2.65\ \mathrm{Mpc}$. The violet and pink arrows extending from $r_1$ to $r_2$ indicate the spatial orientation of the detected filaments. A2029 (primary), A2033 (secondary) and SIG (secondary) are marked with a yellow star, pentagon, and triangle, respectively. Spectroscopic member galaxies $(0.07 < z < 0.09)$ in the field are marked with green circles.}
\label{fig:A2029_results}
\end{figure*}

Abell 2029 (A2029; $z = 0.0766$) is one of the most massive galaxy clusters in the local universe. Together with its close neighbor, Abell 2023 (A2033; $z = 0.0817$), located approximately $3.16 \ \mathrm{Mpc}$ (or $\sim37 \ \mathrm{arcmin}$) to the north, the system has been extensively studied in the X-rays. Deep \textit{Chandra} imaging reveals a sloshing spiral pattern centered on A2029, indicative of complex dynamical evolution \citep{paterno2013}. Early studies with ROSAT and \textit{Suzaku} observations \citep{walker2012} reported excess emission along the axis connecting A2029 and A2023. More recently, \citet{mirakhor2022} confirmed this feature at a significance of $6.5-7\sigma$, providing strong evidence for an intercluster gas filament. The complex structure of the system has also been widely analyzed in the optical regime through lensing measurements \citep{mccleary2020, sohn2019, fu2024}. Using intensive spectroscopy in combination with WL and X-ray measurements, \citet{sohn2019} demonstrated that A2033 and the Southern Infalling Group (SIG) are dynamical substructures bound to be accreted onto A2029. Building on previous lensing studies, \citet{fu2024} reported a high lensing signal for A2029 ($7.6 \sigma$), a moderate signal for A2033 ($\sim3\sigma$), and a tight alignment of $3$-$4\sigma$ contours with the intercluster direction. The SIG is detected at a lower significance of $\sim1.8\sigma$.

\begin{figure}[htb!]
\hspace*{-0.75cm} 
\includegraphics[width=\linewidth]{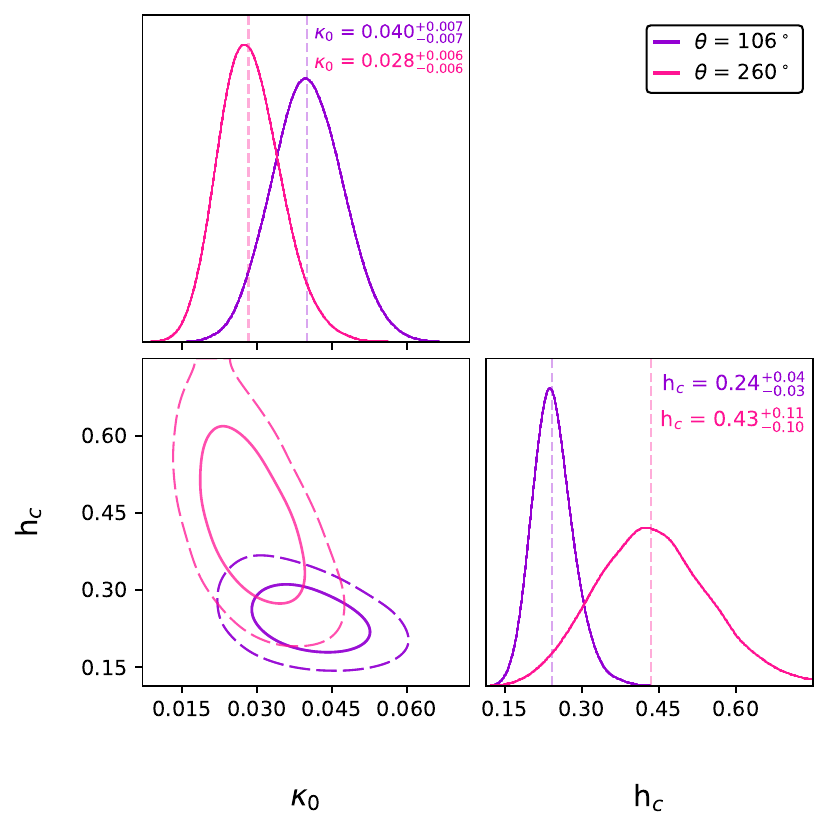}
\caption{A2029: Posterior distributions of the filament parameters obtained via MCMC sampling. The inner (solid) and outer (dashed) contours in the joint distribution enclose the  68\% $(1\sigma)$ and 95\% $(2\sigma)$ credible regions, respectively. Dashed vertical lines indicate the median values of the marginalized distributions. The legend in the upper-right corner shows the color coding for the two detections.}
\label{fig:A2029_mcmc}
\end{figure}

Figure~\ref{fig:A2029_results} summarizes our results. We detect two prominent filaments connecting A2029 to the SIG $(\theta = 260 ^\circ;\  \text{south})$ and A2033 $(\theta = 106 ^\circ; \ \text{north})$, with detection significances of $5.2 \sigma$ and $4.9 \sigma$, respectively. Notably, the southern (S) filament is detected at a higher significance despite the SIG having a lower WL mass than A2033 \citep{sohn2019}. This trend is also reflected in our parameter estimation analysis (see Figure~\ref{fig:A2029_mcmc} and Table \ref{tab:A2029_results_table}), which yields a much higher characteristic width for the S filament $(h_{\mathrm{c}} = 0.43^{+0.11}_{-0.10})$ than the N filament $(h_{\mathrm{c}} = 0.24^{+0.04}_{-0.03})$. The maximum convergences, however, follow the opposite trend: $\kappa_0 = 0.028^{+0.006}_{-0.006}$ (S) and $\kappa_0 = 0.040^{+0.007}_{-0.007}$ (N). Both filamentary structures are spatially consistent with the distribution of the spectroscopic members $(0.07<z<0.09)$ presented in Figure~\ref{fig:A2029_results}(c) and with the red-sequence galaxy distribution presented in \citet{fu2024}. A less prominent peak ($1.8 \sigma$) is observed in the north-west direction, coincident with an overdensity in the spec-$z$ member distribution. The absence of a pronounced decline in the cross statistic, however, indicates that this signal is not associated with a filamentary structure. Although the radial cutoffs derived for this system ($r_1 = 0.98 \ \mathrm{Mpc}$, $r_2 = 2.65 \ \mathrm{Mpc}$) include the SIG within the search space, owing to its shear profile remaining below the threshold of $\gamma_+ = 0.02$ (see Appendix~\ref{sec:appendix_c} for details), the filament detection results are largely unchanged when the analysis is repeated with revised cutoffs ($r_1 = 0.98 \ \mathrm{Mpc}$, $r_2 = 1.98 \ \mathrm{Mpc}$) that explicitly exclude the SIG: $4.4\sigma$ (N) and $5.7\sigma$ (S).

\begin{deluxetable}{cccc@{\hspace{15pt}}c}[h!]
\tablecaption{Filament Properties - A2029 \label{tab:A2029_results_table}}
\tablehead{
\colhead{Direction} & \colhead{Orientation} & \colhead{Significance} & \colhead{$\kappa_0$} & \colhead{$h_{\mathrm{c}}$} \\
\colhead{} & \colhead{(deg)} & \colhead{($\sigma$)} & \colhead{} & \colhead{(Mpc)}
}
\startdata
N & 106  & 4.9 & $0.040^{+0.007}_{-0.007}$ & $0.24^{+0.04}_{-0.03}$ \\
S & 260  & 5.2 & $0.028^{+0.006}_{-0.006}$ & $0.43^{+0.11}_{-0.10}$ \\
\enddata
\end{deluxetable}

\subsection{Abell 3558}

\begin{figure*}[ht!]
\centering
\hspace*{-0.75cm} 
\gridline{\fig{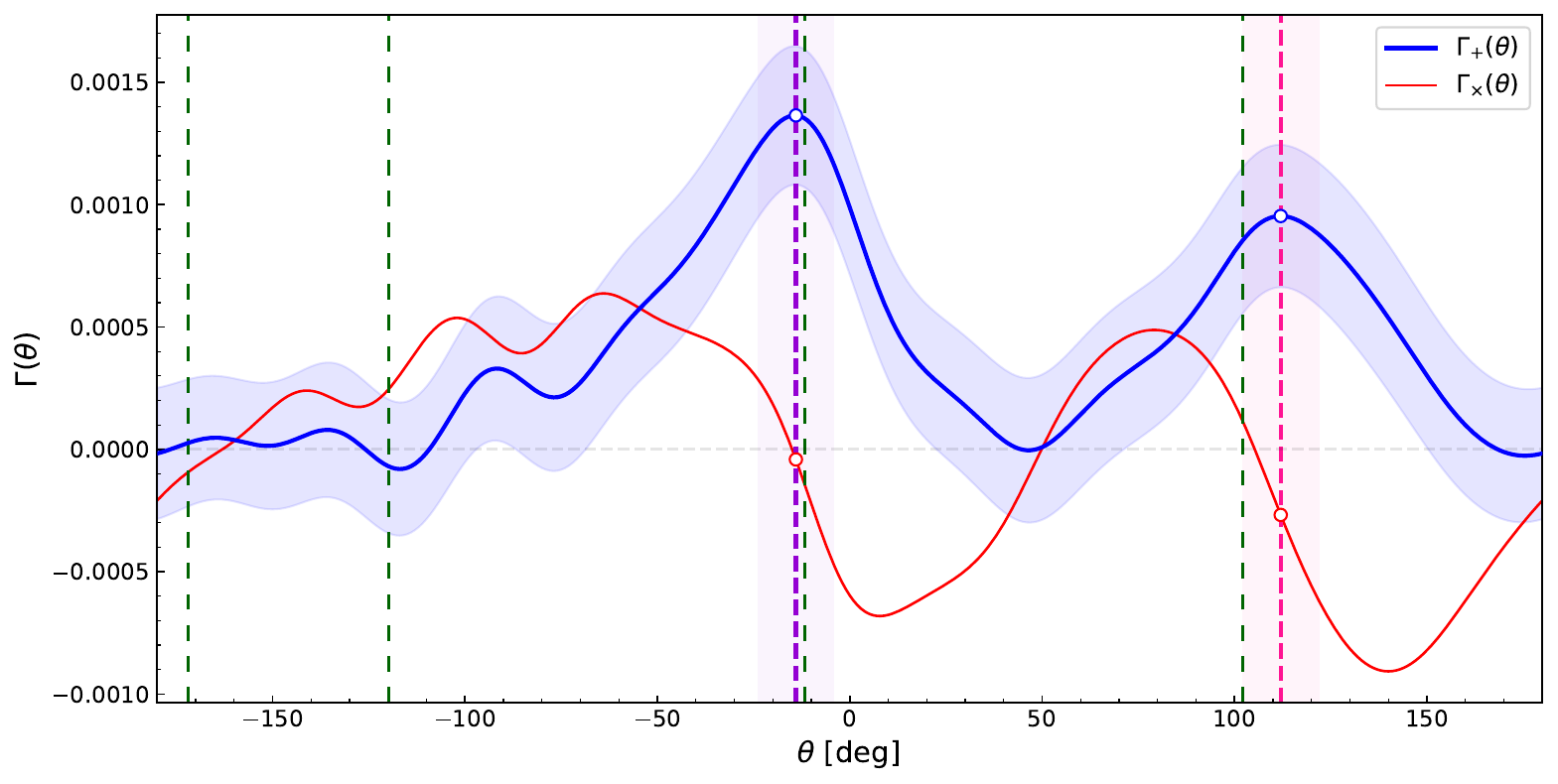}{0.92\textwidth}{(a) Linear plot}}
\gridline{\fig{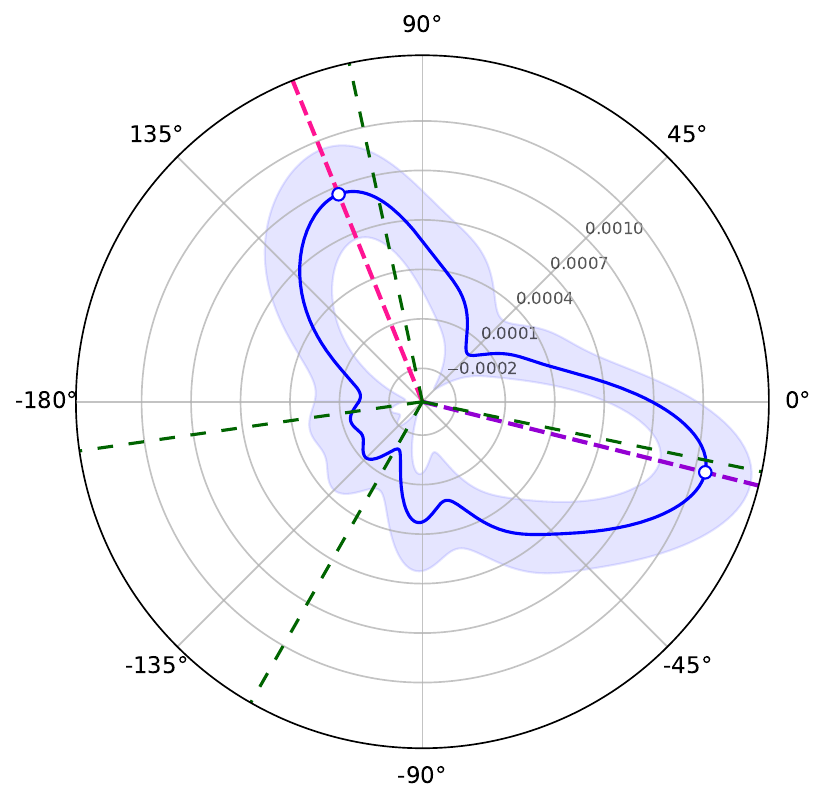}{0.49\textwidth}{(b) Polar plot}
         \fig{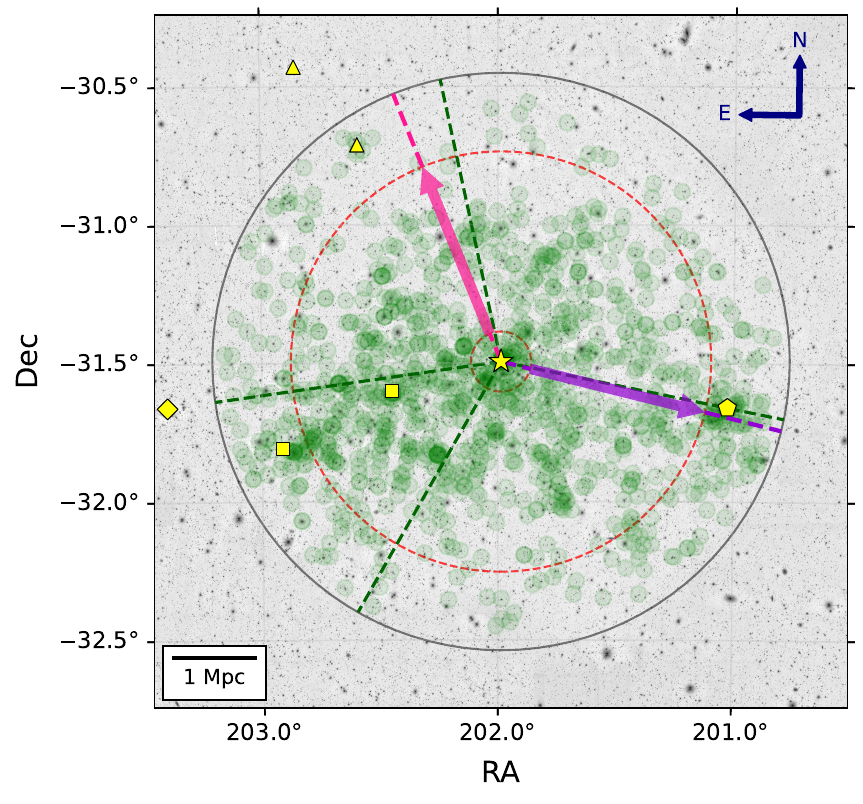}{0.49\textwidth}{(c) Overlaid plot}}
\caption{Filament detection results for the A3558 system. \textit{Top:} Matched-filter statistics $\Gamma_{+}(\theta)$ and $\Gamma_{\times}(\theta)$ as a function of the search angle $\theta$. The violet and pink vertical dashed lines mark the tangential signal peaks detected at $\theta = -14^{\circ}$ and $\theta = 112^{\circ}$, respectively. The corresponding light shade indicates a $\pm10^{\circ}$ window centered on the detected peak, provided for comparison with orientations of other clusters in the field. The loosely dashed green lines mark the orientations of A3562, A3560, A3556 and A3559 relative to A3558, ordered counterclockwise from $-180^\circ$ to $180^\circ$. \textit{Bottom-left:} Polar representation of the top panel. \textit{Bottom-right:} Detected filament orientations overlaid on the inverted $r$-band coadded image used for shape measurement. The dashed red circles represent the radial cutoffs $r_1 = 0.36\ \mathrm{Mpc}$ and $r_2 = 2.56\ \mathrm{Mpc}$. The violet and pink arrows extending from $r_1$ to $r_2$ indicate the spatial orientation of the detected filaments. A3558 (primary), A3556 (secondary), A3562 (secondary), SC 1327-312/SC 1329-313 (secondary) and FG1/FG2 (secondary) are marked with a yellow star, pentagon, diamond, squares, and triangles, respectively. Spectroscopic member galaxies $(0.04 < z < 0.06)$ in the field are marked with green circles.}
\label{fig:A3558_results}
\end{figure*}

Abell 3558 (A3558; $z = 0.048$) lies at the heart of the Shapley Supercluster (SSC) -- the largest conglomeration of Abell clusters in the local universe. Together with Abell 3562 (A3562; $z = 0.049$) to the east, Abell 3556 (A3556; $z = 0.048$) to the west, and two poor clusters embedded in between (SC 1327-312 and SC 1329-313), they form the Shapley Supercluster Core -- a continuous linear structure spanning $\sim 8 \ \mathrm{Mpc}$ (or $ \sim 2\ \mathrm{deg}$) permeated by hot gas. In contrast to the gas bridges reported in previous systems, direct detection of filaments in the SSC Core is sparse and contested \citep{kull1998, ursino2015, mitsuishi2012}. In particular, the bridge between A3558 and A3556 remains ambiguous: reported X-ray excesses have been attributed to cluster outskirts or unresolved point sources.

Optical studies, in comparison, have been more successful at finding these elusive features. Spectroscopic studies reveal that all 11 clusters in the SSC are inter-connected and lie embedded within a cosmic sheet at $z \sim 0.048$, in the process of collapsing into A3558. \citet{haines2018} find a filament-like overdensity extending from A3562 through A3558 to A3556, with a sharp low-density edge between A3558 and A3556. They also report evidence for a second stream of galaxies extending northward from A3558 towards A3559, with two small groups, Filament Group 1 \& 2 (FG1 \& FG2), embedded between them. WL maps reveal the continuous structure of the SSC Core, with a prominent peak at the BCG of A3558 and secondary peaks centered at A3556 and A3562 \citep{merluzzi2015, higuchi2020, fu2024}. Consistent with spectroscopic studies, \citet{higuchi2020} detected WL contours at the $1\sigma$ level, providing complementary evidence for filamentary structures linking A3558 with neighboring clusters (A3556, A3559 and A3562).

\begin{figure}[htb!]
\hspace*{-0.75cm} 
\includegraphics[width=\linewidth]{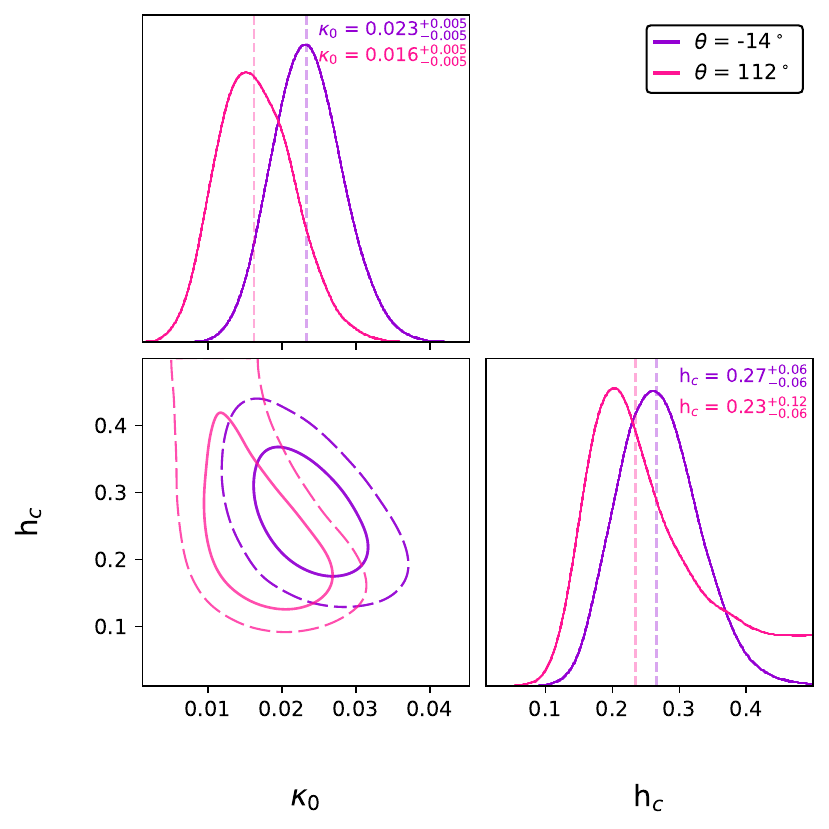}
\caption{A3558: Posterior distributions of the filament parameters obtained via MCMC sampling. The inner (solid) and outer (dashed) contours in the joint distribution enclose the  68\% $(1\sigma)$ and 95\% $(2\sigma)$ credible regions, respectively. Dashed vertical lines indicate the median values of the marginalized distributions. The legend in the upper-right corner shows the color coding for the two detections.}
\label{fig:A3558_mcmc}
\end{figure}

Figure~\ref{fig:A3558_results} summarizes our results. We detect two peaks in the tangential signal at $\theta = -14 ^\circ$ (west; W) and $\theta = 112 ^\circ$ (north; N) with significances of $4.8 \sigma$ and $3.3 \sigma$, respectively. The W detection is exactly aligned with A3556, while the N detection roughly follows the FG1-FG2 axis (eventually leading to A3559). A marginal tangential signal ($1.1\sigma$), accompanied by a substantial negative cross-gradient, is measured to the south. This is consistent with a $1-2 \sigma$ contour between A3558 and A3560 in the WL map reported by \citet{higuchi2020}. However, no filamentary structure is detected in the direction of A3562, despite the presence of a linear overdensity in the spec-$z$ member distribution $(0.04<z<0.06)$, as shown in Figure~\ref{fig:A3558_results}(c). We attribute the absence of a straight filament between A3558 and A3562 to the complex morphology of the region, which includes the smaller clusters SC 1327–312 and SC 1329–313. The W filament has a higher maximum convergence $(\kappa_0 = 0.023^{+0.005}_{-0.005})$ than the N filament $(\kappa_0 = 0.016^{+0.005}_{-0.005})$. Both filaments exhibit similar inferred widths, with $h_{\mathrm{c}} = 0.27^{+0.06}_{-0.06}$ for the W filament, and $h_{\mathrm{c}} = 0.23^{+0.12}_{-0.06}$ for the N filament (see Figure~\ref{fig:A3558_mcmc} and Table \ref{tab:A3558_results_table}).

\begin{deluxetable}{cccc@{\hspace{15pt}}c}[h]
\tablecaption{Filament Properties -- A3558 \label{tab:A3558_results_table}}
\tablehead{
\colhead{Direction} & \colhead{Orientation} & \colhead{Significance} & \colhead{$\kappa_0$} & \colhead{$h_{\mathrm{c}}$} \\
\colhead{} & \colhead{(deg)} & \colhead{($\sigma$)} & \colhead{} & \colhead{(Mpc)}
}
\startdata
W & -14  & 4.8 & $0.023^{+0.005}_{-0.005}$ & $0.27^{+0.06}_{-0.06}$ \\
N & 112  & 3.3 & $0.016^{+0.005}_{-0.005}$ & $0.23^{+0.12}_{-0.06}$ \\
\enddata
\end{deluxetable}

\subsection{Quantifying Terminal Cluster Contribution via Mock Simulations} \label{subsec:terminal_cluster_contribution}

\begin{figure*}[htb!]

\gridline{
\raisebox{-0.5\height}{
\fig{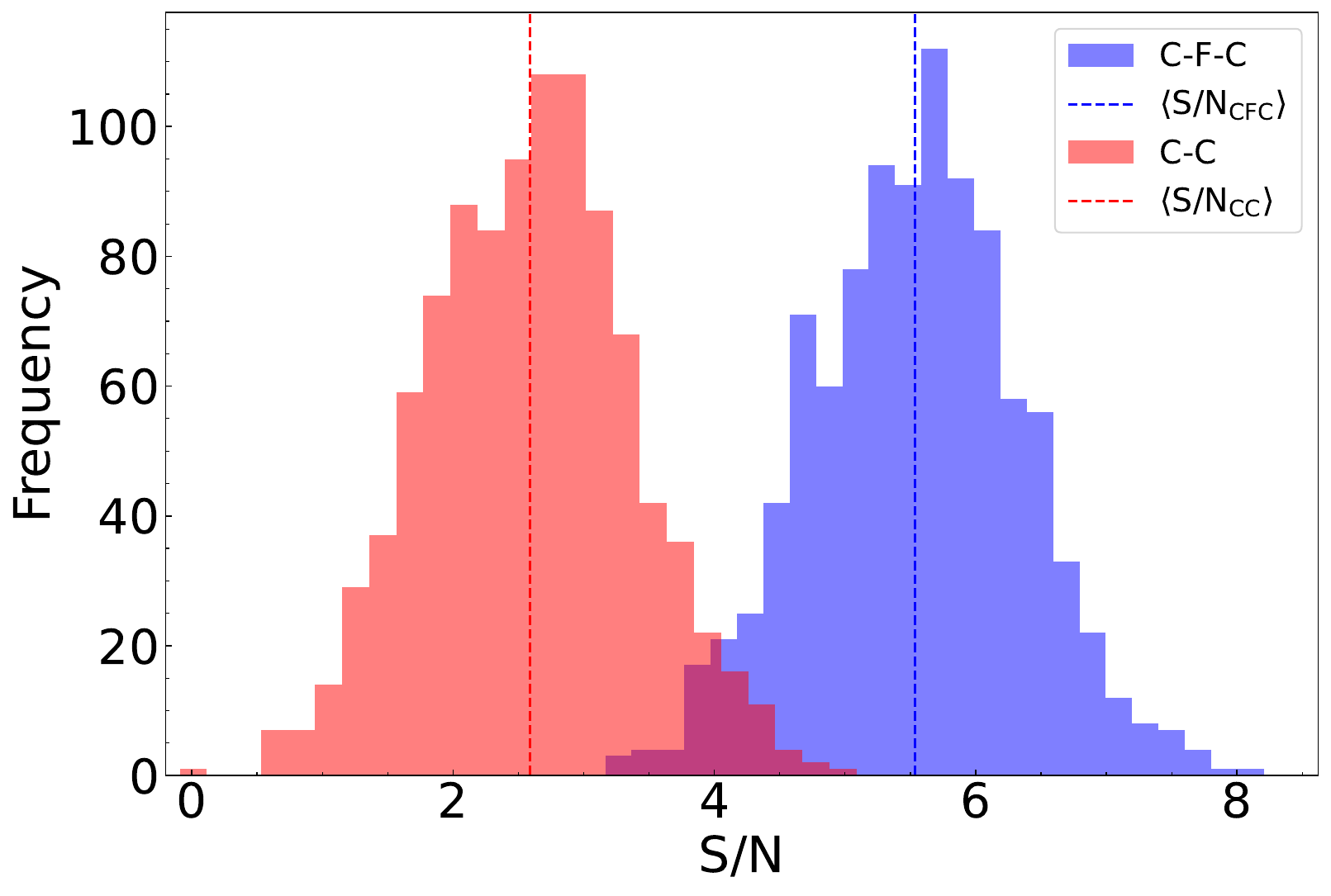}{0.4\textwidth}{(a) A}}
\raisebox{-0.49\height}{
\fig{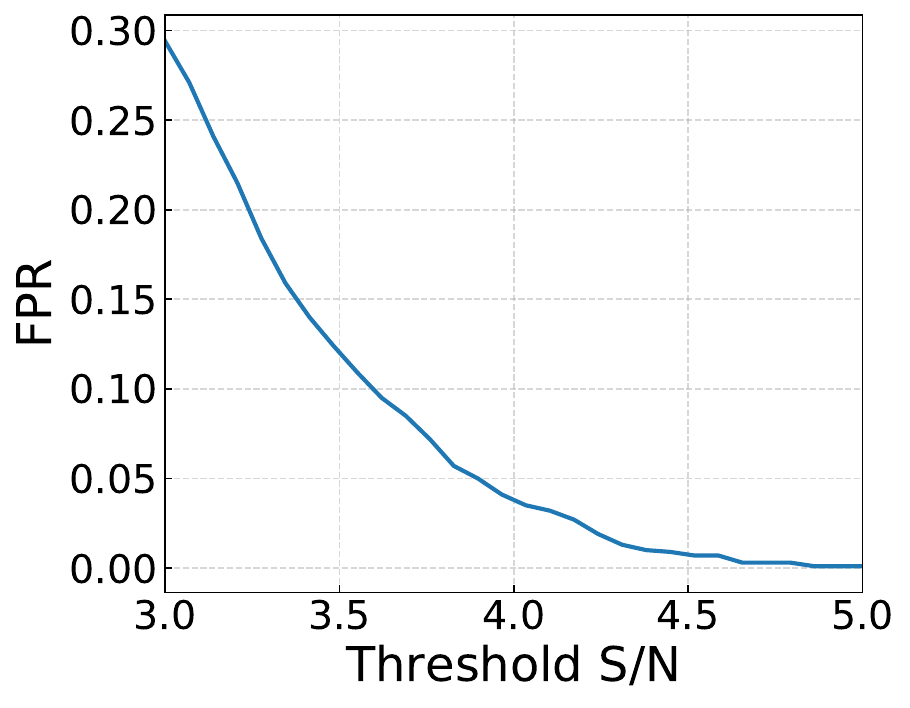}{0.28\textwidth}{(b) B}}
\raisebox{-0.49\height}{
\fig{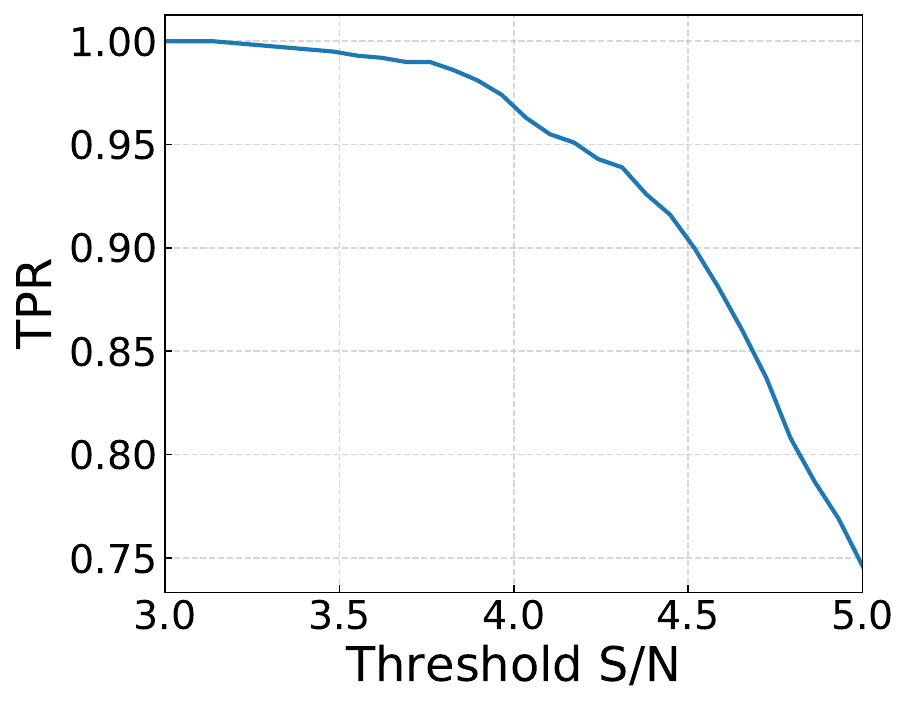}{0.28\textwidth}{(c) C}}
}
\caption{Mock simulation results quantifying the contribution of terminal clusters to filament detection significance. \textit{Left:} (a) S/N distributions from 1000 independent mock shear catalog realizations for the primary cluster--filament--secondary cluster (C-F-C, blue) and primary cluster--secondary cluster (C-C, red) configurations. Dashed blue and red vertical lines indicate the mean detection significances, $\langle S/N \rangle_{\rm C\text{-}F\text{-}C} = 5.5$ and $\langle S/N \rangle_{\rm C\text{-}C} = 2.6$, respectively. \textit{Right:} (b) False-positive rate (FPR) as a function of threshold S/N, showing the probability that terminal cluster contamination alone produces a detection exceeding a given significance. (c) True-positive rate (TPR) as a function of threshold S/N, showing the fraction of genuine filament-containing configurations (C-F-C) detected above a given threshold.}
\label{fig:cluster_contamination_test}
\end{figure*}

Some of the systems studied above, such as Abell 401 and Abell 2029, exhibit filaments connecting closely separated pair of clusters with angular separations of $\sim 0.6^\circ$. Although we examined the false-positive contribution of intervening secondary halos embedded within the search space in Section~\ref{sec:combined_statistic}, the tangential alignment of the shear induced by terminal clusters can also significantly bias the signal measured for the connecting filament. This effect becomes increasingly consequential when the terminal clusters are minimally separated.

We assess the signal bias introduced by terminal clusters using a mock setup designed to resemble features of the A401 and A2029 systems, following the procedure outlined in Section~\ref{subsec:mock_catalog}. We consider two configurations involving terminal clusters and a possible intervening filament: (i) primary cluster -- filament -- secondary cluster (C-F-C), and (ii) primary cluster -- secondary cluster with no connecting filament (C-C), which serves as the null case. A third configuration, primary cluster -- filament (C-F), could also be considered; however, the cluster bias in this case is expected to be smaller than that of the C-F-C configuration. We therefore exclude it from the analysis.

To reduce computational overhead and simultaneously sample both the null and filament-containing configurations within a single simulated field, we generate mock shear catalogs for a system at redshift $z = 0.07$ consisting of a primary cluster ($M_{\mathrm{200c}} = 8 \times 10^{14}\,\mathrm{M}_\odot$) and a secondary cluster ($M_{\mathrm{200c}} = 4 \times 10^{14}\,\mathrm{M}_\odot$) located at an separation of $r = 2.93 \ \mathrm{Mpc}$ ($37 \ \mathrm{arcmin} $) and $\phi = 45 ^ \circ$. This combination of primary and secondary cluster masses is representative of the systems studied in this work. Another identical secondary cluster ($M_{\mathrm{200c}} = 4 \times 10^{14}\,\mathrm{M}_\odot$) is placed at the same separation $r = 37 \ \mathrm{arcmin}$ away but at $\phi = 225 ^ \circ$. A filament ($\kappa_0 = 0.03$, $h_{\mathrm{c}} = 0.25 \ \mathrm{Mpc}$) is included between the primary cluster and the secondary cluster at $\theta_{\mathrm{f}} = 45^\circ$, corresponding to the C-F-C configuration . No filament is included between the primary cluster and the secondary cluster at $\phi = 225^\circ$ to serve as the null configuration (C-C).  We adopt a background source density of $n_g = 8 \ \mathrm{arcmin}^{-2}$ and add zero-mean Gaussian noise with $\sigma_{\gamma}$ = 0.32 to mimic the LoVoCCS shear dataset. To mitigate the impact of noise-induced fluctuations in the detection significance, we generate $N_\mathrm{iterations} = 1000$ independent realizations of the mock shear catalog and perform a statistical analysis of the resulting significances.

For each mock realization, we run the filament detection analysis described in Section \ref{sec:implementation} using an optimal matched filter analogous to that used for the A401 and A2029 systems ($h_{\mathrm{c, filter}} = 0.15 \ \mathrm{Mpc}, \ r_1 = 0.88 \ \mathrm{Mpc}$, $\ r_2 = 2.49 \ \mathrm{Mpc}$). To facilitate comparison with the results presented in previous sections, we include the mean LSS noise uncertainty across all systems ($\langle\sigma_{\mathrm{LSS}}\rangle = 1.21 \times 10^{-4}$) in the signal-to-noise ratio calculation. We record the S/N measured at $\theta_{\mathrm{}} = 45^\circ$ (C-F-C) and $\theta_{\mathrm{}} = 225^\circ$ (C-C) for all 1000 realizations. 

Figure~\ref{fig:cluster_contamination_test}(a) shows the resulting S/N distributions for the filament-containing and null configurations. Both distributions are unimodal with mean detection significances of $\langle S/N \rangle_{\rm C\text{-}C}= 2.6$ and $\langle S/N \rangle_{\rm C\text{-}F\text{-}C}= 5.5$, respectively. These results indicate that, on average, massive terminal clusters in close proximity can alone contribute to an apparent filament detection of $\mathrm{S/N} \simeq 2-3$. However, the probability of obtaining false-positive detections with $\mathrm{S/N} > 3$ decreases rapidly with increasing signal-to-noise ratio. Consequently, detections with $\mathrm{S/N} \gtrsim 5$ are most likely to be produced by genuine filaments rather than contamination from the terminal clusters. To further quantify this distinction, we compute the false-positive rate (FPR) and true-positive rate (TPR) associated with the distributions as a function of a chosen threshold signal-to-noise ratio $(\mathrm{SNR}_{\rm thresh})$. 
\begin{align*}
    \mathrm{FPR}(\mathrm{SNR}_{\rm thresh}) &= \frac{N\!\left(\mathrm{SNR}_{\rm C\text{-}C} > \mathrm{SNR}_{\rm thresh}\right)}{N_{\rm iterations}}, \\ 
    \mathrm{TPR}(\mathrm{SNR}_{\rm thresh}) &= \frac{N\!\left(\mathrm{SNR}_{\rm C\text{-}F\text{-}C} > \mathrm{SNR}_{\rm thresh}\right)}{N_{\rm iterations}},
\end{align*}
where $N(\cdot)$ denotes the number of iterations satisfying the condition. 

Figure~\ref{fig:cluster_contamination_test}(b) and \ref{fig:cluster_contamination_test}(c) present the resulting FPR and TPR curves respectively. Figure~\ref{fig:cluster_contamination_test}(b) shows the declining probability of obtaining a false-positive filament detection from the null (C-C) configuration at high signal-to-noise ratios. Specifically, the false-positive rate is 29\% at $\mathrm{S/N} = 3$, 12\% at $\mathrm{S/N} = 3.5$, 4\% at $\mathrm{S/N} = 4$, and only 0.1\% at $\mathrm{S/N} = 5$. This implies that the $\mathrm{S/N} \sim 3$ detections reported in our analysis should be interpreted with some caution, as we cannot entirely rule out terminal cluster contribution. For detections with $\mathrm{S/N} \gtrsim 4$, however, the false-positive rate is negligible, suggesting that such detections are highly likely to correspond to genuine filaments.

\subsection{Null Test: Abell 2351} 

\begin{figure*}[htb!]
\gridline{\fig{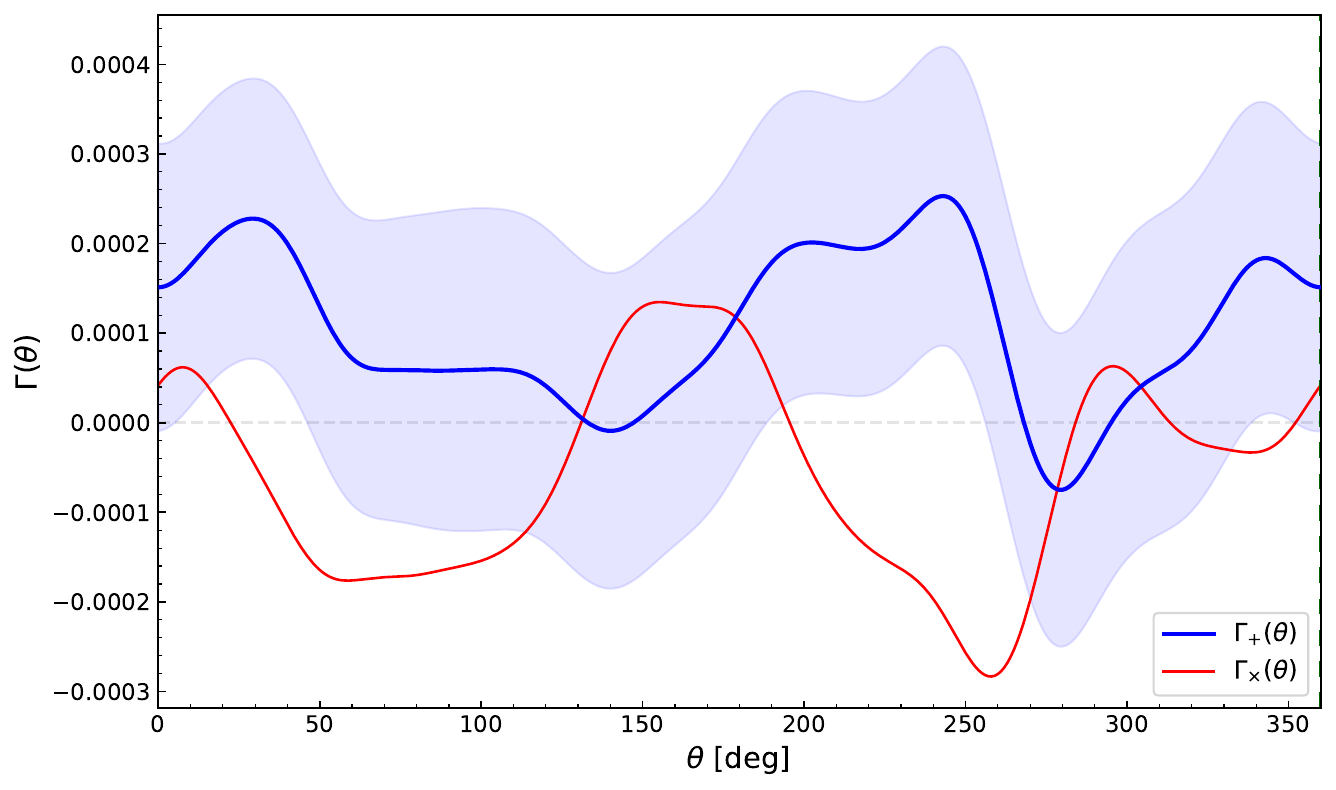}{0.6\textwidth}{}
          \fig{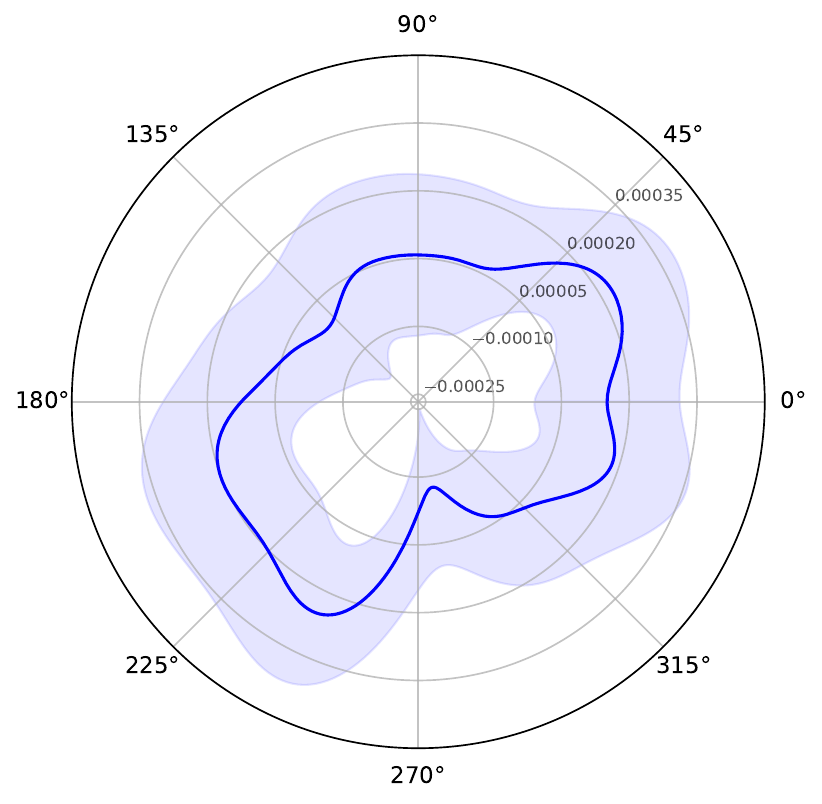}{0.39\textwidth}{}}
\caption{Filament detection results for A2351. \textit{Left:} Matched-filter statistics $\Gamma_{\times}(\theta)$ and $\Gamma_{+}(\theta)$ as a function of the search angle $\theta$. The blue and red solid lines correspond to the tangential and cross components, respectively, with the light blue shade indicating the $1\sigma$ uncertainty in the tangential signal. \textit{Right:} Polar representation of the left panel.}
\label{fig:A2351_results}
\end{figure*}

Compared to the systems studied above, Abell 2351 (A2351; $z =  0.0897$) is relatively less massive, with an X-ray mass of $M_\mathrm{X500c} = 2.75\times10^{14}\,\mathrm{M}_\odot$ \citep{piffaretti2011} and a WL mass of $M_\mathrm{WL200c} = 2.8\times10^{14}\,\mathrm{M}_\odot$ (LoVoCCS Collaboration; private communication; Oct 1, 2025).

We therefore select A2351 as a control system because of its low mass, limited substructure in both the WL mass map and the red-sequence member distribution \citep{fu2024}, and consequently the absence of a close companion cluster. 

The choice to treat A2351 as a null test is motivated by the the well-established mass dependence of cosmic web connectivity. More massive clusters tend to occupy higher-connectivity nodes of the cosmic web and are fed by a larger number of dense filaments, whereas lower-mass systems typically reside in environments with fewer and weaker filamentary connections \citep{aragon2010, cautun2014, cordis2018, sarron2019}. Furthermore, N-body simulations show that close cluster pairs with separations $\leq 5\,h^{-1}\mathrm{Mpc}$ are connected by at least one filament \citep{colberg2005}. Given its low mass and lack of external substructure, A2351 is not expected to host a rich, detectable filamentary network. The null result for A2351 thus reflects a physically motivated expectation rather than a limitation of our methods.

Figure \ref{fig:A2351_results} presents our results. We find no significant detections $(> 1.5\sigma)$ in the tangential statistic $\Gamma_+(\theta)$. The absence of spurious peaks in A2351 suggests that the reported detections in previous systems are unlikely to be artifacts introduced by the analysis methods.
Furthermore, as demonstrated in Section~\ref{subsec:terminal_cluster_contribution}, terminal cluster contamination alone is insufficient to produce significant false-positive detections ($>4\sigma$). This indicates that the matched-filter statistic is robust against terminal cluster contamination, and that the signals recovered in the higher-mass systems are more plausibly associated with genuinely filamentary presence.

\section{Conclusions} \label{sec:conclusions}
In this study, we applied a matched-filter method to detect the weak-lensing signal of intercluster filaments in three massive systems centered on Abell 401, Abell 2029, and Abell 3558. To improve its performance in the presence of noise, the matched filter was optimized using mock simulations of the observed shear field. We also investigated the use of the B-mode signal of filaments to enhance detection significance and distinguish genuine filamentary signal from cluster contamination.

Our main results are as follows:

(1) In each system, we detect two prominent filaments with significances in the range $3.3\sigma-5.8\sigma$. In particular, we report the \emph{first} definitive WL detections $(\gtrsim5\sigma)$ of the intercluster bridges in the following cluster pairs: A401/399, A2029/2033, A2029/SIG, and A3558/3556.

(2) Except for A3558/3562, we identify intercluster filaments connecting each primary-secondary cluster pair within $10^\circ$ of the intercluster direction measured from the primary cluster.

(3) The orientation of detected intercluster filaments is consistent with the spatial distribution of spectroscopic cluster members drawn from NED \citep{nasa_ned, nasa_ned2} and DESI DR1 \citep{desi_dr1}, as well as the red-sequence galaxy distribution reported by \citet{fu2024}.

(4) We measured the physical properties of detected filaments via MCMC sampling, obtaining maximum convergences in the range $\kappa_0 \sim 0.016 - 0.040$ and characteristic widths spanning $h_{\mathrm{c}} \sim 0.23 - 0.43 \ \mathrm{Mpc}$. Among all systems studied in this paper, A2029 exhibits the strongest filament detections with the most tightly constrained physical parameters.

These findings are consistent with the \textit{cosmic web} picture of structure formation and provide direct observational evidence to support predictions from N-body simulations \citep{colberg2005} that massive clusters are typically embedded in a rich network of filaments with high-density contrast.

We note several limitations of the present work. While our sample of three cluster systems is sufficient to demonstrate the efficacy of the matched-filter method for filament detection, a substantially larger sample is required to characterize the ensemble properties of filaments. In a follow-up paper, we plan to extend our analysis to additional clusters from the LoVoCCS survey to develop a more comprehensive understanding of low-redshift $(z \lesssim 0.1)$ filamentary structures. Given the relatively low source density in our survey after applying selection cuts, careful characterization of the shape noise is imperative for effective filter optimization. In this study, we approximate the noise as Gaussian; however, future efforts will aim to model the spectral properties of shear noise in our observations in greater detail. To mitigate the effect of shear contamination from nearby clusters, we apply radial cuts to the matched filter. In future work, we plan to incorporate adjacent clusters directly into the mass model and account for the redshift distribution of background sources to enable an accurate estimation of the linear mass density of filaments. 

\begin{acknowledgments}
We thank the anonymous reviewer for their comments that greatly improved our manuscript. We are grateful to Shenming Fu for his constructive feedback on this paper. We would like to thank the members of the Observational Cosmology, Gravitational Lensing, and Astrophysics Group at Brown University, in particular Anthony Englert, Zacharias Escalante and Shenming Fu, for their contribution in processing the data for all galaxy clusters analyzed in this work using the LoVoCCS pipeline.

We acknowledge support from the National Science Foundation (No. AST-2108287; Collaborative Research; LoVoCCS).

This project used data obtained with the Dark Energy Camera (DECam), which was constructed by the Dark Energy Survey (DES) collaboration. Funding for the DES Projects has been provided by the DOE and NSF (USA), MISE (Spain), STFC (UK), HEFCE (UK), NCSA (UIUC), KICP (U. Chicago), CCAPP (Ohio State), MIFPA (Texas A\&M), CNPQ, FAPERJ, FINEP (Brazil), MINECO (Spain), DFG (Germany) and the Collaborating Institutions in the Dark Energy Survey, which are Argonne Lab, UC Santa Cruz, University of Cambridge, CIEMAT-Madrid, University of Chicago, University College London, DES-Brazil Consortium, University of Edinburgh, ETH Z\"urich, Fermilab, University of Illinois, ICE (IEEC-CSIC), IFAE Barcelona, Lawrence Berkeley Lab, LMU M\"unchen and the associated Excellence Cluster Universe, University of Michigan, NSF NOIRLab, University of Nottingham, Ohio State University, OzDES Membership Consortium, University of Pennsylvania, University of Portsmouth, SLAC National Lab, Stanford University, University of Sussex, and Texas A\&M University.

Based on observations at NSF Cerro Tololo Inter-American Observatory, NSF NOIRLab (NOIRLab Prop. ID 2019A-0308; PI: I. Dell’Antonio), which is managed by the Association of Universities for Research in Astronomy (AURA) under a cooperative agreement with the U.S. National Science Foundation.

This research used data obtained with the Dark Energy Spectroscopic Instrument (DESI). DESI construction and operations is managed by the Lawrence Berkeley National Laboratory. This material is based upon work supported by the U.S. Department of Energy, Office of Science, Office of High-Energy Physics, under Contract No. DE–AC02–05CH11231, and by the National Energy Research Scientific Computing Center, a DOE Office of Science User Facility under the same contract. Additional support for DESI was provided by the U.S. National Science Foundation (NSF), Division of Astronomical Sciences under Contract No. AST-0950945 to the NSF’s National Optical-Infrared Astronomy Research Laboratory; the Science and Technology Facilities Council of the United Kingdom; the Gordon and Betty Moore Foundation; the Heising-Simons Foundation; the French Alternative Energies and Atomic Energy Commission (CEA); the National Council of Humanities, Science and Technology of Mexico (CONAHCYT); the Ministry of Science and Innovation of Spain (MICINN), and by the DESI Member Institutions: \href{www.desi.lbl.gov/collaborating-institutions}{www.desi.lbl.gov/collaborating-institutions}. The DESI collaboration is honored to be permitted to conduct scientific research on I’oligam Du’ag (Kitt Peak), a mountain with particular significance to the Tohono O’odham Nation. Any opinions, findings, and conclusions or recommendations expressed in this material are those of the author(s) and do not necessarily reflect the views of the U.S. National Science Foundation, the U.S. Department of Energy, or any of the listed funding agencies.

The Hyper Suprime-Cam (HSC) collaboration includes the astronomical communities of Japan and Taiwan, and Princeton University. The HSC instrumentation and software were developed by the National Astronomical Observatory of Japan (NAOJ), the Kavli Institute for the Physics and Mathematics of the Universe (Kavli IPMU), the University of Tokyo, the High Energy Accelerator Research Organization (KEK), the Academia Sinica Institute for Astronomy and Astrophysics in Taiwan (ASIAA), and Princeton University. Funding was contributed by the FIRST program from the Japanese Cabinet Office, the Ministry of Education, Culture, Sports, Science and Technology (MEXT), the Japan Society for the Promotion of Science (JSPS), Japan Science and Technology Agency (JST), the Toray Science Foundation, NAOJ, Kavli IPMU, KEK, ASIAA, and Princeton University. 

This paper makes use of software developed for the Large Synoptic Survey Telescope. We thank the LSST Project for making their code available as free software at  \href{http://dm.lsst.org}{http://dm.lsst.org}

This paper is based in part on data collected at the Subaru Telescope and retrieved from the HSC data archive system, which is operated by the Subaru Telescope and Astronomy Data Center (ADC) at National Astronomical Observatory of Japan. Data analysis was in part carried out with the cooperation of Center for Computational Astrophysics (CfCA), National Astronomical Observatory of Japan. The Subaru Telescope is honored and grateful for the opportunity of observing the Universe from Maunakea, which has the cultural, historical and natural significance in Hawaii. 

This research has made use of the NASA/IPAC Extragalactic Database (NED), which is operated by the Jet Propulsion Laboratory, California Institute of Technology, under contract with the National Aeronautics and Space Administration.

This research uses services or data provided by the Astro Data Lab, which is part of the Community Science and Data Center (CSDC) Program of NSF NOIRLab. NOIRLab is operated by the Association of Universities for Research in Astronomy (AURA), Inc. under a cooperative agreement with the U.S. National Science Foundation.

This research was conducted using computational resources and services at the Center for Computation and Visualization, Brown University.
\end{acknowledgments}

\facilities{Blanco, Astro Data Archive, Astro Data Lab}

\software{\texttt{astropy} \citep{astropy2013, astropy2018, astropy2022},
          \texttt{galsim} \citep{galsim2015},
          \texttt{numpy} \citep{numpy},
          \texttt{scipy} \citep{scipy},
          \texttt{emcee} \citep{emcee},
          \texttt{pandas} \citep{pandas1, pandas2},
          \texttt{matplotlib} \citep{matplotlib},
          \texttt{corner} \citep{corner}. \\ 
    }

The filament detection code used to produce the figures presented in this paper is publicly available at: \dataset[doi:10.5281/zenodo.21146400]{https://doi.org/10.5281/zenodo.21146400} and\dataset[GitHub]{https://github.com/rahulshinde98/filament-detection}.

\appendix
\section{Optimal Cutoff Frequency}\label{sec:appendix_a}

\begin{figure*}[htb!]
\gridline{\fig{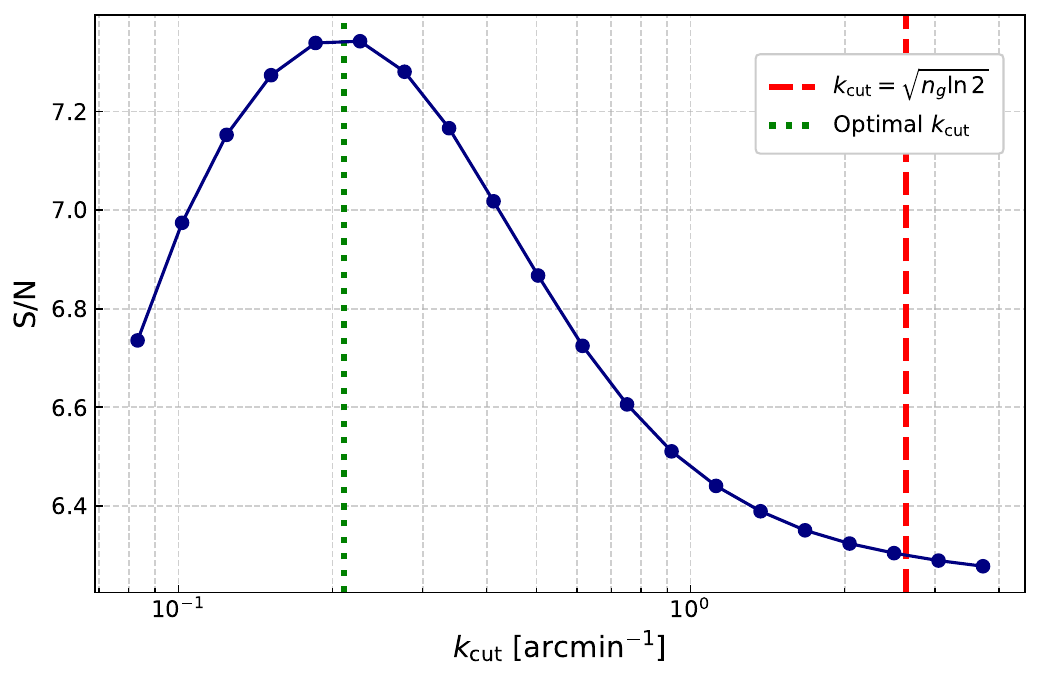}{0.49\textwidth}{(a) Filter S/N Performance}
         \fig{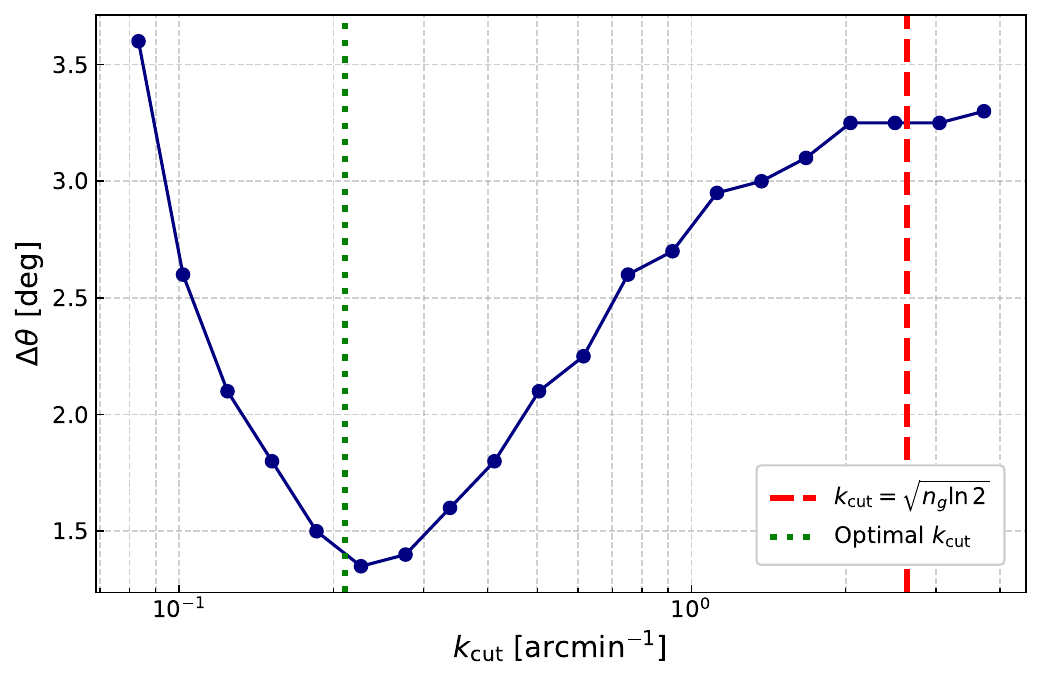}{0.49\textwidth}{(b) Orientation Error}}
\caption{Performance of the optimal matched filter as a function of the cutoff frequency, $k_{\mathrm{cut}}$, evaluated using multiple realizations of mock shear data. \textit{Left:} Mean signal-to-noise ratio (S/N) of filament detections plotted against the cutoff frequency. The S/N profile exhibits a maximum at $k_{\mathrm{cut}} = 0.21 \ \mathrm{arcmin}^{-1}$ (dotted green line), indicating the optimal cutoff frequency used in our analysis. The red dashed line indicates the cutoff frequency obtained from the definition $k_{\mathrm{cut}} = \sqrt{n_{\mathrm{g}} \ln 2}$. \textit{Right:} Absolute orientation offset between the detected and true filament orientations, $\Delta\theta(k_{\mathrm{cut}}) = |\theta_{\mathrm{det}} - \theta_{\mathrm{f}}|$, as a function of the cutoff frequency. The filter tuned to the optimal cutoff frequency identified in the left panel ($k_{\mathrm{cut}} = 0.21 \ \mathrm{arcmin}^{-1}$) also minimizes the orientation error.}
\label{fig:optimal_cutoff_frequency}
\end{figure*}

As noted in Section \ref{subsubsec:shape_noise}, the definition $k_{\mathrm{cut}} = \sqrt{n_{\mathrm{g}} \ln 2}$ yields a suboptimal filter for our dataset. Here we describe how mock data can be used to determine the optimal cutoff frequency. Following the procedure outlined in Section \ref{subsec:mock_catalog}, we generate multiple realizations of mock shear data for a configuration with a filament centered on a primary cluster $(M_{\mathrm{200c}} = 5 \times 10^{14} \ \mathrm{M}_{\odot})$ and oriented at an angle $\theta_{\mathrm{f}} = 135^\circ$. The filament convergence and characteristic width are fixed to $\kappa_0 = 0.03$ and $h_{\mathrm{c}} = 0.25 \ \mathrm{Mpc}$. We add zero-mean Gaussian noise with $\sigma_{\gamma} = 0.32$ to simulate the LoVoCCS dataset. The unoptimized matched filter with width $h_{\mathrm{c, filter}} = 0.15 \ \mathrm{Mpc}$ is restricted to the annulus $r_1 < r < r_2$, with $r_1 = 0.91 \ \mathrm{Mpc}$ and $r_2 = 3.64 \ \mathrm{Mpc}$. 

For each mock realization, we run the filament-detection analysis described in Section \ref{sec:implementation} with a range of cutoff frequencies $(0.08 \ \mathrm{arcmin}^{-1} < k_{\mathrm{cut}} < 3.72 \ \mathrm{arcmin}^{-1})$ used to optimize the filter. We record the peak S/N of the detections corresponding to filament branches at $\theta = 135^\circ$ and $\theta = 315^\circ$ for each run. This process is repeated for 10 mock realizations (20 filament branches in total) to compute the mean S/N as a function of the cutoff frequency. 

As shown in Figure~\ref{fig:optimal_cutoff_frequency}(a), we obtain a unimodal profile for $\mathrm{S/N}(k_{\mathrm{cut}})$ with a maximum at $k_{\mathrm{cut}} = 0.21 \ \mathrm{arcmin}^{-1}$ and a steep decline in both directions. For the average source density in our dataset, $\langle n_{\mathrm{g}}\rangle \sim 10 \ \mathrm{arcmin}^{-2}$, the earlier definition, $k_{\mathrm{cut}} = \sqrt{n_{\mathrm{g}} \ln 2}$ corresponds to a cutoff frequency of $k_{\mathrm{cut}} = 2.63 \ \mathrm{arcmin}^{-1}$. Lowering the cutoff frequency from $k_{\mathrm{cut}} = 2.63  \ \mathrm{arcmin}^{-1}$ to $k_{\mathrm{cut}} = 0.21 \ \mathrm{arcmin}^{-1}$ reduces the angular resolution of the filter in real space, thereby effectively suppressing noise at low $k$-modes and improving the performance by $\Delta \mathrm{S/N} \sim 1$. 

Another performance metric is the filter's ability to recover the input filament orientation. Figure~\ref{fig:optimal_cutoff_frequency}(b) displays the orientation offset, $\Delta\theta(k_{\mathrm{cut}}) = |\theta_{\mathrm{det}} - \theta_{\mathrm{f}}|$, plotted as a function of the cutoff frequency $k_{\mathrm{cut}}$. Here, $\theta_{\mathrm{det}}$ and $\theta_{\mathrm{f}}$ denote the detected and true filament orientations, respectively. The filter constructed with $k_{\mathrm{cut}} = 0.21 \ \mathrm{arcmin}^{-1}$ (optimal cutoff frequency identified above) consistently yields small offsets across the examined frequency range, demonstrating its reliability. Thus, throughout this paper, we adopt a cutoff frequency of $k_{\mathrm{cut}} = 0.21 \ \mathrm{arcmin}^{-1}$ to optimize the matched filter. 

\section{LSS Noise Contribution: Analytical Estimation}\label{sec:appendix_b}

To validate the numerical estimation of the LSS noise uncertainty ($\sigma_{\mathrm{LSS}}$) provided in Section~\ref{subsubsec:lssnoise}, we present an alternative analytical approach here. The underlying principle is to leverage the shear power spectra of the large-scale structure derived from the $\kappa$TNG simulations and relate it directly to the variance of the tangential shear statistic used for filament detection. While we previously expressed the tangential statistic as a discrete sum over galaxy positions,
\begin{equation*}
    \Gamma_{+}(\theta) = \sum_{i} \tilde{\Psi}_{i}(\theta)\,\gamma_{+,i}(\theta),
\end{equation*}
the analysis in this section is conducted using a fine-grid that extends over a flat patch of the sky. In this equivalent formulation, we introduce a binary occupancy or density matrix $D_{ij}$ that is convolved with the normalized filter to yield the tangential statistic,
\begin{equation*}
\Gamma_+(\theta) = \sum_{ij} D_{ij}\tilde{\Psi}_{ij}(\theta)\,\gamma_{+,ij}(\theta).
\end{equation*}
To simplify the expression, we define the combined weight function $W_{ij}(\theta) = D_{ij} \cdot \tilde{\Psi}_{ij}(\theta)$ such that the tangential statistic takes the concise form:
\begin{equation*}
\Gamma_+(\theta) = \sum_{ij} W_{ij}(\theta)\,\gamma_{+,ij}(\theta),
\end{equation*}
where, 
\begin{equation*}
\gamma_{+,ij}(\theta) = -\cos(2\theta) \, \gamma_{1,ij} - \sin(2\theta) \, \gamma_{2,ij}.
\end{equation*}
We can define similar statistics corresponding to the $\gamma_1$ and $\gamma_2$ fields, 
\begin{equation}
\Gamma_a(\theta) = \sum_{ij} W_{ij}(\theta)\,\gamma_{a,ij} \, \qquad a\in\{1,2\}.
\label{eq:Gamma_a_decomp}
\end{equation}
Thus we obtain, 
\begin{equation*}
\Gamma_+(\theta) = -\cos(2\theta)\,\Gamma_1(\theta) - \sin(2\theta)\,\Gamma_2(\theta)
\end{equation*}
Since the LSS shear fields are zero-mean ($\langle \gamma_1 \rangle = \langle \gamma_2 \rangle = 0$), it follows that $\langle \Gamma_a(\theta)\rangle = 0$, and consequently $\langle \Gamma_+(\theta)\rangle = 0$. Hence, the variance of the tangential statistic can be written as:
\begin{equation}
\boxed{
\mathrm{Var}(\Gamma_+) = \cos^2(2\theta)\,\mathrm{Var}(\Gamma_1) + \sin^2(2\theta)\,\mathrm{Var}(\Gamma_2) + 2\cos(2\theta)\sin(2\theta)\,\mathrm{Cov}(\Gamma_1, \Gamma_2).
}
\label{eq:variance_Gamma+_decomp}
\end{equation}
The problem of estimating the variance then reduces to evaluating $\mathrm{Var}(\Gamma_1)$, $\mathrm{Var}(\Gamma_2)$, and $\mathrm{Cov}(\Gamma_1, \Gamma_2)$.

Using Eq.~(\ref{eq:Gamma_a_decomp}) and the assumption $\langle \gamma_a \rangle = 0$ we get, 
\begin{equation}
\mathrm{Var}(\Gamma_a) = \langle \Gamma_a^2\rangle = \sum_{ij}\sum_{kl}W_{\theta,ij}\,W_{\theta,kl}\,\langle\gamma_{a,ij}\,\gamma_{a,kl}\rangle, 
\label{eq:Var_Gamma_a_double_sum}
\end{equation}
where, $W_{\theta}$ is the compact notation for $W(\theta)$. For a homogeneous random field, we can express the two-point correlation function as the inverse Fourier transform of the 2-D power spectrum $P_{ab}(\boldsymbol\ell)$,
\begin{equation*}
\langle\gamma_a(\boldsymbol{x}_1)\,\gamma_b(\boldsymbol{x}_2)\rangle = \int \frac{d^2\ell}{(2\pi)^2}\,P_{ab}(\boldsymbol{\ell})\,e^{i\boldsymbol{\ell}\cdot(\boldsymbol{x}_1 - \boldsymbol{x}_2)}.
\end{equation*}
Substituting this into Eq.~(\ref{eq:Var_Gamma_a_double_sum}) allows the spatial sums to factorize into the discrete Fourier transform of the real filter weights, $\widetilde W_\theta(\boldsymbol{\ell}) = \sum_{ij} W_{\theta,ij}\,e^{-i\boldsymbol{\ell}\cdot\boldsymbol{x}_{ij}}$. This yields the variance of the statistic in Fourier space:
\begin{equation}
\mathrm{Var}(\Gamma_a) = \int \frac{d^2\ell}{(2\pi)^2}\,P_{aa}(\boldsymbol{\ell})\,|\widetilde{W}_\theta(\boldsymbol{\ell})|^2  \qquad a\in\{1,2\}
\label{eq:Var_Gamma_a_v1}
\end{equation}
Similarly, the covariance can be written as,
\begin{equation}
\mathrm{Cov}(\Gamma_1, \Gamma_2) = \int \frac{d^2\ell}{(2\pi)^2}\,P_{12}(\boldsymbol{\ell})\,|\widetilde{W}_\theta(\boldsymbol{\ell})|^2
\label{eq:Cov_Gamma_12_v1}
\end{equation}

The 2-D power spectra of the shear components, in terms of the standard E-mode and B-mode power spectra $C_\ell^{EE}, C_\ell^{BB}$, are
\begin{align}
P_{11}(\boldsymbol{\ell}) &= C_\ell^{EE}\cos^2(2\varphi_\ell) + C_\ell^{BB}\sin^2(2\varphi_\ell), \nonumber \\
P_{22}(\boldsymbol{\ell}) &= C_\ell^{EE}\sin^2(2\varphi_\ell) + C_\ell^{BB}\cos^2(2\varphi_\ell), \label{eq:2D_power_spectra} \\
P_{12}(\boldsymbol{\ell}) &= \left(C_\ell^{EE} - C_\ell^{BB}\right)\cos(2\varphi_\ell)\sin(2\varphi_\ell). \nonumber
\end{align}
where $\varphi_\ell = \tan^{-1}(\ell_y/\ell_x)$ is the polar angle of $\boldsymbol\ell$. This follows from the E/B decomposition in Fourier space:
\begin{align*}
\widetilde\gamma_1(\boldsymbol\ell) &= \cos(2\varphi_\ell)\widetilde E(\boldsymbol\ell) - \sin(2\varphi_\ell)\widetilde B(\boldsymbol\ell), \\ 
\widetilde\gamma_2(\boldsymbol\ell) &= \sin(2\varphi_\ell)\widetilde E(\boldsymbol\ell) + \cos(2\varphi_\ell)\widetilde B(\boldsymbol\ell),
\end{align*}
together with the statistical independence of E and B modes, $\langle\widetilde E\widetilde E^*\rangle = C_\ell^{EE}$, $\langle\widetilde B\widetilde B^*\rangle = C_\ell^{BB}$, $\langle\widetilde E\widetilde B^*\rangle = 0$.

Substituting Eq.~(\ref{eq:2D_power_spectra}) into Eq.~(\ref{eq:Var_Gamma_a_v1}) and Eq.~(\ref{eq:Cov_Gamma_12_v1}) we obtain,  
\begin{align*}
\text{Var}(\Gamma_1) &= \int \frac{d^2\ell}{(2\pi)^2} \left[C_\ell^{EE}\cos^2(2\varphi_\ell) + C_\ell^{BB}\sin^2(2\varphi_\ell)\right] |\widetilde{W}_\theta(\boldsymbol{\ell})|^2. \\ 
\text{Var}(\Gamma_2) &= \int \frac{d^2\ell}{(2\pi)^2} \left[C_\ell^{EE}\sin^2(2\varphi_\ell) + C_\ell^{BB}\cos^2(2\varphi_\ell)\right] |\widetilde{W}_\theta(\boldsymbol{\ell})|^2. \\ 
\text{Cov}(\Gamma_1, \Gamma_2) &= \int \frac{d^2\ell}{(2\pi)^2} \left[\left(C_\ell^{EE} - C_\ell^{BB}\right)\cos(2\varphi_\ell)\sin(2\varphi_\ell)\right] |\widetilde{W}_\theta(\boldsymbol{\ell})|^2.
\end{align*}
Using the trigonometric identities, 
\begin{equation*}
\cos^2(2\varphi) = \frac{1}{2} + \frac{1}{2}\cos(4\varphi), \ \
\sin^2(2\varphi) = \frac{1}{2} - \frac{1}{2}\cos(4\varphi), \ \  \sin(2\varphi)\sin(2\varphi) = \frac{1}{2} \sin(4\varphi), 
\end{equation*}
we get,
\begin{align*}
\mathrm{Var}(\Gamma_1) = \;& \frac{1}{2}\int \frac{d^2\ell}{(2\pi)^2} \left[C_\ell^{EE} + C_\ell^{BB}\right] |\widetilde{W}_\theta(\boldsymbol{\ell})|^2 + \frac{1}{2}\int \frac{d^2\ell}{(2\pi)^2} \left[C_\ell^{EE} - C_\ell^{BB}\right] \cos(4\varphi_\ell)\, |\widetilde{W}_\theta(\boldsymbol{\ell})|^2 \\ 
\mathrm{Var}(\Gamma_2) = \;& \frac{1}{2}\int \frac{d^2\ell}{(2\pi)^2} \left[C_\ell^{EE} + C_\ell^{BB}\right] |\widetilde{W}_\theta(\boldsymbol{\ell})|^2 - \frac{1}{2}\int \frac{d^2\ell}{(2\pi)^2} \left[C_\ell^{EE} - C_\ell^{BB}\right] \cos(4\varphi_\ell)\, |\widetilde{W}_\theta(\boldsymbol{\ell})|^2 \\ 
\mathrm{Cov}(\Gamma_1, \Gamma_2) = \;& \frac{1}{2}\int \frac{d^2\ell}{(2\pi)^2} \left[C_\ell^{EE} - C_\ell^{BB}\right] \sin(4\varphi_\ell)\, |\widetilde{W}_\theta(\boldsymbol{\ell})|^2
\end{align*}

To make the expressions less cumbersome we introduce three helper integrals,
\begin{align*}
V_{\mathrm{sum}}  &= \int \frac{d^2\ell}{(2\pi)^2} \left[C_\ell^{EE} + C_\ell^{BB}\right] |\widetilde{W}_\theta(\boldsymbol{\ell})|^2, \\
V_{\mathrm{diff}} &= \int \frac{d^2\ell}{(2\pi)^2} \left[C_\ell^{EE} - C_\ell^{BB}\right] \cos(4\varphi_\ell)\, |\widetilde{W}_\theta(\boldsymbol{\ell})|^2, \\
V_{\mathrm{cross}} &= \int \frac{d^2\ell}{(2\pi)^2} \left[C_\ell^{EE} - C_\ell^{BB}\right] \sin(4\varphi_\ell)\, |\widetilde{W}_\theta(\boldsymbol{\ell})|^2.
\label{eq:V_definitions}
\end{align*}
Thus,
\begin{equation}
\mathrm{Var}(\Gamma_1) = \frac{1}{2}(V_{\mathrm{sum}} + V_{\mathrm{diff}}), \qquad
\mathrm{Var}(\Gamma_2) = \frac{1}{2}(V_{\mathrm{sum}} - V_{\mathrm{diff}}), \qquad
\mathrm{Cov}(\Gamma_1, \Gamma_2) = \frac{1}{2}V_{\mathrm{cross}}.
\label{eq:variances_in_I}
\end{equation}

Plugging Eq.~(\ref{eq:variances_in_I}) back into Eq.~(\ref{eq:variance_Gamma+_decomp}) gives,
\begin{equation*}
\text{Var}(\Gamma_+) = \cos^2(2\theta) \left(\frac{V_{\mathrm{sum}} + V_{\mathrm{diff}}}{2}\right) + \sin^2(2\theta)\left(\frac{V_{\mathrm{sum}} - V_{\mathrm{diff}}}{2}\right) + 2\cos(2\theta)\sin(2\theta) \left(\frac{V_{\mathrm{cross}}}{2}\right).
\end{equation*}

Which finally gives the expression: 
\begin{equation}
\boxed{\mathrm{Var}(\Gamma_+(\theta)) = \frac{1}{2}\left[V_{\mathrm{sum}}(\theta) + \cos(4\theta)\,V_{\mathrm{diff}}(\theta) + \sin(4\theta)\,V_{\mathrm{cross}}(\theta)\right]}.
\label{eq:Var_Gamma_plus_final}
\end{equation}

\begin{figure*}[htb!]
\gridline{\fig{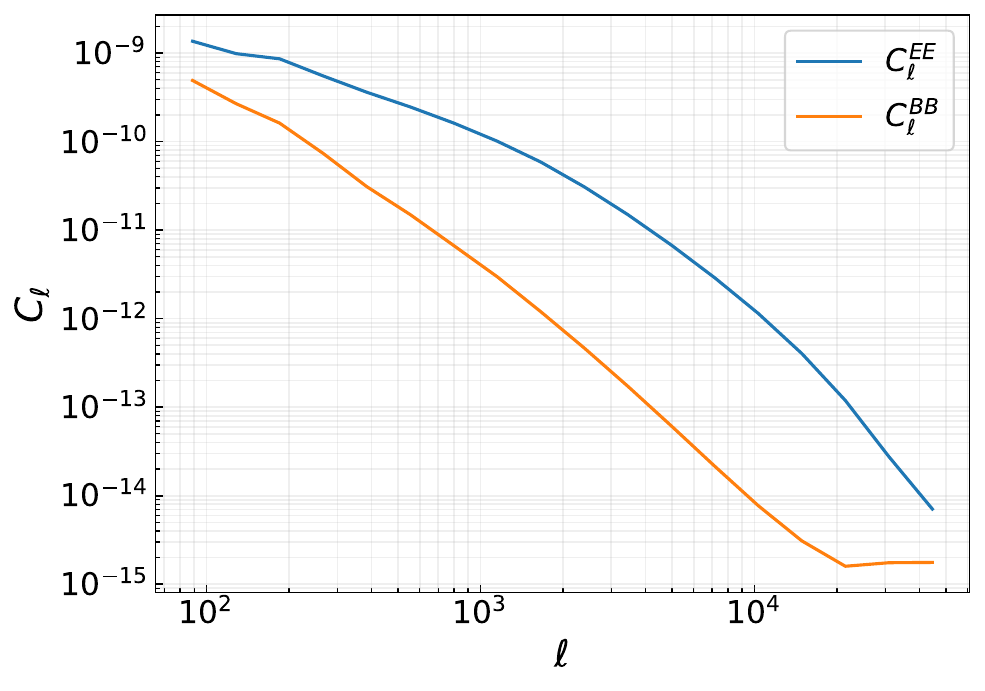}{0.44\textwidth}{(a)}
         \fig{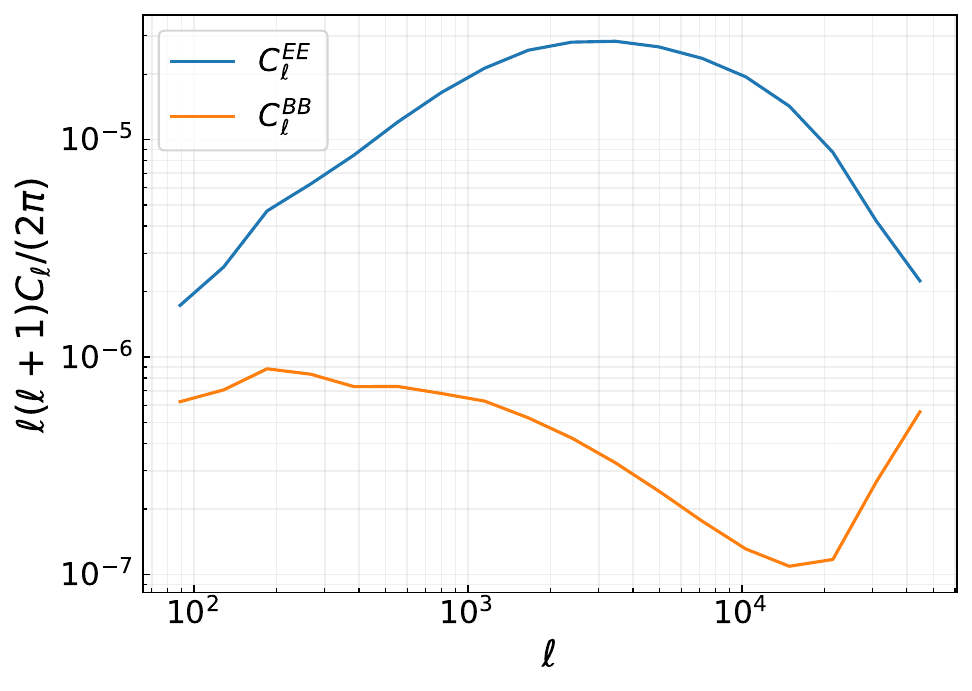}{0.44\textwidth}{(b)}}
\caption{E-mode and B-mode shear power spectra measured from 250 realizations of the mock $\kappa$TNG shear maps corresponding to a source redshift of $z_s = 0.51$. \textit{Left:} (a) Standard power spectra $C_\ell^{EE}$ (blue) and $C_\ell^{BB}$ (orange) as a function of multipole moment $\ell$. The B-mode amplitude is suppressed by up to two orders of magnitude relative to the E-mode across the full range of scales probed. \textit{Right:} (b) Corresponding dimensionless power spectra $\ell(\ell+1)C_\ell/(2\pi)$.}
\label{fig:ktng_power_spectrum}
\end{figure*}

To implement the expression in Equation~(\ref{eq:Var_Gamma_plus_final}) and evaluate the uncertainty of the tangential statistic ($\sigma_+ = \sqrt{Var(\Gamma_+)}$) contributed by the large-scale structure, we require the corresponding E-mode and B-mode shear power spectra $C_\ell^{EE}$ and $C_\ell^{BB}$. We measure the average power spectra across 250 realizations of the mock $\kappa$TNG shear maps (source redshift $z_s = 0.51$) used for numerical noise estimation in Section~\ref{subsubsec:lssnoise}. Figure~\ref{fig:ktng_power_spectrum}(a) and Figure~\ref{fig:ktng_power_spectrum}(b) show the measured E-mode and B-mode shear power spectra in their standard and dimensionless forms, respectively. As expected for weak gravitational lensing, the B-mode amplitude is suppressed by up to two orders of magnitude relative to the E-mode. We find that the residual B-mode power, however, is not negligible for our analysis, and setting $C_\ell^{BB} = 0$ leads to a systematic underestimate of $\sigma_+(\theta)$ by $\sim20\%$. Therefore, we retain both E and B-mode spectra throughout the variance calculation.

\begin{figure*}[htb!]
\gridline{\fig{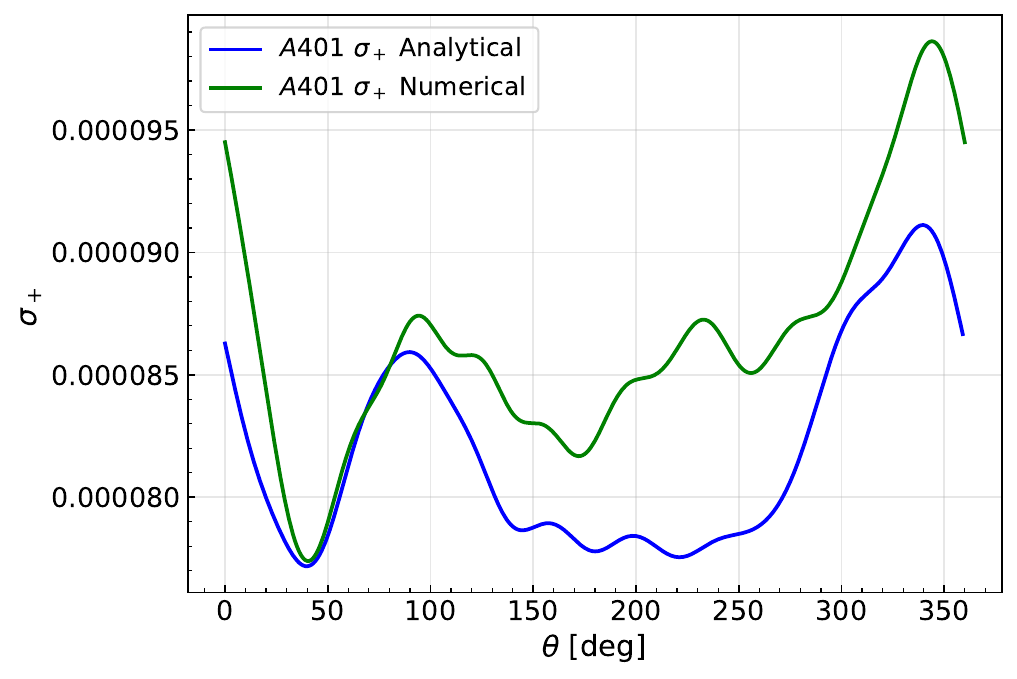}{0.49\textwidth}{(a)}
         \fig{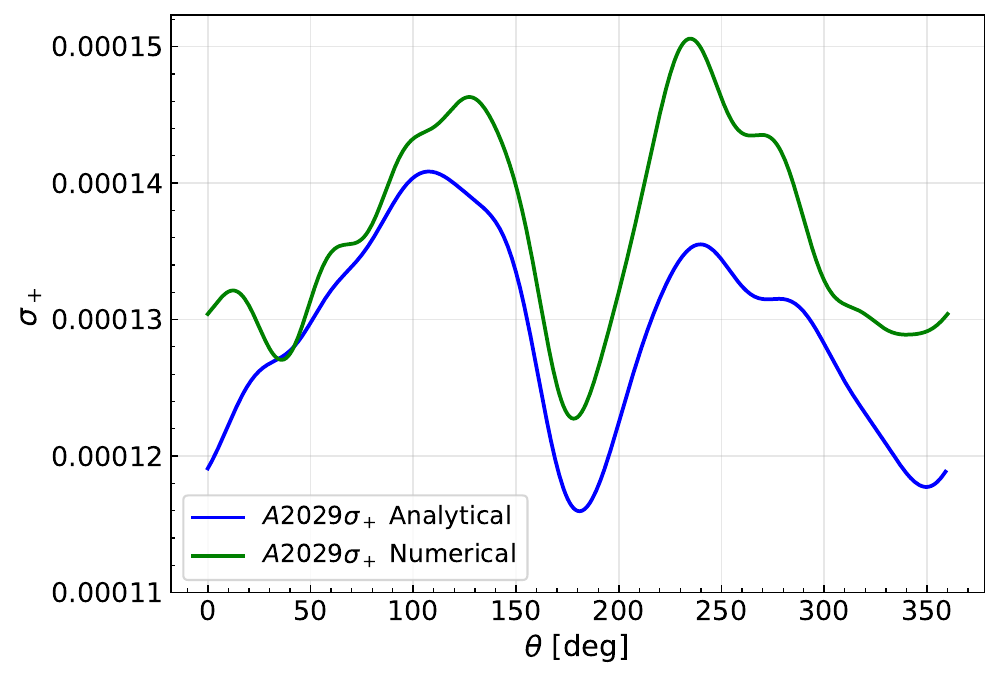}{0.49\textwidth}{(b)}}
\caption{Comparison of the analytical and numerical estimates of the LSS-induced uncertainty $\sigma_{\mathrm{+,\ LSS}}(\theta)$ as a function of filter angle $\theta$ for (a) Abell 401 ($n_g = 7\ \mathrm{arcmin}^{-2}$) and (b) Abell 2029 ($n_g = 12\ \mathrm{arcmin}^{-2}$). The analytical estimate (blue), derived from the E-mode and B-mode shear power spectra via Eq.~(\ref{eq:Var_Gamma_plus_final}), and the numerical estimate (green), obtained from mock $\kappa$TNG shear catalogs (See Section~\ref{subsubsec:lssnoise}), agree to within $\lesssim 5\%$ across the full range of filter angles for both systems. The small systematic offset, with the analytical estimate being lower than the numerical estimate, is attributed to non-Gaussianity in the LSS shear fields that is not captured by the two-point power spectrum.}
\label{fig:lss_analytical}
\end{figure*}

With the power spectra measured, we can calculate the helper integrals $V_{\mathrm{sum}}(\theta)$, $V_{\mathrm{diff}}(\theta)$, and  $V_{\mathrm{cross}}(\theta)$ at each filter angle $\theta$ and thereby obtain the LSS uncertainty $\sigma_{+,\mathrm{LSS}}(\theta)$. Here we present the results for the Abell 401 and Abell 2029 systems. Figure~\ref{fig:lss_analytical} compares the analytical uncertainty estimate to the numerical estimate from Section~\ref{subsubsec:lssnoise}. On average, the two approaches agree to within $\lesssim 5\%$ across the full range of filter angles for both systems, despite their differing source densities and survey geometries ($n_{g, \, \mathrm{A401}} = 7 \ \mathrm{arcmin}^{-2}$ and $n_{g, \, \mathrm{A2029}} = 12 \ \mathrm{arcmin}^{-2}$). However, we find that the analytical estimate is generally slightly lower than the numerical estimate. We attribute this difference to the non-Gaussianity of the LSS shear fields, which is not fully captured by the two-point power spectrum $C_\ell^{EE}$. Although we defer a more detailed analysis of this effect to future work, it is important to note that the discrepancy between the two estimates is minimal in comparison to the shape noise contribution $\sigma_{\mathrm{shape}}(\theta)$, which is on average $>50\%$ larger than the LSS contribution. In conclusion, the close agreement between the analytical and numerical estimates provides an independent validation of the variance budget reported in Section~\ref{subsubsec:lssnoise} and further strengthens our results.

\section{Radial Cutoffs}\label{sec:appendix_c}

\begin{deluxetable*}{l c c c c c | c c | c c}[htb!]
\tabletypesize{\small}
\tablecaption{Radial cutoffs evaluated from tangential shear profiles to exclude cluster-induced shear contamination. \label{tab:radial_cutoffs_table}}
\tablehead{
\colhead{System} & \colhead{Role} & \colhead{Cluster} &
\colhead{$M_{\mathrm{WL200c}}$} &
\colhead{$r_t$} & \colhead{$R_i$} & \colhead{$r_1$} & \colhead{$r_2$} & \colhead{$r_1$} & \colhead{$r_2$} \\
\colhead{} & \colhead{} & \colhead{} &
\colhead{($10^{14}\,\mathrm{M}_\odot$)} &
\colhead{(Mpc)} & \colhead{(Mpc)} & \colhead{(Mpc)} & \colhead{(Mpc)} & \colhead{(deg)} & \colhead{(deg)}
}
\startdata
\multirow{2}{*}{1. Abell 401} & p & Abell 401 & $9.2$ & $0.92$ & $\cdot$ & \multirow{2}{*}{$0.92$} & \multirow{2}{*}{$2.58$} & \multirow{2}{*}{$0.18$} & \multirow{2}{*}{$0.52$} \\
& s & Abell 399 & 4.2 & $0.49$ & $3.07$ & & & & \\
\hline
\multirow{3}{*}{2. Abell 2029} & p & Abell 2029 & $9.8$ & $0.98$ & $\cdot$ & \multirow{3}{*}{$0.98$} & \multirow{3}{*}{$2.65$} & \multirow{3}{*}{$0.19$} & \multirow{3}{*}{$0.51$} \\
 & s & Abell 2033 & $4.1$ & $0.51$ & $3.16$ & & & & \\
 & s & SIG        & $0.4$ & $-$ & $2.16$ & & & & \\
\hline
\multirow{4}{*}{3. Abell 3558 (SSC)} & p & Abell 3558 & $5.6$ & $0.36$ & $\cdot$ & \multirow{4}{*}{$0.36$} & \multirow{4}{*}{$2.56$} & \multirow{4}{*}{$0.11$} & \multirow{4}{*}{$0.76$} \\
 & s & Abell 3556 & $3.9$ & $0.25$ & $2.81$ & & & & \\
 & s & Abell 3562 & $1.9$ & $0.09$ & $4.09$ & & & & \\
 & s & Abell 3560 & $1.0$ & $-$ & $6.37$ & & & & \\
\hline
4. Abell 2351 (Control) & p & Abell 2351 & $2.3$ & $0.40$ & $\cdot$ & $0.40$ & $3.57^{*}$ & $0.07$ & $0.60$ \\
\enddata
\tablenotetext{}{\footnotesize $M_{\mathrm{WL200c}}$ = weak lensing mass derived from best-fit NFW profile, p = primary cluster, s = secondary cluster, --- = shear profile does not exceed threshold $\gamma_t = 0.02$, $\cdot$ = Not Applicable, $*$ = arbitrarily chosen due to the absence of a secondary cluster constraint.}
\end{deluxetable*}

\begin{figure*}[htb!]
\gridline{
    \fig{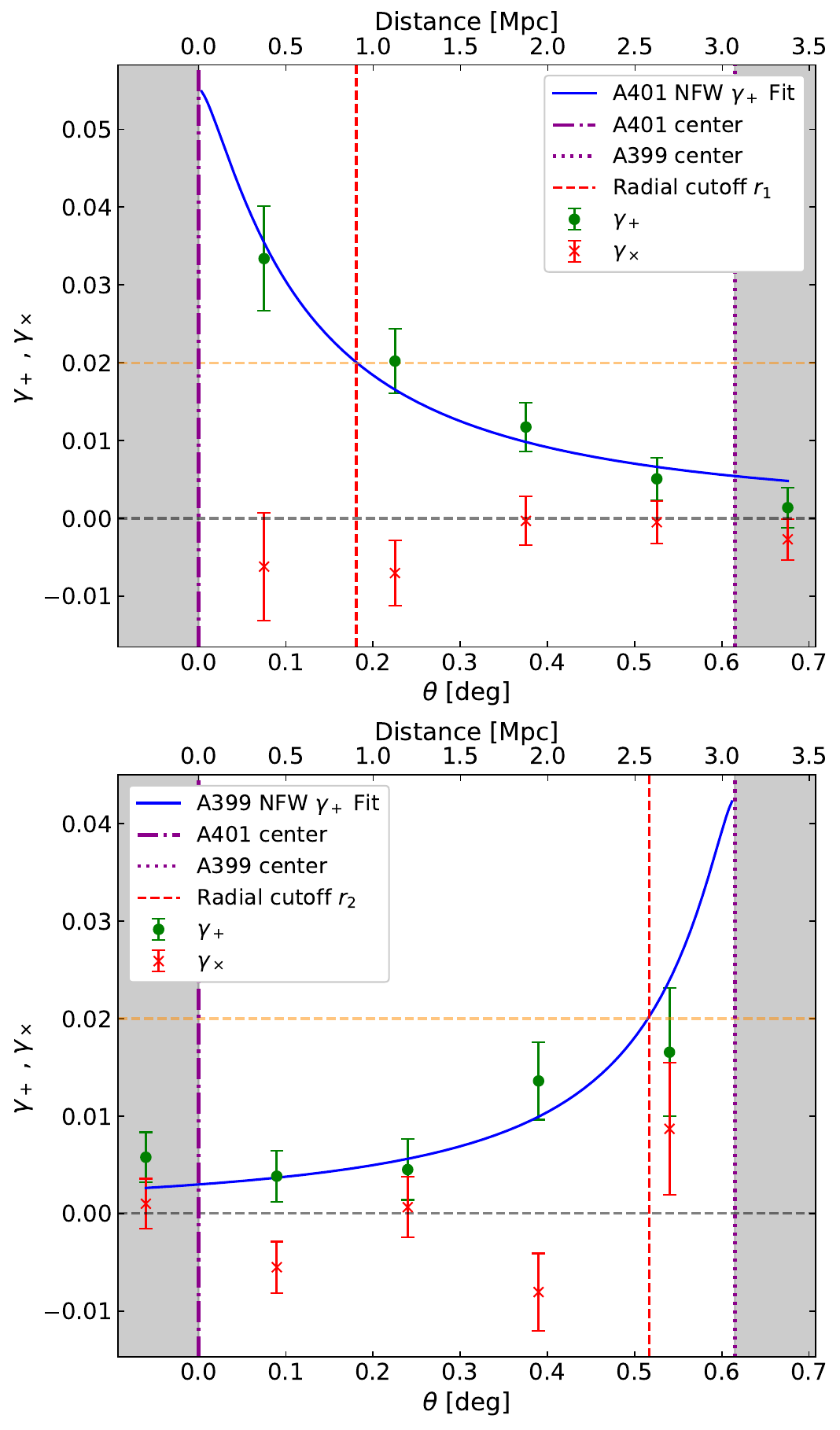}{0.325\textwidth}{(a) A401 - A399}
    \fig{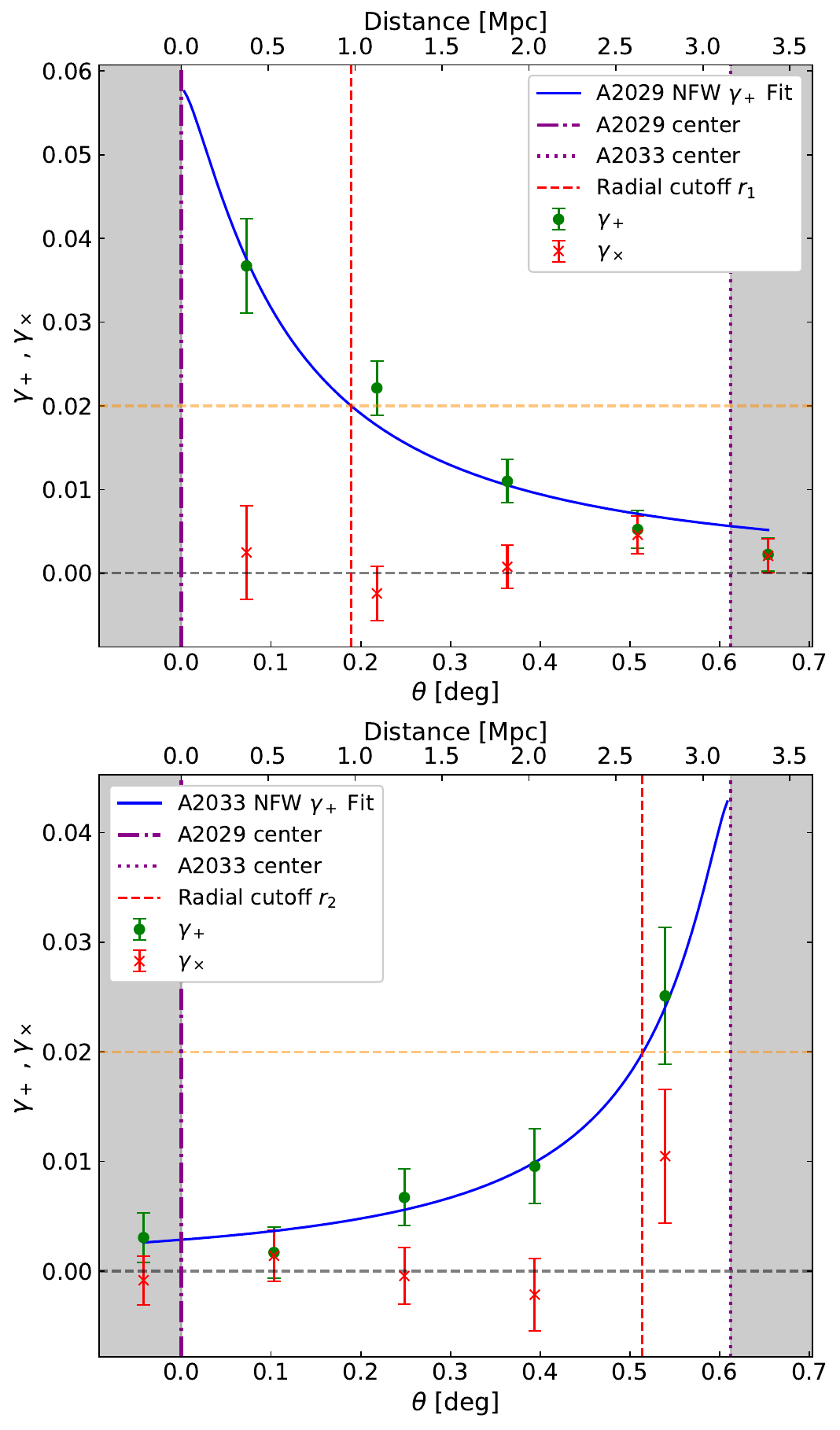}{0.325\textwidth}{(b) A2029 - A2033}
    \fig{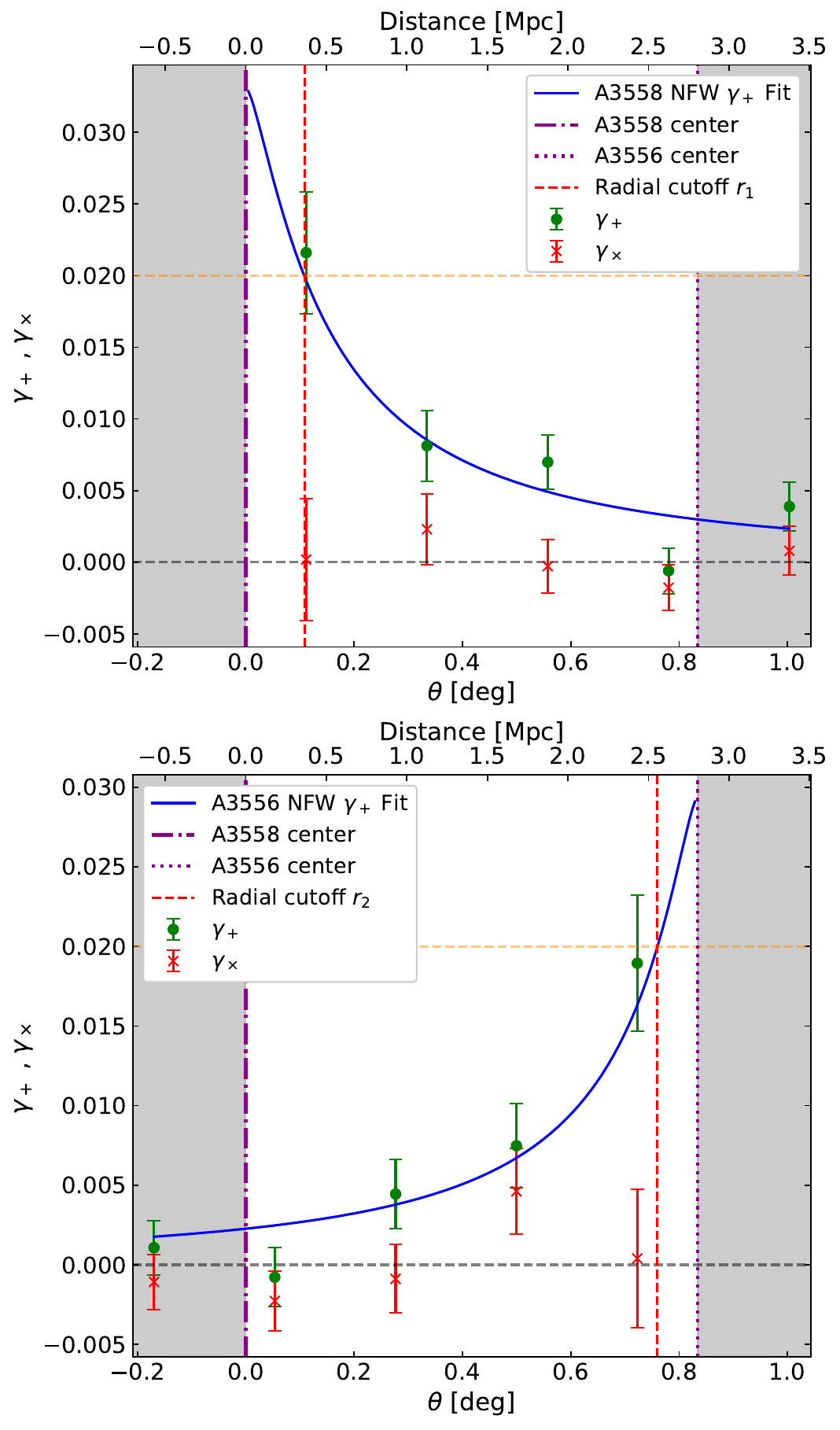}{0.325\textwidth}{(c) A3558 - A3556}
    }
\gridline{
    \fig{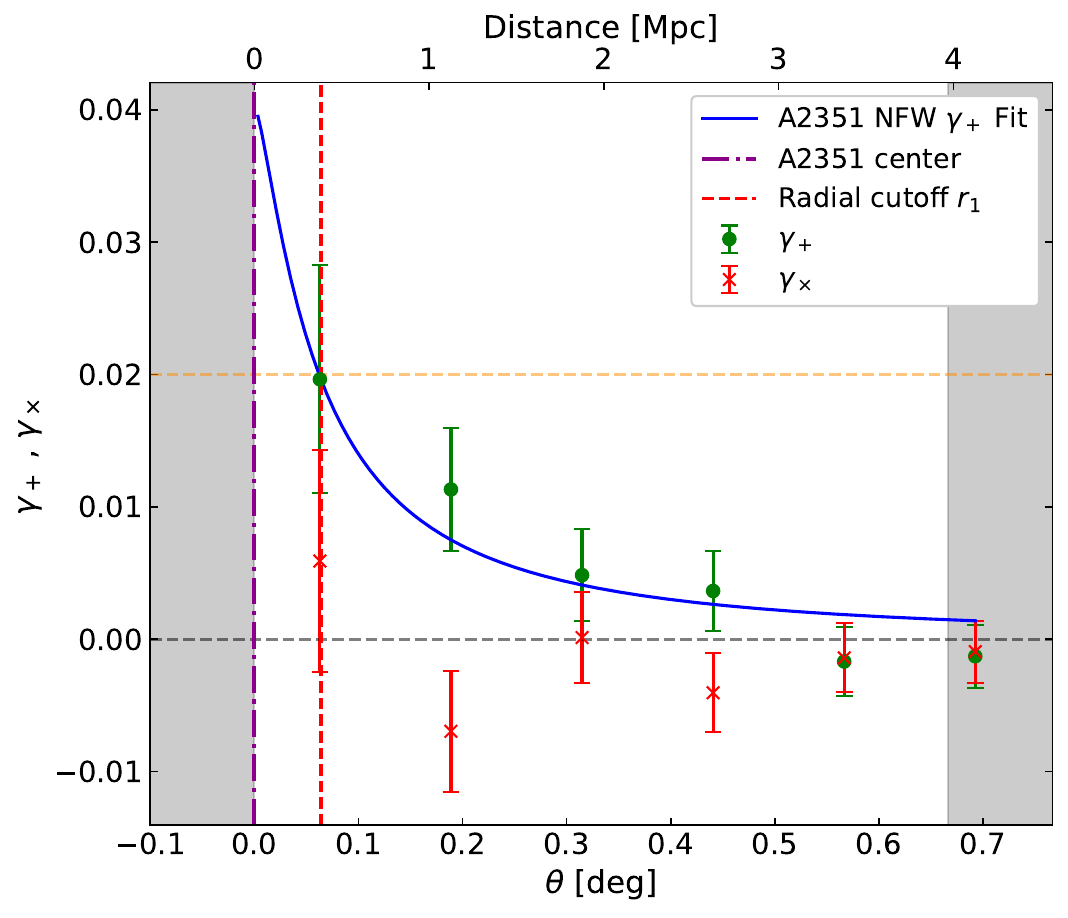}{0.33\textwidth}{(d) A2351}
}
\caption{Binned shear profiles ($\gamma_+$, green) and cross shear profiles ($\gamma_\times$, red) for dominant clusters in each system, used to determine the radial cutoffs $r_1$ and $r_2$. Each pair of panels shows the shear profiles relative to the primary cluster (top) and the relevant secondary cluster (middle) for the systems: (a) A401-A399 (b) A2029-A2033 and (c) A3558-A3556. The shear profiles for (d) A2351 are separately showcased in the bottom panel. In each panel, the solid blue curve represents the NFW fit to the binned tangential shear profile. The vertical purple dashdotted and dotted lines indicate the primary and secondary cluster centers, respectively, and the orange horizontal dashed line marks the shear threshold $\gamma_+ = 0.02$ used to define the radial cutoffs. The radial cutoffs $r_1$ (top panel) and $r_2$ (middle panels) are indicated by vertical red dashed lines, and the shaded grey shaded regions mark the boundary of the intercluster space between the primary and secondary clusters. The lower $x$-axis represents the angular separation (in deg) from the primary cluster while the  upper $x$-axis shows the corresponding physical distance (in Mpc) at the cluster redshift.}
\label{fig:shear_profiles}
\end{figure*}

Shear contamination from the primary and secondary clusters can bias the inferred filament signal. We mitigate this effect by imposing radial cutoffs $r_1$ and $r_2$ on the matched filter to exclude shear-dominated regions near the clusters. We determine the radial cutoffs by evaluating binned tangential shear profiles for all clusters present in the field, fitting them with an NFW shear model, and requiring that the fitted shear profile remains below the threshold $\gamma_+ < 0.02$ within the search space. For each cluster, we first decompose the shear field $(\gamma_1, \gamma_2)$ about the cluster center into its tangential and cross components $(\gamma_t, \gamma_\times)$. We then bin the decomposed shear field in uniform annuli of width $750~\mathrm{kpc}$ centered on the relevant cluster and compute the mean tangential and cross shear values in each bin along with their associated uncertainties. Setting the annular bin width to $750~\mathrm{kpc}$ also achieves the purpose of excluding the unstable, fast-varying core region where the shear data may diverge from the NFW model. The resulting binned tangential shear profile is fit with an NFW model characterized by the mass $M_{\mathrm{200c}}$, redshift $z = 0.51$ (corresponding to the effective source redshift adopted in this study), and concentration parameter $c = 4$. With tangential shear profiles determined for all clusters, we define the minimum threshold radius $r_{\mathrm{t}}$ as the radial distance from the cluster center at which the fitted profile reaches the threshold $\gamma_+ = 0.02$. The primary cluster determines the inner cutoff $r_1$, while the secondary clusters determine the outer cutoff $r_2$. Let $R_i$ be the distance to the $i$-th secondary cluster from the primary cluster. The cutoffs are then defined as 
\begin{align*}
    r_1 &= r_{\mathrm{t},\mathrm{p}} \ , \\ 
    r_2 &= \min_{i}\{R_i - r_{\mathrm{t}, i}\},
\end{align*}
where $r_{\mathrm{t},\mathrm{p}}$ and $r_{\mathrm{t}, i}$ are the threshold radii for the primary and  $i$-th secondary cluster, respectively. As a consequence, only the closest and most massive clusters in each system determine the radial cutoffs. Table \ref{tab:radial_cutoffs_table} summarizes the evaluated cutoff radii $r_1$ and $r_2$ for all systems studied in the paper, along with all associated quantities. Figure \ref{fig:shear_profiles} illustrates the tangential and cross shear profiles for the dominant clusters in each system that eventually determine the radial cutoffs $r_1$ and $r_2$ and constrain the optimal filter used for filament detection.

\bibliography{citations}{}
\bibliographystyle{aasjournalv7}



\end{document}